\documentclass[twocolumn]{aastex631}

\usepackage{amsmath}

\newcommand\clm{}

\revised{\today}
\shorttitle{Bubble Overlap in a Protocluster}
\shortauthors{Martin et al.}

\newcommand{\et}{{\rm et\thinspace al.}\ }   

\newcommand{\xion}{\mbox {$\xi_{ion}$}}

\newcommand{\xh}{\mbox {$X_{HI}$} }
\newcommand{\barxh}{\mbox {$\bar{X}_{HI}$} }
\newcommand{\dv}{\mbox {$\Delta v(Ly\alpha)$}}

\newcommand{\oiii}{\mbox {[O III]}}

\newcommand{\oii}{\mbox {[O II]}}

\newcommand{\Ha}{\mbox {H$\alpha$}}
\newcommand{\Hb}{\mbox {H$\beta$}}
\newcommand{\Hg}{\mbox {H$\gamma$}}

\newcommand{\fism}{\mbox {$f_{ism}^{Ly\alpha}$}}
\newcommand{\fesclya}{\mbox {$f_{esc}^{Ly\alpha}$}}
\newcommand{\fesc}{\mbox {$f_{esc}$}}
\newcommand{\lya}{\mbox {Ly$\alpha$}}

\newcommand{\kms}{\mbox {$ {\rm km~s}^{-1}$}}

\newcommand{\msun}{\mbox {$ {\rm ~M_\odot}$}}
\newcommand{\zsun}{\mbox {$ {\rm ~Z_\odot}$}}

\newcommand{\msunyr}{\mbox {$ {\rm ~M_\odot}$~yr$^{-1}$}}
\newcommand{\Muv}{\mbox {$ {\rm ~M_{UV}}$}}

\newcommand{\um}{\mbox {$ \mu {\rm m}$}}
\begin{document}

\title{Galaxy Protoclusters as Drivers of Cosmic Reionization:  \\
I. Bubble Overlap at Redshift $z \sim 7$ in LAGER-z7OD1
}

\correspondingauthor{Crystal Martin}
\email{cmartin@ucsb.edu}

\author[0000-0001-9189-7818]{Crystal L. Martin}
\affiliation{Department of Physics,
University of California Santa Barbara,
Santa Barbara, CA 93106, USA }

\author{Weida Hu}
\affiliation{Department of Physics and Astronomy,
Texas A \& M University,
College Station, TX 77843-4242, USA}

\author{Isak G. B. Wold} 
\affiliation{Goddard Space Flight Center,
Greenbelt, MD 20771, USA}

\author{Andreas Faisst}
\affiliation{Caltech/IPAC,
MS314-6,
1200 E. California Boulevard,
Pasadena, CA 91125, USA }

\author{Crist\'{o}bal Moya-Sierralta} 
\altaffiliation{Las Campanas Observatory, 
Carnegie Institution for Science, 
Ra\'{u}l Bitr\'{a}n 1200, La Serena, Chile}
\affiliation{Instituto de Astrofisica,
Pontificia Universidad Cat\'{o}lica de Chile,
Santiago 7820436, Chile}

\author{Sangeeta Malhotra} 
\affiliation{Goddard Space Flight Center,
Greenbelt, MD 20771, USA}

\author{James E. Rhoads} 
\affiliation{Goddard Space Flight Center,
Greenbelt, MD 20771, USA}

\author{Luis Felipe Barrientos} 
\affiliation{Instituto de Astrofisica Facultad de F\'{i}sica,
Pontificia Universidad Cat\'{o}lica de Chile,
Santiago 7820436, Chile}

\author[0000-0002-6047-430X]{Yuichi Harikane}
\affiliation{
Institute for Cosmic Ray Research, 
The University of Tokyo, 
5-1-5 Kashiwanoha, Kashiwa, 
Chiba 277-8582, Japan}

\author{Leopoldo Infante}
\altaffiliation{Las Campanas Observatory, 
Carnegie Institution for Science, 
Ra\'{u}l Bitr\'{a}n 1200, La Serena, Chile}
\affiliation{Instituto de Astrofisica Facultad de F\'{i}sica,
Pontificia Universidad Cat\'{o}lica de Chile,
Santiago 7820436, Chile}

\author[0000-0002-6610-2048]{Anton M. Koekemoer}
\affiliation{
Space Telescope Science Institute, 3700 San Martin Drive,
Baltimore, MD 21218, USA
}

\author{Jorge Gonzalez Lopez}
\affiliation{
Carnegie Institution of Washington
}

\author{Masami Ouchi}
\affiliation{
Institute for Cosmic Ray Research, 
The University of Tokyo, 
5-1-5 Kashiwanoha, Kashiwa, 
Chiba 277-8582, Japan
}

\author{Junyan Xu} 
\affiliation{Department of Physics,
University of California Santa Barbara,
Santa Barbara, CA 93106, USA }

\author{Jiayang Yang} 
\affiliation{Department of Physics,
University of California Santa Barbara,
Santa Barbara, CA 93106, USA }

\author[0000-0003-3466-035X]{{L. Y. Aaron} {Yung}}
\affiliation{Space Telescope Science Institute, 
3700 San Martin Drive, Baltimore, MD 21218, USA}

\author[0000-0003-1614-196X]{John R. Weaver}\thanks{Brinson Prize Fellow}
\affiliation{MIT Kavli Institute for Astrophysics and Space Research, 70 Vassar Street, Cambridge, MA 02139, USA}

\author{Henry McCracken}
\affiliation{
CNRS, Institut d'Astrophysique de Paris
}

\author{Zhenya Zheng}
\affiliation{
Chinese Academy of Sciences
}

\author{Junxian Wang}
\affiliation{
Department of Astronomy, 
University of Science and Technology of China, 
Hefei, Anhui 230026, PR China}

\email{cmartin@ucsb.edu,
weidahu@tamu.edu,
isak.wold@gmail.com,
afaisst@ipac.caltech.edu,
cmoya@carnegiescience.edu,
Sangeeta.Malhotra@nasa.gov,
James.E.Rhoads@nasa.gov,
barrientos@astro.puc.cl,
hari@icrr.u-tokyo.ac.jp,
linfante@carnegiescience.edu,
koekemoer@stsci.edu,
jgl@uc.cl,
ouchims@icrr.u-tokyo.ac.jp,
junyan@ucsb.edu,
jiayangyang@ucsb.edu,
yung@stsci.edu,
john.weaver.astro@gmail.com,
hjmcc@iap.fr,
zhengzy@shao.ac.cn,
jxw@ustc.edu.cn 
}



\begin{abstract}
{\clm
Since the launch of JWST, the sample size of reionization-era \lya-emitters (LAEs) has been
steadily growing; yet inferences about the neutral hydrogen fraction in the intergalactic medium exhibit
increasing variance at redshift $z \approx 7$, possibly indicating significant field-to-field
fluctuations in the progression of cosmic reionization. In this paper, we present new JWST/NIRSpec
and Keck/LRIS spectra of nine LAEs in the redshift $z \sim 7$ protocluster, LAGER-z7OD1.  
Measurements of \lya-transmission and \lya-velocity offset along multiple sightlines map
the \lya-damping wing optical depth across the galaxy overdensity. In the standard context
of inside-out ionization, we estimate radii of ionized bubbles, $R_i^{min} = 0.07 - $  0.69~Mpc (physical),
based on the distance from each LAE to the first neutral patch along the sightline.  The resulting 
3D topology reveals three distinct sub-clusters where the ionized bubbles are approaching overlap. 
Five of the nine LAEs plausibly ionized their bubbles, a few bursts of star formation and a
modest escape fraction are sufficient. We demonstrate, however, that the actual ionized
volumes are likely larger, at least $R_i^{ism} = 0.42 - $  1.29 Mpc (physical), based on an empirical
model for interstellar attenuation of \lya. Modeling galactic attenuation of \lya\ significantly 
increases the inferred intergalactic transmission (thus enlarging the ionized pathlength). 
The errorbars on the reddening correction allow fully overlapping bubbles, and our
results are consistent with accelerated reionization in the protocluster.
}
\end{abstract}

\keywords{
--- High-redshift galaxies (734)
--- Lyman-alpha galaxies (978)
--- Galaxy formation (595)
--- Galaxy encounters (592)
--- Galaxy interactions (600)
--- Interacting galaxies (802)
--- Galaxy mergers (608)
--- Disk galaxies (391)
--- Cosmological phase transitions (342)
}

\section{Introduction} \label{sec:intro}

The first 380,000 years of cosmic expansion cooled the intergalactic medium (IGM) 
sufficiently for hydrogen atoms to recombine \citep{Planck2020}. Later, the 
emergence of luminous objects in the universe fully ionized the IGM. 
This transition was already underway when the universe was just 300~Myr 
old \citep{Witstok2024-z13} yet not quite complete 700~Myr later at redshift 
$z \sim 5.7$ \citep{Becker2015}. Understanding how reionization occurred provides 
direct insight into the early growth of galaxies and supermassive black holes 
\citep{Robertson2022, Madau2024}.

The Lyman-$\alpha$ line of hydrogen ($\lambda = 1215.67$ \AA; hereafter \lya) is an 
important probe of the transition from a neutral IGM to a mostly ionized IGM 
\citep{Malhotra2006}.  High 
equivalent-width \lya\ emission emerges from low mass, low metallicity galaxies 
which leak ionizing radiation \citep{Izotov2024}. During cosmic reionization, 
neutral hydrogen in the IGM scatters the resonant photons, decreasing the \lya\ 
equivalent widths of galaxies \citep{Loeb1999}. Well before the launch of the 
James Webb Space Telescope (JWST), declining detection rates of \lya\ 
emission in spectra of Lyman-break galaxies \citep{Pentericci2014}, the evolution 
of the \lya\ luminosity function \citep{Ouchi2010}, and the clustering of \lya\ 
emitters \citep{Ouchi2018}, all indicated an increasing fraction of neutral hydrogen 
in the $z > 5$ IGM. 

By $z \approx 7$,  the reduction in the \lya\ equivalent width distribution suggests an
average neutral hydrogen fraction $\barxh = 0.59^{+0.11}_{-0.15}$ \citep{Mason2018a}. 
A volume-averaged fraction of neutral hydrogen, $\barxh \approx  0.5 $, is
in remarkable agreement with the evolution of the \lya\ forest \citep{Greig2017},
and the electron-scattering optical depth to the cosmic microwave background \citep{Planck2020}. 
{\clm
Now, with JWST, \lya\ emission has been detected out to redshift $z \approx 14$ 
\citep{Witstok2024-z13,Tang2025-z13}. 
}
The late and sharp brightening of both the \lya\ luminosity function \citep{Kageura2025}, and the
\lya\ equivalent width distribution \citep{Nakane2024,Tang2024,Jones2024,Jones2025},  are not easily
explained by numerous, faint galaxies \citep{Finkelstein2019} or rare, luminous galaxies \citep{Naidu2020}.

Interestingly, as the number of neutral fraction measurements grows, the variance in \barxh\
has been increasing between redshift 6.7 and 7.3; see \citet[][Figure 11]{Wold2022}, \citet{Bolan2022},
\citet[][Figure 15]{Nakane2024}, \citet[][Figure 10]{Umeda2024}, and \citet[][Figure 11]{Kageura2025}.
Large field-to-field fluctuations over this short ($ \sim 100$ Myr) transition period
likely indicate signicant differences in the ionized pockets. Mapping ionized bubbles is of particular
interest during this period of rapid evolution because it allows identification of the ionizing
agents (inside the ionized regions).

Transmission of \lya\ emission during reionization requires locally ionized regions around galaxies
\citep{Haiman2002,Gnedin2004,Dijkstra2014}. In the pre-overlap phase, reionization models predict the 
emergence of ionized bubbles 
around galaxies which leak Lyman continuum (LyC) photons \citep[][]{Furlanetto2004,Iliev2006}. While the 
global neutral fraction remains high, the largest ionized bubbles grow around galaxy overdensities 
\citep[][]{Wyithe2007,McQuinn2007,Sobacchi2015,Lu2024}. The accelerated assembly of galaxies in
protoclusters, the most overdense regions in the universe, may therefore drive the timing and
topology of cosmic reionization \citep{Chiang2017}.  Bubble overlap eventually triggers rapid growth in 
the volume of ionized pockets, a process known as percolation \citep[][and references therein]{Neyer2024}.
Throughout this process the IGM outside these ionized pockets maintains a high neutral fraction due to
the small mean free path of one Rydberg photons \citep{Verner1996,Gnedin_Madau2022}.

The Lyman Alpha Galaxies in the Epoch of Reionization, LAGER, survey built a statistical
sample of LAEs over the $\sim 3$~deg$^2$ COSMOS field \citep{Zheng2017}. A custom 
narrowband filter captured \lya\ emission from galaxies in a narrow redshift slice 
(30 to 45 comoving Mpc deep) at redshift 6.9.  LAGER-z7OD1 is the largest photometrically 
identified overdensity \citep{Hu2021}. In this paper, we measure the sizes of the ionized
bubbles in LAGER-z7OD1 and discuss their ionizing agents.

We present new NIRSpec spectroscopy covering rest-frame optical emission lines and extract
\lya\ line profiles from new ground-based spectroscopy. The near-infrared sensitivity of JWST
detects hydrogen Balmer lines (and bright forbidden lines) from 9 galaxies in the overdensity.
These emission lines define the systemic redshift, allowing us to measure the \lya velocity 
offset. We measure the emergent \lya\ escape fraction from \Ha\ (or \Hb) luminosities using 
recombination theory. Our analysis is the first study to simultaneously consider aperture 
corrections reddening corrections, and interstellar transmission in the reionization-era. 
The resolved \lya\ line profiles, broad optical emission-line wings, gas-phase metallicity,
and galaxy morphologies are the focus of additional  papers in the series.

In this paper, we estimate the distance to the first neutral patch along our sightline from the intergalactic 
attenuation of the \lya\ emission emergent from each galaxy \citep{Haiman2002,Mason2020}. 
This is possible because the expansion of the universe redshifts the \lya\ line profile which 
emerges from a galaxy. A frequency shift away from the resonance, and into the  damping wing of 
the  line profile (in the frame of the intergalactic hydrogen atoms), strongly decreases the 
scattering cross section for \lya\ photons. Differentiating intergalactic and interstellar 
attenuation of \lya\ poses a significant challenge which we address using the nearest local analogs.
We interpret these pathlengths as the radii of ionized bubbles centered on the LAEs, discuss
whether each galaxy can ionize its bubble, and examine the topology of ionized pockets in the protocluster.

The presentation is organized as follows. Sec.~\ref{sec:observations} describes the
data acquistion and reduction.  In Sec.~\ref{sec:results}, quantitative descriptions of the   
morphological structure (\S~\ref{sec:imaging}), 
the rest-frame optical spectral lines (\S~\ref{sec:spectral_measurements}), and
the \lya\ line profiles (\S~\ref{sec:line_profile}) lead to
 measurements of \lya\ velocity offset and escape fraction in \S~\ref{sec:voff}. 
In Sec.~\ref{sec:discussion}, we  estimate the
sizes of the ionized bubbles (\S~\ref{sec:rion}), express the required number of 
ionizing photons in terms of LyC escape fraction (\S~\ref{sec:Nph}),
and model the number of ionizing photons produced by each LAE (\S~\ref{sec:sed}). 
Sec.~\ref{sec:budget} discusses balancing the ionization budget, including factors
such as bubble overlap and fainter galaxies. Finally, Sec.~\ref{sec:implications}
describes the implication for the three-dimensional topology of ionized region
of the protocluster. Sec.~\ref{sec:summary} summarizes the main conclusions and
lists ways that future studies might improve the accuracy of mapping ionized bubbles.

We adopt the Planck cosmology  \citep{Planck2020} throughout this paper:
 Hubble constant $H_0 = 67.4 \pm 0.5$~km~s~Mpc$^{-1}$, and density parameters
$\Omega_m = 0.315 \pm 0.007$, $\Omega_{\Lambda} = 0.685 \pm 0.007$, and
$\Omega_b = 0.04931 \pm 0.0006$.
At $z=6.93$, the universe is just 771 Myr old.
The angular diameter distance
is 1108 Mpc, and the luminosity distance is 69.70 Gpc.

We adopt a helium mass fraction $Y_P=0.245 \pm 0.004$ \citep{Aver2015}.
For the reader seeking a quick read through this work, we recommend starting with
the main measurements in \S~\ref{sec:voff}, examining the implications for ionized
bubble volumes in Sections~\ref{sec:rion} and \ref{sec:Nph}, and then moving straight
to the ionization budget and implications in Sections \ref{sec:budget} and
\ref{sec:implications}, respectively.

\section{Observations} \label{sec:observations}

LAGER-z7OD1 is an elongated overdensity of $z \sim 7$  LAEs.  This structure 
spans roughly $67.6 {\rm ~cMpc} \times 30.7  {\rm ~cMpc}$ on the sky \citep[][Figure~2]{Hu2021}. 
The mean density of LAEs is six times higher than the average over the CHORUS \citep{Itoh2018} 
and LAGER \citep{Zheng2017} narrowband imaging surveys.  The field lies slightly east 
of the COSMOS-Webb footprint  \citep{Casey2023_web}. Here we present new JWST and Keck
observations of the western half of LAGER-z7OD1.

Table~\ref{tab:observations} lists the coordinates and  position angles of our five
JWST NIRSpec pointings. Table~\ref{tab:clumps} identifies the nine LAEs by their coordinates
in the NIRCam pre-imaging, which provides accurate astrometry.  We did not detect
the object which \citet{Hu2021} called LAE-16 in the NIRCam images. Their original
narrowband detection of LAE-16 was marginal, and we will not discuss this object further. 
We did, however,  detect a previously unpublised narrowband excess source, hereafter LAE-22,
which we include in this analysis. Our F150W2 and F444W NIRCam images easily detect LAE-15. 
{\clm
The emission line was discovered in CHORUS narrowband imaging \citep{Itoh2018}
but was later rejected as \lya\ on the basis of a detection in the blue-side broadband
image \citep{Kikuta2023}. The broadband detection appears to have been a statistical
fluctuation which slightly exceed 2-sigma (pvt. comm.).
}

\begin{table}[ht]
\caption{NIRSpec Pointings} \label{tab:observations}
\begin{center}
\begin{tabular}{lllll}
\hline
\hline
NIRSpec   &  RA     & DEC     & PA\_APER           \\
Mask          & (J2000) & (J2000) & ($^{\circ}$)        \\
\hline
NRS I   & 10:02:02.3161 & +02:07:45.21  &  246.790       \\
NRS II  & 10:01:53.0263 & +02:06:31.23  &  245.572       \\
NRS III & 10:02:35.5200 & +02:07:30.00  &  245.629       \\
NRS IV  & 10:02:25.6310 & +02:06:22.12  &  245.647       \\
NRS V   & 10:02:27.5340 & +02:07:45.11  &  245.730       \\
\hline
\end{tabular}
\end{center}
\end{table}

A unique aspect of our program is that we configured the
NIRSpec microshutter array (MSA) to obtain spectra of individual clumps in galaxies 
composed of multiple components. We compute the aperture corrections for the hydrogen 
Balmer line luminosities clump-by-clump (Appendix~\ref{sec:appendix}), 
sum the clumps to obtain total luminosities,
and then calculate their Case~B intrinsic \lya\ emission. Comparison to the integrated 
\lya\ luminosity, measured previously from narrowband imaging, yields a \lya\ escape 
fraction without an aperture bias.  The 0\farcs2 wide MSA shutters, in contrast, attenuate 
scattered \lya\ emission \citep{Jung2024_CEERS_laes}, likely biasing the \lya\ escape 
fraction \citep{Saxena2023} and  modifying the shape of the \lya\ line profile \citep{Hu2023}.
Our high-resolution \lya\ line profiles were obtained with a wide (1\farcs2) slit, 
largely eliminating aperture losses due to \lya\ scattering. We measure \lya\ velocity offsets 
relative to \oiii\ and Balmer  emission lines in the NIRSpec  spectra.

\subsection{Broadband Imaging of LAEs} \label{sec:nircam}

{\clm
NIRCam takes images simultaneously in two cameras.  
}
We configured the long wavelength
camera (LW) with the F444W filter (3880 - 4990 nm HPBW) and the short wavelength
camera (SW) with the F150W2 (1008-2334 nm HPBW) filter. At  $z = 6.93$, the broad 
half-power bandwidth of F150W2 (1.01 to 2.38 micron) covers
the  rest-frame ultraviolet spectrum from just longwards of \lya\ 
through 2990 \AA; the pivot wavelength corresponds to about 2100 \AA\
in the rest-frame of LAGER-z7OD1. 
The F444W bandpass includes the rest-optical continuum and
bright emission lines, including \Hb\ and \oiii\ $\lambda \lambda 4960.30, 5008.24$.
The total exposure time was 4810~s.

Reduction of these data produced an output mosaic with absolute astrometry  
$\sim25$ mas for the LAEs and 6~mas for alignment stars.
We modeled the point spread function (PSF) using WebbPSF \citep{Perrin2015}.
The Gaussian cores of the F150W2  and F444W  PSFs have full widths at half
maximum intensity (FWHM) of 0\farcs0504 and 0\farcs1416, respectively. 
We refer the reader to Appendix~\ref{sec:weida} for a complete
description of NIRCam data reduction and Appendix~\ref{sec:bagpipes}
for the LAE photometry.

The deep stripes of the public UltraVISTA DR6 release \citep{McCracken2015} 
detect most of our targets. The UltraVISTA Y-band lies entirely longward of \lya, 
so the YJHKs filters constrain the UV spectral slope photometrically from 
rest-frame 1230 \AA\ to 2900 \AA. We matched the resolution to the seeing-limited 
UltraVISTA Y image by convolving each image with a Gaussian kernel. The resulting
point-spread function has a width of 0\farcs82 (FWHM). We measured the
flux in 2\farcs0 diameter circular apertures,  encircling 98.4\% of the energy
from a point source. The standard deviation of the background level, 
measured in 2\farcs0 aperature off the source, defines the photometric uncertainty.
To flatten the background level near LAE-13 prior to photometry, we fit
foreground galaxies with GalFit \citep{Peng2002_galfit, Peng2010_galfit} and
subtracted them. Combining NIRCam and UltraVISTA photometry in Appendix~\ref{sec:bagpipes}
improves our measurements of the UV luminosities and \lya\ equivalent widths of the
LAEs.

\subsection{NIRSpec Multi-object Spectroscopy} \label{sec:nirspec_intro}

We configured NIRSpec with the G395H grating and F290LP blocking filter. This configuration
provides $R = 2700$ resolution (110~\kms\ FWHM) from roughly 2.87 to 5.2 \um, just covering
the \Ha\ emission line at $z = 6.93$. The coadded, two-dimensional spectra have an average 
dispersion of 6.5 \AA\ pix$^{-1}$ , and the pixel sampling of the spatial profile is 
0\farcs10 per pixel. The dispersion solution is expected to be accurate 
to 10 \kms\ \citep{Boker2023}. And NIRSpec spectrocopy has an approximately 
15\% flux accuracy \citep{Bunker2023}.

Unlike traditional slit and fiber spectrographs, individual targets are
generally not centered in a shutter. The data reduction pipeline sets the
wavelength scale based on the position of the primary source. We 
apply wavelength shifts for individual clumps based on their offset
from the primary source in the dispersion direction; the maximum
correction is $\pm 22$~\kms\ with the  high-dispersion grating. Since
our primary goal was to measure the velocity offset between the \lya\ 
line and bright rest-frame optical lines, this wavelength accuracy 
was deemed more important than spectral coverage.

The visibility periods for our field require MSA position angles between either
232$^{\circ}$ and 252$^{\circ}$ (April/May) or 64$^{\circ}$ - 84$^{\circ}$ 
(November - January).  We adopted a position angle of 68$^{\circ}$ for proposal
planning purposes but were assigned position angles in the spring window.  
Flipping the mask design 180$^{\circ}$ forced us to modify our dithering 
strategy because there were more collisions between primary targets and stuck
shutters. Although we had planned an 80 shutter shift in the dispersion direction 
to close the wavelength gap, it was not possible to perform this dither while also
keeping all the primary targets in columns with \Ha\ coverage and rows that avoided
stuck shutters. We elected not to dither in the dispersion direction because 
the gap never consumes all three bright lines -- \Hb, 
 \oiii\  $\lambda 4960$, and \oiii\ $\lambda  5008$. This strategy  generally
provided uniform integration time over the rest of the  bandpass from [OII] $\lambda \lambda 3727, 3730$ 
through \Ha\ for the LAEs.  
The shutters assigned to LAE-8 and LAE-15 do not provide wavelength coverage of \Ha, and
we use their \Hb\ fluxes in our analysis.
Target acquistion was completed by opening shutters on alignment stars.

We dithered the exposures spatially in an ABBA pattern, using a step size 
of 1 or 2 shutters depending on the pattern of stuck shutters near primary targets. 
Pixel-to-pixel subtraction using the ABBA sequence turned out to not be suitable for  some
of the LAEs due to the spatial extent of their line emission along the slit direction.
Our final reduction therefore uses background measurements from non-conflicted shutters,
which were opened to allow construction of a master background.
Appendix~\ref{sec:weida} describes the steps in the full spectral reduction, 
including the contruction of the master background.
{\clm
The Stage~3 pipeline spectra have been shifted to a barycentric reference frame
and have vacuum wavelengths.
}

The accuracy of the analysis presented here is limited by the error bars
on the \Hg\ and \Hb\ line fluxes. 
{\clm
Systematic errors in the calibration, described in Appendix~\ref{sec:master_background},
currently dominate the uncertainty. 
}
We anticipate reducing the flux errors
in a future reprocessing of the data. In this paper, we simply  propagate their
large uncertainties. 
{\clm Systemic errors
aside, the 7120~s exposure times detect \Hg\ emission lines 
as faint as $2. 9 \times 10^{-19}$~ergs~s$^{-1}$~cm$^{-2}$ ($4 \sigma$).
}

\subsection{Keck/LRIS  Multislit Spectroscopy} \label{sec:lris}

Using the Keck~I telescope, Low Resolution Imaging Spectrograph (LRIS) spectroscopy of
LAGER-z7OD1 galaxies were obtained under clear conditions on 2022 January 29th and 2024 January 14th.
The mosaic of fully depleted, high resistivity CCD detectors provide high throughput and
reduced fringing beyond 800 nm \citep{Rockosi2010}, ideal characteristics for detecting
\lya\ emission at redshift $z \approx 6.93$.

We configured the red channel of the spectrograph with the 1200 line per millimeter
grating blazed at 9000 \AA. With our 1\farcs2 wide slitlets, this configuration 
provided a spectral resolution of 56 \kms\ FWHM (full width half maximum). We
binned the detector by two pixels in the spectral direction, obtaining a dispersion
of 0.8 \AA\  pix$^{-1}$, which Nyquist samples the 1.8 \AA\ FWHM linewidth. We
also binned the detector by two pixels in the spatial direction, and the resulting
0\farcs27 pixels sample the spatial profile of these seeing-limited observations well.

Spectra were obtained through two slit masks, one on each night.
These observations provide the first spectroscopic detections of LAE-8,
LAE-14, and LAE-22. For the brighter LAEs, the new LRIS spectra provide
higher resolution and  better S/N ratio than were previously available.  
We adopt measurements from \cite{Hu2021} for LAE-15.

Integrations totaling 24,300~s on 2022 January 29 confirmed 
LAE-8 and LAE-14; they also provided high S/N ratio line profiles for LAE-1 and LAE-2. 
Exposures with the second mask on 2024 January 14 accumulated 12,600~s; the better
seeing, 0\farcs7 FWHM in 2024 versus 1\farcs2 FWHM in 2022, partially compensated 
for the shorter integration time. This second mask provided new \lya\ spectra of 
LAE-10, LAE-11, LAE-13, and LAE-22.  The spectrum of LAE-21 revealed a pair of 
emission lines which we identified as \oiii\ $\lambda \lambda 4960, 5008$ emission; we do not
discuss this foreground, $z \sim 0.9$, interloper further in this paper.\footnote
   {The coordinates for LAE-21 in Table 1 of \cite{Hu2021} contain an error. 
    We observed the galaxy at the corrected position (RA,DEC)$_{J2000}$ = (10:02:51.6, +02:06:54.1). 
    }

The slit position angle was 35.0$^{\circ}$ on the 2022 mask and -75.0$^{\circ}$ on the 2024 mask. 
The parallactic angle swung across the slitlets during these long observations, and
the atmospheric dispersion corrector at the Keck I Cassegrain ensured that the targets
remained well centered in their slitlets in both blue and red spectrographs.

We read out the detector every 900~s because the thin CCD records a high rate of 
cosmic ray hits. The individual frames were corrected for fixed pattern noise and
wavelength calibrated using the the Python Spectrosopic Data Reduction Pipeline, 
PypeIt \citep{pypeit2005,pypeit2020}.  We identified and masked cosmic rays when
we stacked these frames. The PypeIt coadding task produced a rectified frame with
vacuum wavelengths on a heliocentric scale.  The \lya\ emission was clearly visible
on these stacked 2D spectra. We extracted 1D \lya\ spectra using custom python scripts.

\section{Results} \label{sec:results}

\subsection{NIRCam Imaging of LAEs} \label{sec:imaging}

\begin{figure*}[ht]
 \centering
     \includegraphics[scale=0.85,angle=0,trim = 20 460 0 0]{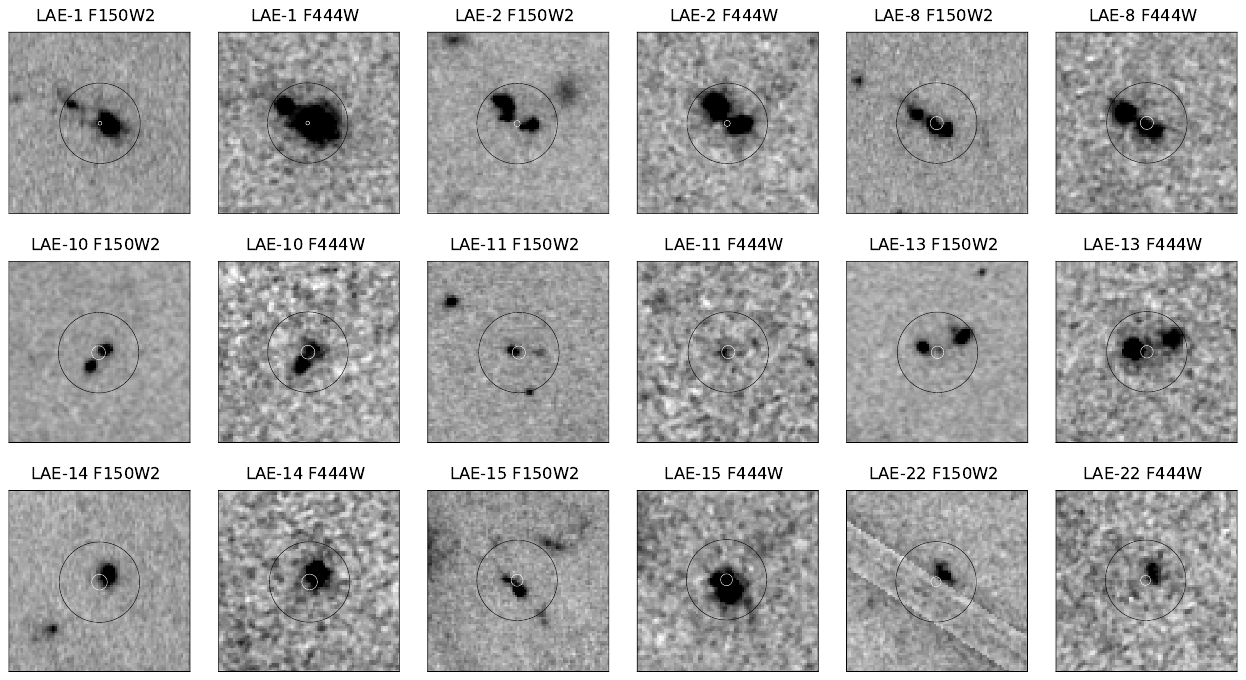}
     \caption{NIRCam images of LAGER-z7OD LAEs. The centroid of the \lya\ nebulae
              (tiny white circles) lie between the clumps discovered in the rest-UV
              and rest-optical images. Each cutout is 2\farcs25
            by 2\farcs25; north is up and east is the left. The 1\farcs0
            diameter black circles illustrate the PSF of
            the LAGER narrowband images. The smaller white circle is the SDOM.
            }
      \label{fig:nircam} \end{figure*}

The new NIRCam images resolve six of the nine LAGER-z7OD1 LAEs into multiple clumps. 
Figure~\ref{fig:nircam} shows that the same sub-components appear in the 
rest-frame optical and rest-frame UV images, so the underlying stellar
mass distribution has multiple components. The clump separations vary from a
few hundred parsecs up to 2.5~kpc. Based on the presence of these  close companions,
galaxy interactions may fuel the starburst activity \citep{Witten2024-NatAs}.
This appears to be common at redshift $z \sim 7$, where  roughly
70\% of bright galaxies \citep{Harikane2024_clumpy_compact} and nearly
all LAEs \citep{Witten2024-NatAs} exhibit clumpy morphologies.  
The presence of multiple sub-components in starburst galaxies, 
may also assist LyC leakage \citep{Martin2024,Kostyuk2024,Mascia2025}. 

The clumps which we placed in NIRSpec shutters are identified by 
their coordinates in Table~\ref{tab:clumps} and Fig.~\ref{fig:clumps} 
of Appendix~\ref{sec:appendix}.  
Table~\ref{tab:clumps} lists the fitted effective radii of the individual clumps. 
The median radius of 0\farcs0704 subtends 403 pc, so most of the clumps are larger
than compact galaxies, defined as having effective radii less than 200~pc
\citep{Bunker2023,Tacchella2023}. 
The angular size of the smallest clumps
approaches the  core of the F150W2 PSF, which subtends about 270 pc FWHM and is
an order of magnitude larger than the sizes of individual star clusters \citep{Adamo2024-Nature}.
The relationship between the sub-components and the \lya\ nebulae is examined next.

\subsubsection{Registration of \lya\ and NIRCam Images}

Figure~\ref{fig:nircam} compares the center of each \lya\ nebula to the positions 
of individual clumps. While the narrowband point-spread function is wider than the 
angular separation of the sub-components, the centroid of the \lya\ emission is 
known to higher precision.\footnote{
         The point-spread function of the narrowband image has a Gaussian core,
         fit by a standard deviation of $\sigma = $0\farcs43.
         The positional uncertainties are given by
         the $SDOM =  \sigma / \sqrt{SNR}$. 
         The centroid uncertainty scales inversely with the signal-to-noise
         ratio of each narrowband detection.
         }

The UV and optical morphologies of LAE-11, LAE-14, and LAE-22 show a single component
in Fig.~\ref{fig:nircam}. Their NIRCam and \lya\ coordinates agree. The positional
offset is between 1 and 2 times the error on the position, the standard deviation
of the mean (SDOM).  Each clump is clearly the power source of the corresponding \lya\ nebula. 

In contrast, the seeing-limited centroids of the LAE-2 and LAE-13 \lya\ nebulae
do not match the positions of any of the UV/optical clumps. The centroid of the
nebular emission lies between each pair of clumps. While less significant statistically,
the \lya\ nebulae in LAE-1, LAE-10, and LAE-15 also have positions between the UV/optical
clumps.  We interpret these positional offsets as evidence that the \lya\ nebulae have 
structure on the scale of the NIRCam sub-components. Convolution with the atmospheric 
seeing kernel would then shift the centroid of the \lya\ image to a location between clumps.
Alternatively, \lya\  emission powered by gravitational energy  would center the nebula
on the minimum in the gravitational potential \citep{Aung2024}. JWST observations could 
resolve the \lya\ nebulae and distinguish these scenarios.

It is clear, even from a qualitative comparison of the F444W and F150W2 images in Figure~\ref{fig:nircam}, 
that the F444W morphologies in LAE-1 and LAE-15 are not simply PSF broadened copies of
the F150W2 structures. We suggest that spatially extended \oiii\ emission enlarges the sizes of
these clumps in F444W. Emission in the F444W band is dominated by strong \oiii\ emission, so it
is a good template for the intrinsic \lya\ emission prior to scattering.  The centroid of each of
these \lya\ nebulae lies well within what appears to an extended \oiii\ nebulae in both cases.

In this paper, we take the morphological sub-components into account in order to accurately measure
integrated rest-frame-optical spectral properties. The spatial offsets illustrated here demonstrate
that measuring \lya\ emission through individual NIRSpec shutters might significantly attenuate the
\lya\ emission. We quantitatively discuss the relationship between morphological sub-components
and their (rest-frame) optical emission-line spectra in a follow-up paper.

\subsubsection{The UV Continuum Level \& W(\lya)}

Table~\ref{tab:jwst_offset} lists the narrowband \lya\ luminosities from \cite{Hu2021},
scaled to the cosmology used in this paper.  
{\clm 
Using the M$_{2100}$ magnitudes, we calculate the \lya\ equivalent widths. 
For a fiducial continuum slope, $\beta = -2.35$ \citep{Bouwens2014}, the median
rest-frame \lya\ equivalent width is 80 \AA, and the individual values range from 28 to 130 \AA.
}
These estimates are consistent with the narrowband imaging sensitivity limit of 10 \AA\ \citep{Hu2019}.
Variations in continuum slope determine the uncertainties on W(\lya). A flat continuum in
$F_{\nu}$ ($\beta = -2.0$), for example,  would increase the equivalent widths by a factor of 
1.20.  A very steep continuum with $\beta = -2.7$ would decrease the equivalent width to 0.82
of the estimate shown.\footnote{ 
              The UV spectral slope $\beta$ is defined by $F_{\lambda} \propto \lambda^{\beta}$.}

\begin{deluxetable*}{llllllcccl}
\tablecaption{NIRSpec and NIRCam Measurements}
\colnumbers
\tablehead{
\colhead{Galaxy} &
\colhead{Mask} &
\colhead{$\log L(\lya)$}   &
\colhead{$z_{\rm sys}$} &
\colhead{$\log L(\Ha)$} &
\colhead{$ {\rm ~M_{2100}}$} &
\colhead{W(\lya)} &
\colhead{O32} &
\colhead{$\log U$} &
\colhead{$E(B-V)^{\rm gas}$} 
\\
\colhead{}        &
\colhead{(NRS)}        &
\colhead{(ergs~s$^{-1}$)} &
\colhead{}        &
\colhead{(ergs~s$^{-1}$)} &
\colhead{(mag)}   &
\colhead{(\AA)}   &
\colhead{} &
\colhead{}   &
\colhead{(mag)}        
}
\startdata
LAE-1   &
I &
$43.54^{+0.03}_{-0.03} $     &
$ 6.9329 \pm 0.0001 $ &  
$43.45 \pm 0.07 $     &
-21.37 &
$92  $ &
\nodata &
\nodata &
$ 0.29^{+0.28}_{-0.29} $ 
\\
LAE-2   &
II &
$ 43.33^{+0.07}_{-0.08} $     &
$ 6.9272 \pm 0.0001 $ &
$ 42.92 \pm 0.04 $     &
-20.86     &
91   &
$11 \pm 3 $&
-1.6 &
$0.52^{+0.22}_{-0.52} $  
\\
LAE-8  &
I &
$42.84^{+0.11}_{-0.24} $   &
$ 6.9194 \pm 0.0001 $ &
$ 42.62 \pm 0.09$\tablenotemark{a}   &
-20.90 &
28     &
$6 \pm 2 $ &
-1.9 &
$0^{+0.60}_{-0} $  
\\
%
%
LAE-10 &
III, V &
$ 42.56^{+0.13}_{-0.18}$  &
$ 6.9194 \pm 0.0004 $ &
$ 42.60 \pm 0.05$  &
-19.84  &
40      &
$> 19$ &
-1.4 &
$0^{+0.47}_{-0} $  
\\
LAE-11 &
III  &
$ 42.69 ^{+0.11}_{-0.15} $ &
$ 6.9611 \pm 0.0003 $ &
$ 42.51 \pm 0.04 $ &
-19.20  &
96      &
$ > 6 $ &
-1.5 &
$0.11 ^{+0.13}_{-0.11} $  
\\
%
%
%
LAE-13 &
IV, V  &
$ 42.68 ^{+0.12}_{-0.16} $ &
$ 6.9298 \pm 0.0008 $ &
$ 43.04 \pm 0.05 $ &
-20.25     &
36   &
$ > 2 $&
-1.9 &
$0.18^{+0.12}_{-0.18} $  
\\
LAE-14  &
I &
$ 42.79^{+0.14}_{-0.21} $ &
$ 6.9325 \pm 0.0002 $ &
$ 42.83 \pm 0.02 $ &
-20.79  &
28      &
\nodata &
\nodata  &
$0.23^{+0.09}_{-0.23} $  
\\
LAE-15  &
IV &
$ 43.38^{+0.10}_{-0.15}$ & 
$ 6.96611 \pm 0.00004 $ &
$ 43.19 \pm 0.08 $\tablenotemark{a}  &
-20.36 &
130    &
$> 130 $\tablenotemark{b} &
-0.33 &
$0.36^{+0.50}_{-0.36} $  
\\
LAE-22  &
V &
$ 42.53^{+0.15}_{-0.20}$ & 
$ 6.9154 \pm 0.0002 $ &
$ 42.33 \pm 0.12 $ &
-19.0    &
80       &
$ > 8 $ &
-1.8 &
$0.14^{+0.30}_{-0.14} $  
\\
\enddata
\tablenotetext{a}{Computed from the \Hb\ flux and an assumed flux ratio 
F(\Ha)/F(\Hb) = 2.75, appropriate for Case B recombination at an electron
temperature and density $2 \times 10^4$~K and 100~cm$^{-3}$, respectively.}
\tablenotetext{b}{Identifies AGN candidate.}
\tablecomments{
{\it (Col 1):}  Galaxy name.
{\it (Col 2):}  NIRSpec MSA mask name.
{\it (Col 3):}  The \lya\ luminosities from Table~1 of  \cite{Hu2021} scaled to the 
                \citep{Planck2020} cosmology
                and \lya\ redshift in Col.~2 of Table~\ref{tab:new_lya}.
{\it (Col 4):}  Systemic redshift based on rest-frame optical emission-line measurements.  The
                uncertainties are larger for LAE-10 and LAE-13 because the redshifts of the two
                clumps have been averaged together.
{\it (Col 5):}  Total \Ha\ luminosity of all spatial and spectral components. 
                Corrections for aperture losses have been applied as described in
                Sec.~\ref{sec:spectral_measurements} and Appendix~\ref{sec:weida},
                but the extinction correction has not been applied yet.
{\it (Col 6):}  Near-UV absolute magnitudes, uncorrected for reddening, were computed
               by summing the F150W2 luminosities of the clumps identified using
               GALFIT \citep{Peng2002_galfit, Peng2010_galfit}.
{\it (Col 7):} Rest-frame \lya\ equivalent width estimated from Col.~3 and Col.~6
               assuming a UV continuum slope $\beta = -2.35$.
               A flat continuum in
               $F_{\nu}$ ($\beta = -2.0$) would increase the
               equivalent widths by a factor of 1.20, and a very steep continuum with $\beta = -2.7$
               would decrease the equivalent width to 0.82 of the estimate shown.
{\it (Col 8):} Reddening corrected emission-line ratio of \oiii\ $\lambda, \lambda 4960, 5008$ to
               \oii\ $\lambda, \lambda 3727,30$.
{\it (Col 9):} Ionization parameter estimated from O32 ratio \citep[][Eqn.~7]{Shen2024}.
{\it (Col 10):} Color excess derived from the Balmer ratio, $F(\Ha)/F(\Hb)$.  When \Ha\ is
                not covered by the NIRSpec spectrum, the $F(\Hb) / F(\Hg)$ is used.  We
                assume intrinsic Balmer ratios as given in \citet{Osterbrock2006} at
                $T_e = 2 \times 10^4$~K and $n_e = 100 $~cm$^{-3}$.
}
%
\label{tab:jwst_offset}
\end{deluxetable*}

LAE-1, LAE-2, LAE-8, and LAE-14 are all UV luminous galaxies. Their brightness
exceeds the knee in the $z \sim 7$ UV luminosity function at $M^*_{UV} = -20.5$  
\citep{Harikane2022}.  Under the conservative assumption of a flat continuum in $F_{\nu}$,
extrapolation of their near-UV absolute magnitudes, $M_{2100}$, to 1500 \AA\ yields
 $M^*_{UV} < M^*_{UV}$ 
The UV-slope
measurement (from \texttt{BAGPIPES} fitting in \S~\ref{sec:bagpipes}) boosts LAE-15 into the
bright galaxy list. LAE-13 only makes the bright category if it has a continuum slope 
$\beta \le\ -2.7$, significantly steeper than the median UV slope ($\beta = -2.35$)
among $z \sim 7$ LAEs in JADES \citep{Witstok2024}.
LAE-10, LAE-11, and LAE-22 appear to be distinctly less luminous galaxies.


\subsection{Measurements of Rest-Optical Emission Lines} \label{sec:spectral_measurements}

The reduced NIRSpec spectra  detect multiple emission lines in the optical bandpass 
between \oii\ $\lambda \lambda $ 3727.73, 3728.76  and \Ha. In this paper, we focus
on redshifts derived from the strong \Ha, \oiii\ $\lambda \lambda 4960, 5008$, and \Hb\
emission lines. We also measure \oii,  \Hg, and \oiii\ $\lambda 4363$ line fluxes in order
to compute line ratios which constrain the nebular properties. We refer the reader to
Moya-Sierralata \et 2025 (in prep), for measurements of the gas-phase O/H abundance ratio.

Following the procedure described in Appendix~\ref{sec:appendix},
we extracted a one-dimensional spectrum for each clump 
independently and calibrated the spectrum based on that clump's pitch in the shutter.
We fit the emission lines with Gaussian components using the non-linear
least squares method.\footnote{We used the \texttt{SciPy} \citep{SciPy-NMeth2020} 
package \texttt{curve fit}.} Multiple components in the spectrum of a single
clump were justified using an F-test.  A linear fit to bandpasses on either side of 
an  emission line set the local  continuum level.

\begin{figure*}[h]
 \centering
   \includegraphics[scale=0.8,angle=0,trim = 0 0 0 0]{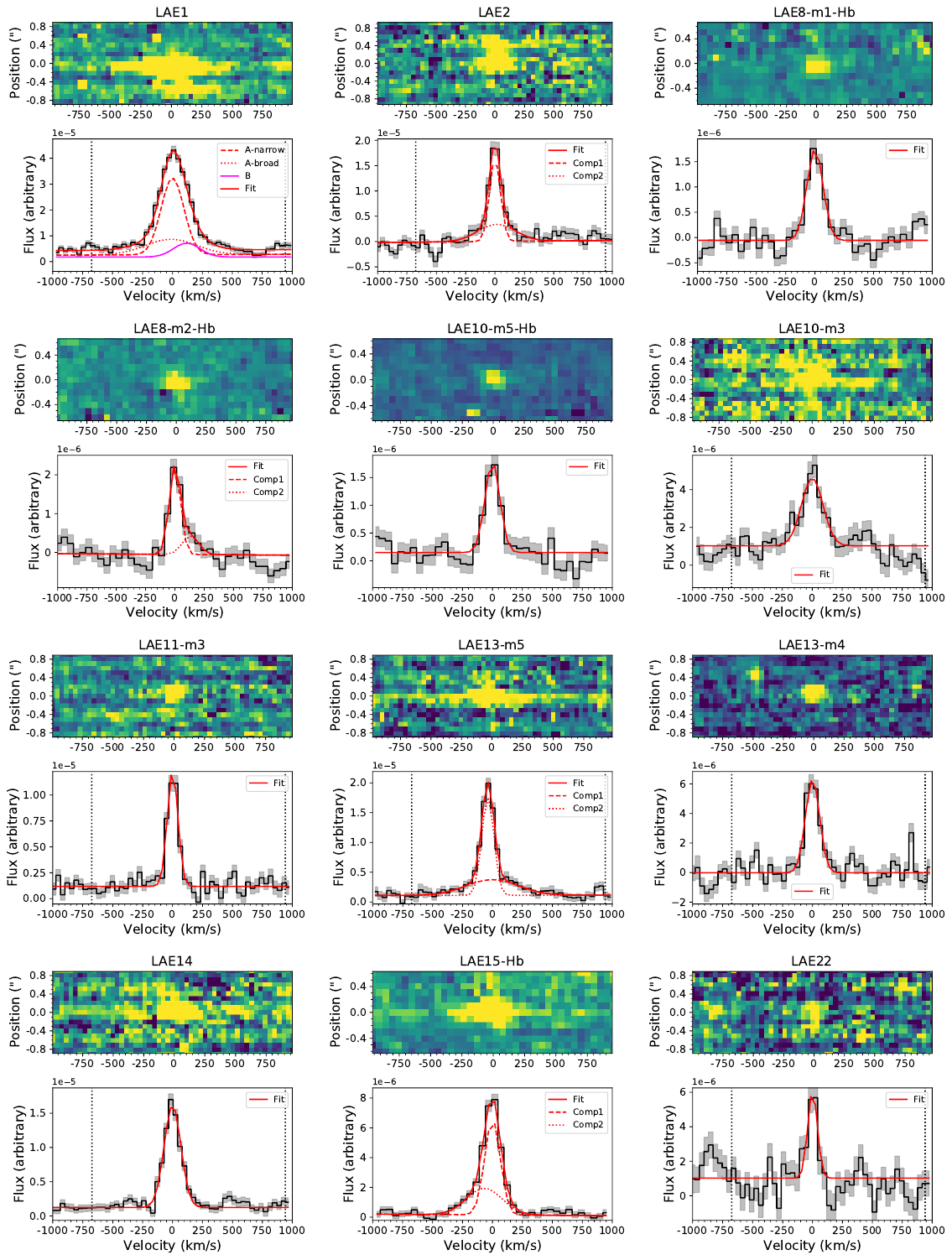} 
   \caption{NIRSpec \Ha\ spectra. The two-dimensional spectra of LAE-1, 
     LAE-2, LAE-8, and LAE-15 detect emission lines from multiple clumps along
     the slit. The sum of their extracted spectra and fitted profiles are
     shown in the lower panel. The two clumps comprising LAE-10 and LAE-13 were 
     observed on different masks; their spectra are shown in separate panels.
     The vertical, dotted lines denote the
     wavelength of [NII] lines. Gray shading shows the on-sigma error.
     Note the substitution of \Hb\ spectra for LAE-8 (m1 and m2), LAE-10 (m5 only),
     LAE-11 (m5 only),  and LAE-15.
     }
    \label{fig:nirspec}
     \end{figure*}

Figure~\ref{fig:nirspec} shows the sum of the \Ha\ spectral components below each 
corresponding 2D-spectrum. Our fit to the brighter clump in the LAE-1 spectrum 
requires a broad component in addition to the narrow component.  The \Ha\ emission 
from the fainter clump is spatially separated from the brighter clump, and we can
see that the fainter clump is redshifted relative to the primary clump.  In the
summed spectrum, the single component emitted by this fainter clump blends with 
the broad component of the brighter clump. A similar spatial decomposition of
the spectrum was performed for LAE-2.

Broad components are detected in the \Ha\ spectra of  LAE-1, LAE-2, LAE-13, and LAE-15. 
These broad components are not mistaken [NII] $\lambda \lambda$ 6550, 6585 emission.
They are also detected in the \Hb\ line and/or the \oiii\ doublet. Unlike  the
very high densities associated with Type I AGN and many Little Red Dots 
\citep{Harikane2023,Greene2024,Matthee2024,Kocevski2024}, the \oiii\ detections indicate
gas densities below the critical density of these forbidden lines.  The gas emitting
these broad wings therefore has a density typical of galactic winds \citep{Peng2025}.

\subsubsection{Redshift Measurements} \label{sec:systemic_redshift}

We adopt the narrow component emitted by each clump as the best measure of its
systemic redshift. We average several strong lines, weighting by their relative 
luminosities.  The standard deviation of the mean determines the redshift uncertainty, 
which is around  4 \kms\ typically. Table~\ref{tab:jwst_offset} lists their
redshifts. 

LAE-8 and LAE-11 were each observed on two masks.
The target has a different shutter pitch in each observation .
Comparison of the redshifts measured on independent
spectra agree within the measurement uncertainties. This
demonstrates consistent wavelength calibration of the MSA spectra. 

The pairs of clumps discovered in the NIRCam pre-imaging of LAE-10 and LAE-13
were observed spectroscopically on different shutters. Table~\ref{tab:jwst_offset} 
gives their \Ha-weighted average redshift. The luminosity-weighted average 
redshift for LAE-13 is higher than the other targets because the relative velocity of 
the two clumps is 50 \kms.

\subsubsection{Hydrogen Balmer Line Luminosities \& Ratios}  \label{sec:balmer_ratio}

The \lya\ luminosity measurements have no aperture losses because they come 
from narrowband imaging, but this seeing-limited \lya\ imaging does not
uniquely identify which clump(s) emit \lya. Correcting the \Ha\ fluxes for aperture
losses is therefore essential for accurately estimating the global \lya\ 
escape fraction. In addition, since multiple clumps make significant
contributions to the \Ha\ flux in most LAEs, measuring the global
\fesclya\ requires extracting NIRSpec spectra clump-by-clump and applying
aperture correction computed for each individual clump. 
Using the fitted morphological properties listed in Table~\ref{tab:clumps}, 
we computed NIRSpec aperture corrections using the forward modeling software
\texttt{msafit} \citep{deGraaff2024}.  Appendix~\ref{sec:appendix} outlines the
procedure used to compute the aperture correction. 

Table~\ref{tab:jwst_offset} lists the total \Ha\ luminosities computed from the 
sum of these aperture-corrected, clump fluxes.  Relative to the narrow component, 
each broad component has lower amplitude and lower integrated flux.  We include
the broad component in the total Balmer line luminosity. Outflowing gas in local analogs 
produces broad components with similar properties; the substantial luminosity of 
these components indicates that photoionization powers the line emission  \citep{Peng2025}.
Since we will use the Balmer line luminosity to measure the recombination rate in 
the ISM, we include the luminosity of the broad component in the total for each galaxy.

The NIRSpec spectra of LAE-1, LAE-2, LAE-11,
LAE-13, LAE-14, and LAE-22 detect \Ha\ and \Hb\ line emision.
These spectra, and the spectra without \Ha\ coverage --
LAE-8, LAE-10, and LAE-15, detect the \Hb\ and \Hg\ emission lines.
{\bf
We measured the flux ratios of these  Balmer lines
and compared them to their respective Case B
ratios \citep{Osterbrock2006}:  $F(\Ha) / F(\Hb) = 2.75$ and 
$F(\Hb) / F(\Hg) = 2.11$ at an electron temperture $T_e = 2 \times 10^4$~K 
and density $n_e = 100$ cm$^{-3}$, conditions consistent with the
median values derived from flux measurements of  \oiii\ $\lambda 4363$ 
and \oii\ $\lambda \lambda 3727, 3730$ lines in our NIRSpec spectra 
(Moya-Sierralata \et 2025, in prep). 
}
Each Balmer ratio is consistent with its intrinsic Case~B ratio, so  
the Balmer ratios do not require dust attenuation. 

We emphasize, however, that the large error bars on the
\Hg\ and \Hb\  fluxes allow significant visual extinction.
A typical S/N ratio
of 10 in the \Hb\ line propagates to an uncertainty of $\delta{\rm (A_V)} \approx 0.25$
magnitudes. For the lower S/N ratio typical of the \Hg\ detections, however,
we estimate an uncertainty $\delta{\rm (A_V)} \approx 1.0$. 
These values are based on the \cite{Gordon2003} attenuation law. 
We will present  additional insight about the reddening in Sec.~\ref{sec:sed},
where we discuss the UV continuum.

\subsubsection{Nebular Excitation \& Ionization Parameter } \label{sec:U}

High excitation is generally found to be a necessary, but not a sufficient, condition for
LyC escape \citep{Chisholm2022,Flury2022,Jaskot2024_LzLCS,Flury2024}.
Table~\ref{tab:jwst_offset} lists
our measurements of the flux ratio of \oiii\ $\lambda \lambda 4960, 5008$ to
\oii\ $\lambda \lambda 3727,3730$, hereafter O32,  for seven LAEs.\footnote{  
            For LAE-1 and LAE-14, the \oiii\ doublet falls in the chip gap 
            for reasons explained in \S~\ref{sec:nirspec_intro}.
            }
The lower limit for LAE-15, ${\rm O32} > 110$, stands out from the other LAEs,
possibly indicating an AGN  (\S~\ref{sec:agn}).
The O32 ratios of the other six LAEs are high compared to typical galaxies in the local
universe but indistinguishable from Green Pea galaxies.

At the sub-solar metallicities of our targets (Moya-Sierralata \et 2025, in prep), 
the O32 ratio increases linearly with ionization parameter, a dimensionless ratio
of ionizing photons to hydrogen atoms. Table~\ref{tab:jwst_offset} lists the ionization parameters
estimated by applying \citet[][Eqn.~7]{Shen2024} to our O32 measurements. Omitting LAE-15, 
the O32 ratios for the other six LAEs yield ionization parameters  $\log U = -1.9 $ to -1.4. 
The large O32 ratio of LAE-15 requires extrapolating the O32 - $U$ relation beyond the locus
the photoionization models shown in \citet{Shen2024}.

\subsection{ Measurements of \lya\ Line Profiles} \label{sec:line_profile}

\begin{figure*}[h]
  \includegraphics[scale=1.0,angle=-90,trim = 0 0 180 0]{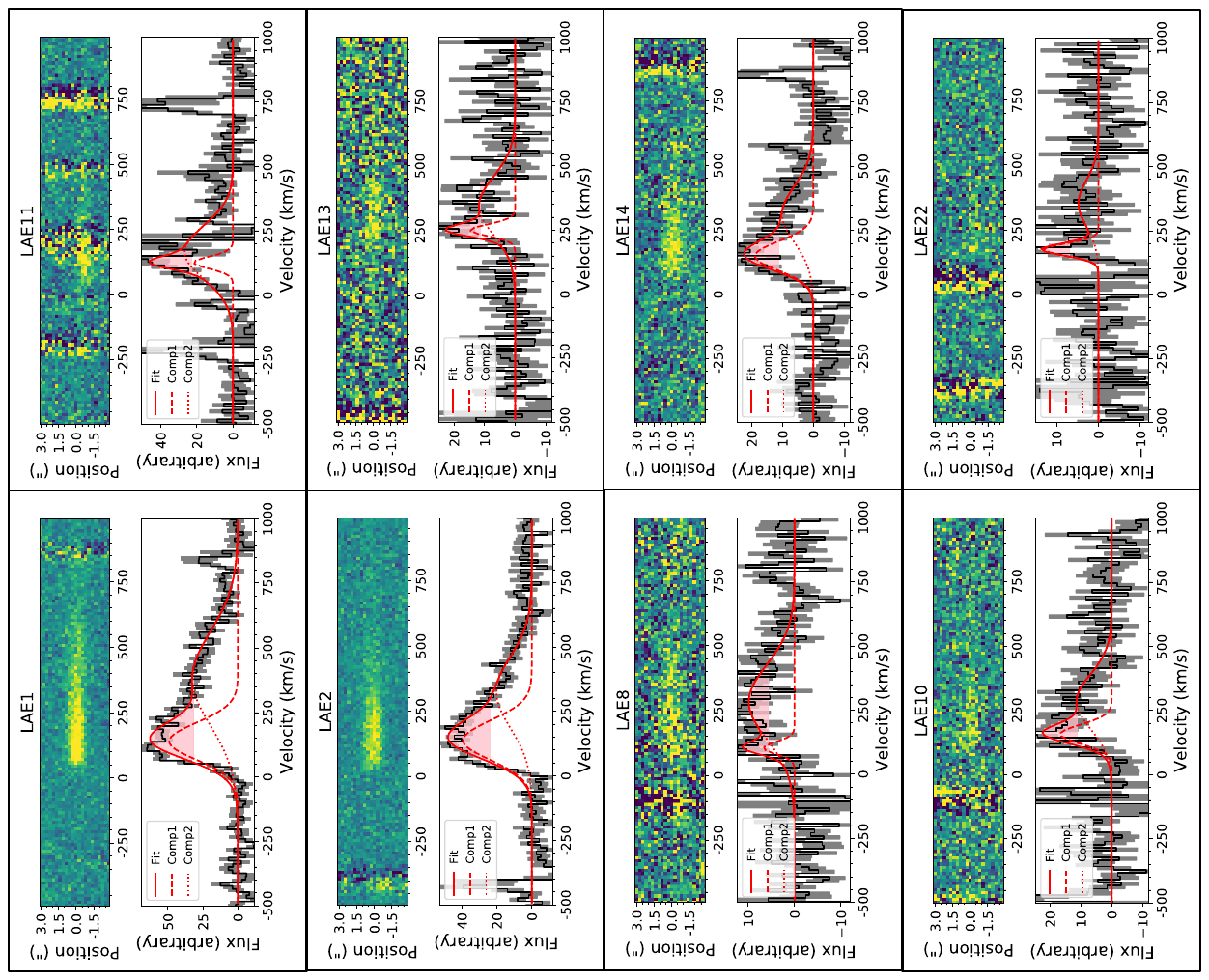}
   \caption{Keck/LRIS \lya\ spectra.
    The systemic redshift is defined by the \oiii\ doublet and H Balmer lines, as
    described in \S~\ref{sec:systemic_redshift}. Red lines show the two-component fit
    described in \S~\ref{sec:line_profile}.  The shaded region illustrates the full width
    at half-maximum intensity.  Gray shading shows the 1-sigma error bars.
    Telluric emission lines create the increased shot noise visible 
    near the blue edge of the LAE-8 and LAE-22 line profiles, and
    also just redward of the narrow component of the LAE-11 line profile.  
    The velocity offsets are given relative to the systemic redshift defined 
    by rest-frame optical emission lines.  
}
    \label{fig:lris}
     \end{figure*}

Figure~\ref{fig:lris} shows our \lya\ spectra relative to the systemic redshift, as defined 
by rest-frame optical emission lines in the previous section. None of the the \lya\ line profiles 
show transmission at the systemic velocity or blueward of it. The \lya\ profiles are generally also 
broad and highly asymmetric.  Very broad, red wings are well-defined in the line profiles of LAE-1, 
LAE-2, LAE-8, LAE-13, and LAE-14. The other \lya\ spectra show red wings, but the lower S/N ratios 
leave significant uncertainty about their width. It is significant however that, even in the lowest
 S/N ratio line profiles, the \lya\ emission is spread over a velocity range exceeding 400 \kms. 
In the brightest target, LAE-1, this asymmetric wing 
extends to at least 700 \kms, where it becomes undetectable due to a telluric emission line. 

Considering the clumpy nature of the LAEs in Fig.~\ref{fig:nircam}, we investigated whether the 
relative Doppler shifts of different morphological components broaden the \lya\ profiles. 
The LRIS slitlet placed on LAE-1 (Sec.~\ref{sec:lris}) contains both of the clumps visible in 
Figure~\ref{fig:nircam}.  The line-of-sight velocity offset of these clumps is significantly
smaller than the velocity offset of the second peak, which is $242 \pm 32$ \kms\ with
respect to the first peak. Even for LAE-8, which 
shows three clumps along the MSA slitlet in Fig.~\ref{fig:clumps},  the velocity offsets 
between these clumps turn out to be far less than width of the \lya\ line profile. Additionally,
LAE-14 contains a single clump,  so its broad, redshifted wing cannot be produced by a
second morphological component. We consistently detect \lya\ emission at velocities larger
than the relative Doppler shifts between clumps. Using integral field spectroscopy,
\citet{Vitte2025} draw a similar conclusion at cosmic noon. Just 10 (out of 248) LAEs
exhibit spatial components which they could identify with peaks in the integrated \lya\ profile.  
The broad, red wings therefore arise from radiative transfer effects.

The broad wings indicate significant column densities of atomic hydrogen around the 
LAGER-z7OD1 LAEs. Comparison to the line profiles shown in \cite[][Figure 5]{Yajima2018} 
suggests  $\log N_{HI} ({\rm ~cm}^{-2}) \approx 19.3 - 20.3$, where the range reflects
the degeneracy between column density and outflow speed. These column densities produce
tension with density-bounded models for LyC leakage. Density bounded models predict
narrow \lya\ line profiles with little asymmetry \citep{Kakiichi2021}.  The broad, 
asymmetric wings favor LyC leakage through channels instead  \citep{Witten2023_low_fesc}.

We expect the linewidth and velocity offset of indivdiual \lya\ components to be related via 
radiative transfer effects. Galactic outflows, for example, often produce multiple \lya\ 
components \citep[][Figure 12]{Verhamme2006}. Line photons backscattered off neutral hydrogen 
(in the receding lobe of the outflow cone or shell) emerge Doppler shifted  by twice the outflow 
speed. Whether galactic outflows shape the broad components of the LAGER-z7OD1 LAEs is not yet clear, 
but we fit double-Gaussian models to each \lya\ profile With this picture in mind. 

The double-Gaussian models in Figure~\ref{fig:lris} fit the high S/N ratio line profiles well.
The fits place the peak intensity firmly in the lower velocity component, and  the more 
redshifted component is distinctly broader. In the LAE-1 fit, for example, the lower redshift, 
$\dv = 138  \pm 6$ \kms, component is narrow, $FWHM = 168 \pm 18$ \kms. The broad component, 
FWHM = $423 \pm 5$ \kms, is Doppler shifted  $V_2 = 242 \pm 32$ \kms\  relative to the 
narrow component.   To aid comparison to other studies, where the spectral
resolution may not separate these components, we also measured the line width without
a two-component fit. In Fig.~\ref{fig:lris}, the shaded area illustrates the profile linewidth at
half the maximum intensity, $FWHM_{NF}$ (no fit).

\begin{figure}[h]
    \includegraphics[scale=0.5,angle=0,trim = 20 0 0 0]{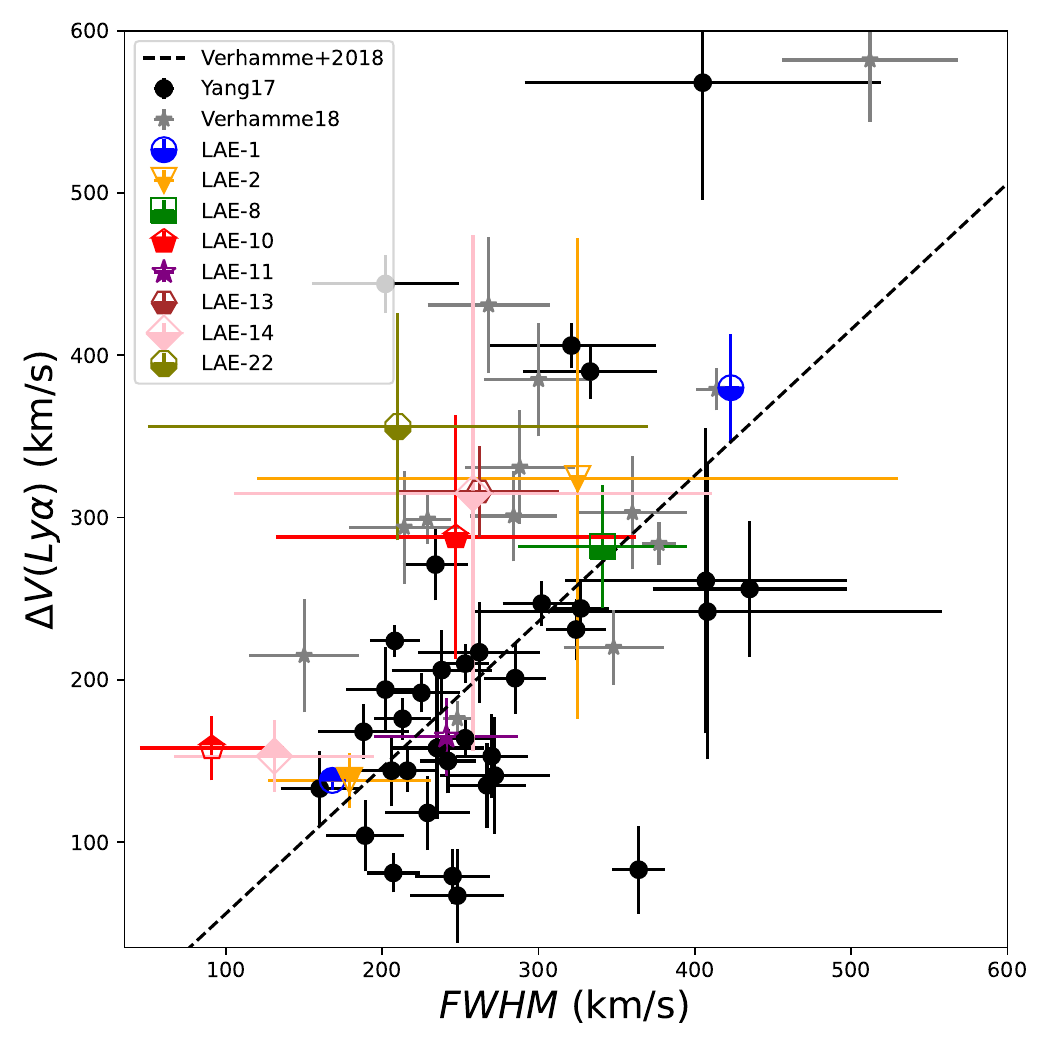}
    \caption{Velocity offsets of \lya\ components versus line width. Top filled symbols
     mark the narrow component defining the velocity offset. Bottom filled symbols indicate
     the broader, higher velocity component. The reionization-era galaxies lie on, or to the
     left, of post-reionization galaxies. The offset may represent the impact of \lya\
     radiative transfer through a partially neutral IGM.
     }
     \label{fig:dv_fwhm_nofit}
      \end{figure}

The combined column density of interstellar and circumgalactic HI determine how
radiative transfer modifies the intrinsic \lya\ line profiles of galaxies. With increasing
column density, the linewidth grows as the \lya\ velocity increases.  \citet{Verhamme2018} 
found fit their  scaling relation to galaxies at cosmic noon. Figure~\ref{fig:dv_fwhm_nofit}
shows that this relation also describes the redshifted component emergent from Green Pea 
galaxies \citep{Yang2017_esc}. Table~\ref{tab:new_lya} lists the fitted velocity offsets
and line widths for the LAGER-z7OD1 LAEs.

The components fitted to LAE-1, LAE-2, and LAE-8 lie on the \citet{Verhamme2018} relation.
These are the highest S/N ratio line profiles. The large error bars on many of the other data
points leave open the possibility of agreement with the \citet{Verhamme2018} relation.
At low S/N ratio, the multi-component fits are subject to parameter degeneracies. The more
well-defined line profiles were all extracted from the deeper observation. More integration
time needs to be acquired through the second mask to clarify the significance of this offset.

It is interesting, however, that  the LAGER-z7OD1 population is offset to the left and upwards in 
Fig.~\ref{fig:dv_fwhm_nofit}.  
Prior to the completion of reionzation, the  \lya\ line profile emergent from  a galaxy is further
modified by radiative transfer in the IGM. Between redshifts  $z=5.7$ and $z=6.6$, 
\lya\ line widths decrease due to the increasing neutral hydrogen fraction in the 
IGM \citep{Songaila2018,Songaila2024,Mukherjee2024}. Attenuation by a more neutral IGM 
is therefore expected to shift the observed line profiles in the observed direction.

\begin{deluxetable*}{lllllllllll}
\tablecaption{Properties of \lya\ Emission}
\colnumbers
\tablehead{
\colhead{Galaxy} &
\colhead{z(\lya)} &
\colhead{$\Delta V(\lya)$} &
\colhead{FWHM$_1$} &
\colhead{V$_2$} &
\colhead{FWHM$_2$} &
\colhead{FWHM$_{NF}$} &
\colhead{$\fesclya\ $} &
\colhead{\fism} &
\colhead{$T_{IGM}^{Ly\alpha} $} &
\colhead{$R_{i}^{min} - R_{i}^{ism}$} 
\\
\colhead{}        &
\colhead{}        &
\colhead{(km~s$^{-1}$)}  &
\colhead{(km~s$^{-1}$)}  &
\colhead{(km~s$^{-1}$)}          &
\colhead{(km~s$^{-1}$)}          &
\colhead{(\kms)}  &
\colhead{}   &
\colhead{}        &
\colhead{}        &
\colhead{(Mpc)}        
}
\startdata
LAE-1   &
$6.93658 \pm 0.00012$ &  
$ 139 \pm 6 $      &
$ 168 \pm  18$ &
$ 380 \pm  33$     & 
$ 423 \pm   5.4$     &
$ 370 \pm 20  $ &
$ 0.14 \pm 0.03  $ &
$ 0.63  $  &
$ 0.22   $ &
$ 0.38 - 0.52   $     
\\
LAE-2   &
$6.93084 \pm  0.00044 $ &
$ 138 \pm 17 $      &
$ 179 \pm  52  $  &
$324 \pm  148 $        & 
$ 325 \pm 205 $        &
$ 310 \pm 50  $  &
$ 0.29 \pm  0.06 $  &
$ 0.63 $ &
$ 0.47 $ &
$ 0.69 - 1.21   $     
\\
LAE-8  &
$ 6.9221 \pm  0.0004 $ &    
$ 102 \pm 25$   &
 56  &
$ 282 \pm  38 $ &  
$ 341 \pm  54 $ &
$ 330 \pm 50  $ &
$ 0.19 \pm 0.07  $ &  
$ 0.94  $  &
$ 0.20  $ &
$ 0.51 - 0.60  $     
\\
LAE-10  &
$6.92352 \pm  0.00033 $ &
$ 158 \pm 20$   &  
$91 \pm  46$    &
$288 \pm  75 $        & 
$ 247 \pm  115$         &
$ 200 \pm 45 $    &
$ 0.10  \pm 0.04  $ & 
$  0.50 $  &
$ 0.21  $ &
$ 0.27 -  0.49   $  
\\
LAE-11  &
$6.9644 \pm 0.0003 $ &  
$ 125 \pm 13$   &  
56 &    
$ 165 \pm 24 $        &  
$ 241 \pm  46 $        &
$ 105  \pm 50  $        &
$ 0.18 \pm 0.06  $ &  
$ 0.72  $  &
$ 0.24   $ &
$ 0.47 - 0.57   $      
\\
LAE-13   &
$6.9357 \pm  0.0002 $ &  
$ 221 \pm 11 $   &  
$ 56$ &
$ 316 \pm 28 $ & 
$ 262 \pm  51 $ &
$ 240  \pm 50 $ &  
$ 0.05  \pm 0.02  $ &  
$  0.25  $  &
$ 0.21  $ &
$ 0.07  - 0.42   $    
\\
LAE-14  &
$ 6.93651 \pm  0.00058 $ &
$ 153 \pm 22$   &
$131 \pm 64 $    &
$315 \pm  159 $        & 
$ 258 \pm  253 $        &
$ 235 \pm 60  $    &
$ 0.10 \pm 0.05  $ &
$  0.53  $  &
$ 0.20  $ &
$  0.28 -  0.48  $     
\\
LAE-15  &
$ 6.971 \pm 0.002 $\tablenotemark{a} &
$ 184 \pm 75$      &
\nodata  &
\nodata &
\nodata  &
\nodata  &
$ 0.18 \pm 0.03  $ &
$  0.38 $  &
$ 0.48  $ &
$ 0.39 -   1.20 $      
\\
LAE-22  &
$ 6.9200 \pm 0.0003 $ &
$ 175 \pm 22$ &
$ 56 $ &
$ 356 \pm 70 $        & 
$ 210 \pm 160 $        &
$ 320 \pm 50 $        &
$ 0.18 \pm 0.05  $ &
$  0.42 $  &
$  0.44  $ &
$  0.41-  1.29 $     
\\
\enddata
\tablenotetext{a}{Measurement from \cite{Hu2021}.}
\tablecomments{
{\it (Col 1):} Galaxy name from \cite{Hu2021}.  
               Supplementary Table 1 of \citet{Hu2021} lists coordinates.
               However, LAE-22 at (RA, DEC) = (10:02:38.755, +02:07:43.59) was
               identified after publication (Hu, pvt. comm.). The latest CHORUS stack 
               \citep{Kikuta2023} places LAE-15 at (10:02:23.383, +02:05:05.10) 
               (Kikuta, pvt. comm.).
{\it (Col 2):} Redshift of the \lya\ emission computed from the Keck LRIS spectra; 
               see Section~\ref{sec:lris}.
{\it (Col 3):} {\clm Velocity offset of the \lya\ emission relative to the systemic redshift in Column 4
               of Table~\ref{tab:jwst_offset}.}
{\it (Col 4):} Width of the narrow \lya\ component in the two-component fit.
               The spectral resolution sets a minimum line width.
{\it (Col 5):} Velocity offset of the broad \lya\ component relative to systemic redshift.
{\it (Col 6):} Width of the broad \lya\ component.
{\it (Col 7):} {\clm Full width at half-maximum intensity.}
{\it (Col 8):} The \lya\ escape fraction defined by Eqn.~\ref{eqn:fesc_lya}. As in
               \citep{Yang2017_esc}, $L(\lya)$ is not corrected for extinction.
               We considered two different values of the extinction correction for
               $L(\Ha)$; the limit of no dust attenuation, $A_V = 0$, is listed here.
{\it (Col 9):}  Estimated ISM escape fraction for \lya\ photons
                based on the measured \dv\ and a
                fiducial $ \rm{E(B-V)} = 0$. Computed from an empirical relation defined by
                Green Pea galaxies \citep{Yang2017_esc}.
{\it (Col 10):} IGM transmission of \lya\ from Eqn.~\ref{eqn:fism}
               including the minimum ISM attenuation of \lya, 
               as defined by Eqn.~\ref{eqn:yang2017}  with $E_{B-V} \approx 0$, which
               is the minimum dust correction consistent with the Balmer decrement. 
{\it (Col 11):} Minimum size of the ionized bubble surrounding each LAE following \citet{Mason2020}.
                The first value adopts Col.~6 for the IGM transmission.
                The second, larger radius assumes that Col.~7 describes the fraction of the
                intrinsic \lya\ emission emerging from the ISM of a galaxy, boosting
                the IGM transmission by $1/\fism$.
}
\label{tab:new_lya}
\end{deluxetable*}

\subsection{\lya\ Velocity Offsets and \lya\ Escape Fractions} \label{sec:voff}

The shape of the \lya\ line profile emerging from a galaxy affects the IGM transmission of \lya.
One challenge in using LAEs to study reionization is that measuring IGM transmission of \lya\ 
requires some knowledge of the \lya\ line profile emergent from the galaxies. The \lya\ line 
profiles of the nearest local analogs provide direct insight into the profile shape emergent
from galaxies in the absence of any IGM attenuation. In this spirit, \citet{Hayes2023} discuss
scaling \lya\ profiles of local galaxies. Yet even the population of low mass, starbursting
galaxies exhibits wide variation in \lya\ line profiles \citep{Hu2023}. 

The local population of Green Pea galaxies offer the nearest spectral match to  the LAGER-z7OD1 
LAEs. The optical emission-line spectra \citep{Jaskot2013,Yang2017_esc,Izotov2018b,Izotov2020a,Jaskot2019}
indicate common physical properties include high ionization parameters, low metallicity 
(Moya-Sierralata \et 2025, in prep), and high specific star formation rates. The \lya\ line 
profiles of most Green Pea galaxies exhibit a blueshifted peak in addition to a stronger, 
redshifted peak \citep{Neufeld1991}. We note that a substantial fraction of their \lya\ 
spectra have non-zero flux at the systemic velocity \citep{Yang2017_esc}.  In contrast, 
P~Cygni line profiles,  are common in more massive galaxies \citep{Shapley2003,Verhamme2006}.

Two obvious differences distinguish the \lya\ line profiles of  LAGER-z7OD1 LAEs and Green Pea galaxies. 
The line profiles of the reionization era galaxies show no blueshifted emission and also no 
emission at the systemic redshift. These differences confirm the expectation that the environment
of reionization-era galaxies further attenuates the emergent \lya\ emission. To constrain the combined
ISM and IGM attenuation, we measure \lya\ velocity offset and \lya\ escape fraction.

We define the
\lya\ velocity offset relative to the systemic redshift,
\begin{eqnarray}
\dv  = c \frac{z_{Ly\alpha} - z_{sys}}{ 1 + z_{sys}}.
\end{eqnarray}
Considering the broad width and complexity of the \lya\ line profile, the physical meaning 
of \lya\ redshift may change with spectral resolution and S/N ratio. In this study, the lower
velocity component defines the \lya\ redshift because it describes the redshift of
the peak \lya\ intensity. For our \lya\ spectra, this definition is therefore not sensitive
to how the line profile is fit. Col.~7 of Table~\ref{tab:new_lya} lists these measurements.

The \lya\ escape fraction describes the attenuation of the intrinsic \lya\ emission.
The extinction-corrected \Ha\ luminosity determines the intrinsic \lya\ luminosity
via Case~B recombination theory \citep{Osterbrock2006}.  At the gas densities and
temperatures of interest \citep{Henry2015}, the fraction of \lya\ photons observed
becomes
 \begin{eqnarray}
 \fesclya\ \equiv \frac{L_{Lya}}{8.7 L_{Ha}}.
  \label{eqn:fesc_lya} \end{eqnarray}
By this definition, the numerator is not corrected for extinction.
Taking the minimum
extinction consistent with the Balmer decrement,  $A_V \approx 0$, we obtain 
upper limits on \fesclya\ between 5\% and 30\%  for the LAGER-z7OD1 LAEs.
These values listed for individual LAEs in Column~8 of Table~\ref{tab:new_lya}.

\begin{figure}[h]
 \centering
    \includegraphics[scale=0.55,angle=0,trim = 0 0 0 0]{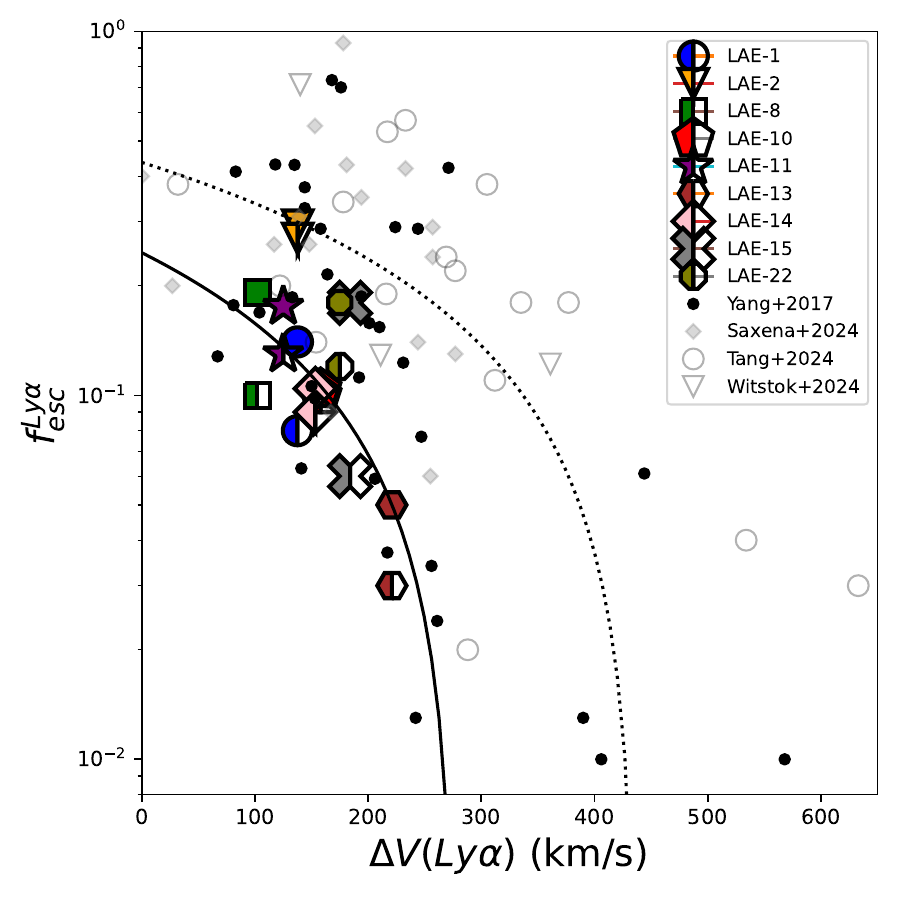}
     \caption{Measurements of \lya\ escape fraction vs. \lya\ velocity offset 
       {\clm
         (col.~3 in Table~\ref{tab:new_lya}). Increasing column densities of {\sc HI} 
         are expected to increase the velocity offset and reduce \fesclya.
       Comparison of left-filled symbols to filled symbols illustrate how the
       \Ha\ extinction correction reduces \fesclya.
       }
       A linear fit to the left-filled symbols has a lower y-intercept than the fit
       to Green Pea galaxies (dotted curve). 
This offset is consistent with, but does not
       require, intergalactic attenuation of \lya\ towards LAGER-z7OD1.  
       See text for description of light gray points.
}
    \label{fig:f_dv}
     \end{figure}

Figure~\ref{fig:f_dv} demonstrates an anti-correlation between \fesclya\  and
\lya\ velocity offset among LAGER-z7OD1 LAEs. This correlation has been
observed among Green Pea galaxies \citep{Yang2017_esc} as well as reionization-era LAEs
\citep{Saxena2024,Tang2024,Witstok2024-z9}. It is naturally explained by radiative transfer
effects. The more times resonance photons scatter, the further they diffuse away from the
line center, and the larger their probability of destruction.

If the factors determining \lya\ escape from the ISM are similar in local analogs and 
reionization-era galaxies, then only radiative transfer effects in reionization-era IGM 
differentiate their observed \lya\ properties. In Fig.~\ref{fig:f_dv}, LAGER-z7OD1 
LAEs are offset to lower \fesclya\ relative to Green Pea galaxies with the same velocity
offset.  This downward offset is expected for a partially neutral IGM.  
{\clm
A linear fit to the left-filled symbols, $\fesclya\ = -0.0009 (\pm 0.006) \dv + 0.25 (\pm 0.10)$, 
has a lower y-intercept than the fit to Green Pea galaxies (dotted curve), 
$\fesclya\ = -0.0009 (\pm 0.0004) \dv + 0.44 (\pm 0.08)$. The statistical significance of 
this offset is just 1.5 standard deviation, so it does not firmly rule out a fully ionized IGM 
toward LAGER-z7OD1. The extinction correction to the \Ha\ luminosities (left-filled symbols)
increases the significance of the offset but has a signficant uncertainty. 
}

{\clm
A comparison to published measurements for redshift 6 to 9 LAEs, light gray points in Fig.~\ref{fig:f_dv}, 
reveals variations in \fesclya\ approaching a dex at fixed velocity offsets. It is not
clear whether this scatter represents physical differences because these studies 
do not use a consistent definition of \fesclya.
The JADES points from \citet[][JADES LAEs $z \approx 6.3$]{Saxena2024} and
\citet[][higher $z$ LAEs]{Witstok2024-z9} correct the \lya\ luminosity 
for dust attenuation.  The large baseline in wavelength between the Balmer 
lines and \lya\ systematically shifts these estimates to higher values than 
the standard definition of \fesclya\ \citep{Yang2017_esc} would give.

Direct comparisons will also require corrections for \lya\ aperture losses.
}

\begin{figure}[h]
 \centering
  \includegraphics[scale=0.7,angle=0,trim = 0 0 0 0]{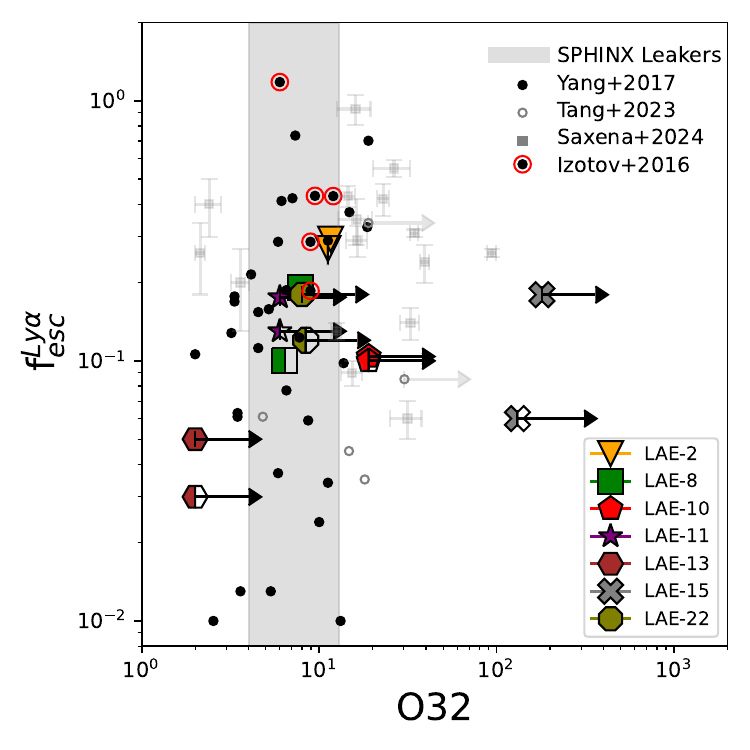}
  \caption{The \lya\ escape fraction versus the O32 emission-line ratio. The nebular
    excitation of the LAGER-z7OD1 LAEs is consistent with conditions necessary for LyC
    leakage, as indicated by Green Pea galaxies with direct LyC detections (red circles,
    \citet[][]{Izotov2016-leakers}) and the most favorable ratios for leakage
    in radiation-hydro simulations (gray band, \citet{Choustikov2024_escape_physics}).
    Comparison of the solid, colored symbols ($A_V = 0$ limit) to left-filled symbols
    ($A_V^*$ or $E(B-V)^{gas}$, as described in the text) illustrates our best 
    correction for dust to date. We have converted to the definition of O32 which 
    sums the \oiii\ $\lambda \lambda 4960, 5008$ intensities.
     }
    \label{fig:fesclya_o32}
     \end{figure}

Green Pea galaxies  provide insight
into additional diagnostics sensitive to the neutral gas and dust content of the ISM. For example,
Figure~\ref{fig:fesclya_o32} shows that over just a factor of four range in O32 ratio, Green Pea
galaxies span two decades in \lya\ escape fraction. In this range of O32 ratios, indicated by gray
shading in Figure~\ref{fig:fesclya_o32}, high \lya\ escape fraction selects the subset of Green
Peas which have direct detections of LyC leakage \citep{Izotov2016-leakers}.   
Their similar locus in the \fesclya\ -- O32 plane  is consistent with  LAGER-z7OD galaxies 
being Lyman continuum leakers.

Local analogs do not reproduce the full range of of physical conditions revealed by 
JWST observations. The influence of short dynamical timescales and cold streams, for example,
on \fesclya\ are not captured by Green Pea scaling relations. Mock spectra generated from 
radiation hydrodynamics simulations provide important insight into the relation between  \fesc,
\fesclya,  and spectral properties.
The gray band in Figure~\ref{fig:fesclya_o32} illustrates the range of 
O32 values ($4 < {\rm O32} < 13$) among simulated galaxies with significant 
LyC leakage ($> 5\%$)  \citep[][]{Choustikov2024_escape_physics}. 
{\clm
\citet[][]{Choustikov2024_escape_physics} find that \fesc\ actually declines when 
O32 values exceed this range. We measured the O32 ratio for seven LAEs in LAGER-z7OD1.
Six of them have O32 values in this preferred range for LyC escape, which is consistent
with the argument that they are LyC leakers.
}

\section{Discussion} \label{sec:discussion}

The short mean free path of ionizing photons in the neutral IGM guarantees a relatively
sharp transition from ionized to neutral gas during reionization \citep{Gnedin_Madau2022}. 
{\clm
Since LyC leakers are often also LAEs \citep{Verhamme2017,Choustikov2024-lya}, these ionized 
pockets can enhance \lya\ transmission during the reionization era. 
}
If the shape of the \lya\ line profile emerging from a galaxy
were known, then comparison to the observed \lya\ profile would determine the distance to
the first neutral patch along our sightline. 

We estimate that distance here using an idealized picture described by \citet{Mason2020}. 
We assume that an ionized bubble surrounds each LAE, and the IGM outside that bubble is entirely neutral.
The expansion of the universe redshifts the \lya\ line profile emergent from a galaxy. Any blueshifted 
component shifts through the resonant frequency, where even the ionized region of the pathlength
is optically thick due to hydrogen recombinations. In contrast, a line component which emerges 
redward of the systemic redshift may be partially transmitted. Using a simple approximation
for each emergent line profile, we calculate how the intergalactic transmission of \lya\
depends on the distance to the first neutral patch.

A distance from each LAE to the first neutral patch can then be defined by our measurement of 
the \lya\  transmission and velocity offset. Importantly, we also illustrate how a correction 
for interstellar 
attenuation of \lya\ reduces the inferred intergalactic transmission (relative  to \fesclya), 
and thereby increases our estimated distance to the first neutral patch. We leave detailed 
modeling of the emergent \lya\ line profile to future work.

Interpreting these ionized pathlengths as the radii of ionized bubbles, we discuss their
ionization requirements in \S~\ref{sec:Nph}. We model the recent star formation history of 
each LAE in Sec.~\ref{sec:sfh} and discuss evidence for AGN in Sec.~\ref{sec:agn}.  
Section~\ref{sec:budget} then discusses how the ionization budget might be balanced.


\subsection{Sizes of Ionized Bubbles} \label{sec:rion}

\begin{figure}[h]
 \centering
  \includegraphics[scale=0.5,angle=0,trim = 0 0 0 0]{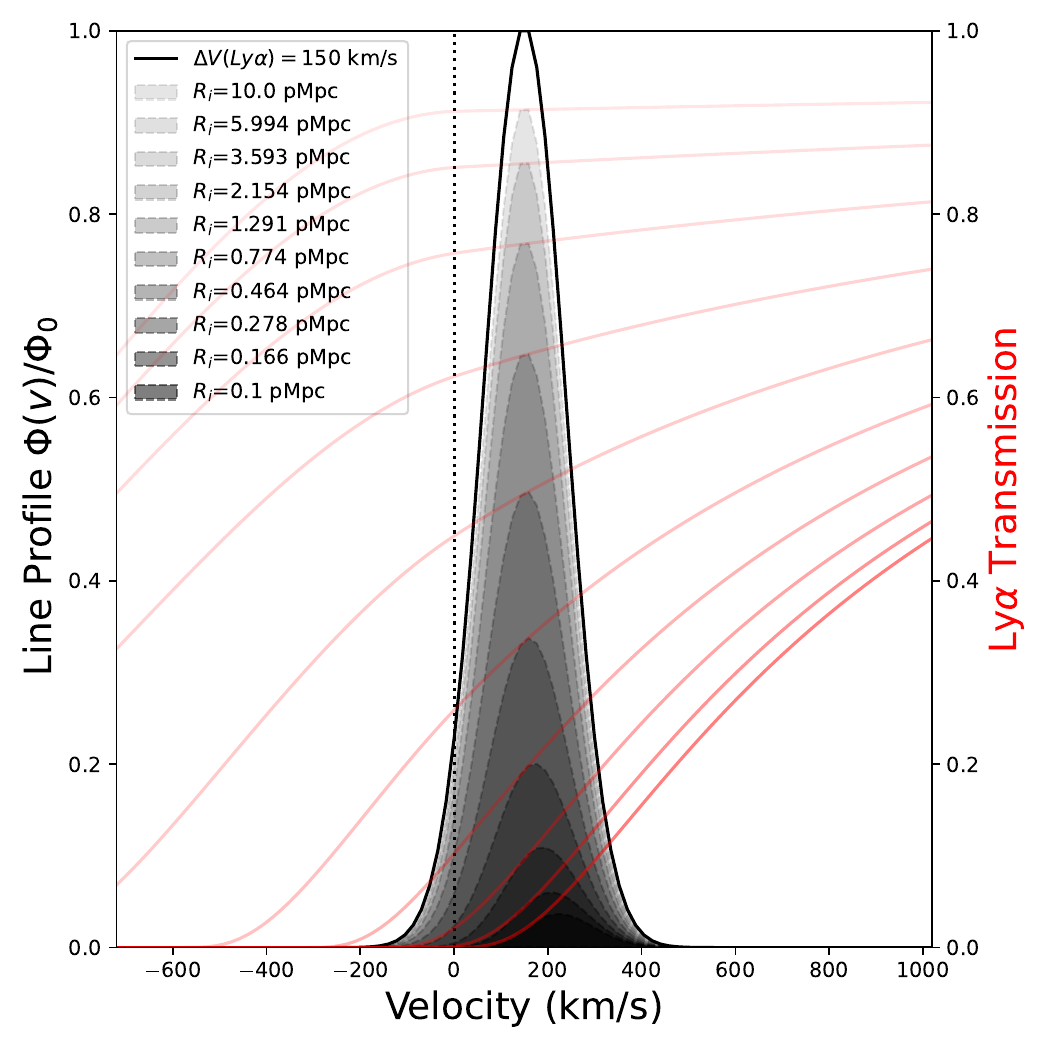}
   \caption{Impact of intergalactic \lya\ transmission (red curve) on the
     \lya\ line profile (black curve). Shading denotes the distance to the
     distance, $R_i$,  from a galaxy to the first neutral patch along the sightline.
     In the example shown, the  line profile emerging from the ISM has a velocity 
     offset of 150 \kms. The IGM significantly attenuates the line flux but only
     slightly shifts the velocity offset (redwards).
     }
    \label{fig:lya_profile}
     \end{figure}

We model the emergent \lya\ profile by a single, redshifted Gaussian profile. The velocity
offset is taken from 
{\clm
Col.~7 of Table~\ref{tab:new_lya},
}
 and the line width calculated from
\citep[][Eqn.~2]{Verhamme2018}.   The solid, black line in Fig.~\ref{fig:lya_profile}
shows an example.

At each observed frequency, we calculate the \lya\ optical depth by integrating along the 
sightline from a LAE to the end of the reionization era. Converting the pathlength to an 
integral over redshift, we obtain
\begin{eqnarray}
\tau(\nu_{obs}) = \int_{z_R}^{z_s} n_{HI}(z) \sigma_L(\nu_{obs}(1+z),T) \times \\
\frac{c}{(1+z) H(z)} dz.
\end{eqnarray}
The ambient hydrogen density increases with redshift as the baryon density, 
$ \rho_b(z) = \bar{\rho}_{b,0}(1+z)^3 \delta $, corrected for the cosmic helium fraction and 
hydrogen ionization fraction. In recognition of a possible density enhancement, we include the local 
overdensity, $\delta$, as a parameter. We present results for $\delta = 1$, however, simply because 
modeling all important parameters --  including infall into the overdense region \citep{Santos2004,Dijkstra2014}
and outflows from galaxies \citep{Haiman2002,Dijkstra2011} -- is well beyond the scope of what the 
dataset can constrain at this time.

We break the integral into two parts. Between the bubble redshift and the LAE redshift, 
recombinations inside each bubble produce a small residual neutral hydrogen. Following 
\citet{Mason2020}, we adopt $\xh(0.1 {\rm ~pMpc}) = 10^{-8}$ near the galaxy and describe
the increasing neutral fraction with distance from the galaxy by $\xh (r) \propto r^2$. 
Outside the bubble, the model requires that the sightline traverse fully neutral IGM 
($\xh = 1$ for at least several Mpc) before encountering  reionized IGM.
We model the \lya\ scattering cross section in the frame of the hydrogen atoms. 
Cosmic expansion redshifts the source photons as they traverse 
the large bubbles. For each source redshift, $z_s$, the optical depth at a velocity 
offset $\Delta v$ in the frame of the atom determines the transmission at an observed wavelength,
\begin{eqnarray}
\lambda_{obs} = \lambda_{\alpha} (1 + z_{abs}) =  (1 + \dv /c) (1 + z_{s}) \lambda_{\alpha}
\end{eqnarray}
where $\lambda_{\alpha} = 1215.67 {\rm ~\AA} $.

The frequency-dependence of the scattering cross section for \lya\ photons
depends on the velocity distribution of the hydrogen atoms as well as the natural
line width. We model it using the Voigt profile approximation from \cite{Tasitsiomi2006}. 
In the rest frame of the gas outside the bubble, the damping wing cross section
decreases smoothly with frequency, $x^{-2} \propto (v / c \Delta \nu_D)^{-2}$,
leading to an $x^{-1}$ decrease in the damping wing optical depth. Since the attenuation
declines slowly across the line profile, the change in  shape of the line profile can be subtle.

Fig.~\ref{fig:lya_profile} illustrates the damping wing attenuation of a redshifted, Gaussian component. 
For the parameters considered, the optical depth is dominated by the \lya\ 
damping wing absorption, which is produced almost entirely from the neutral IGM outside 
the bubble. Resonance absorption from gas inside the bubble contributes relatively little  
optical depth due to the low neutral fraction and velocity offset. 
The main affect is attenuation of the line flux. The damping wing absorption will attenuate $L(\lya)$, 
\fesclya, and W(\lya) by roughly a factor of two when the ionized pathlength is smaller than 2 Mpc.
Yet the (redward) velocity shift of the red peak is barely detectable 
due to the slow variation of the damping wing optical depth with wavelegnth.

\begin{figure}[h]
 \centering
\includegraphics[scale=0.55,angle=0,trim = 0 0 0 0]{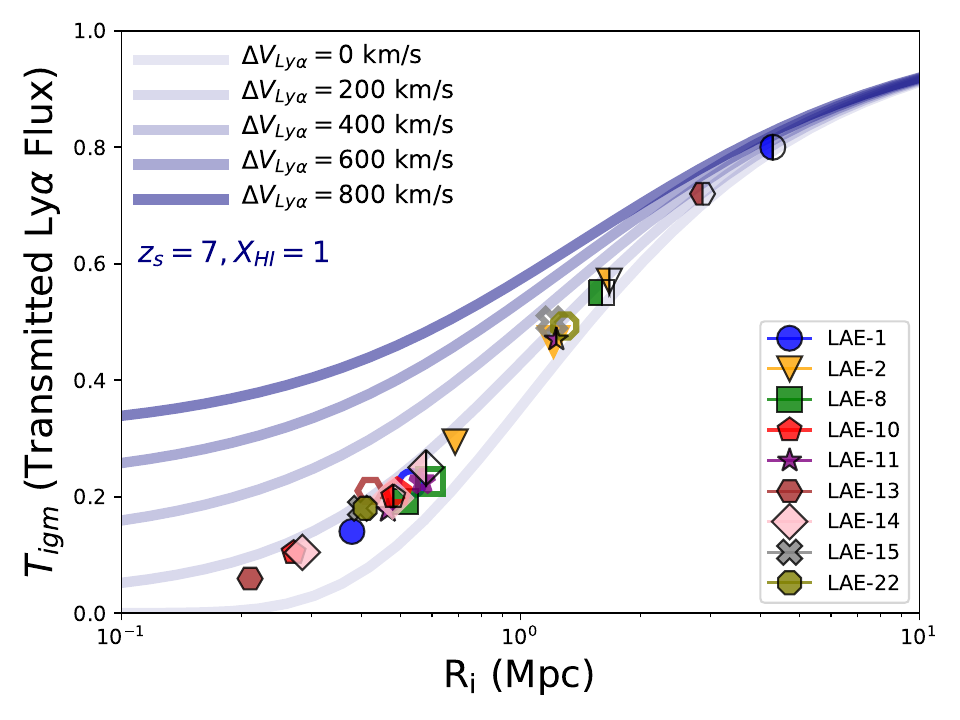}
\caption{Transmission of redshift $z \approx  7$ \lya\ emission by an ionized bubble in a neutral
            IGM.  Tracks illustrate a different velocity offsets of the \lya\ profile emergent
            from the galaxy.  As the proper distance from the source to the first neutral patch 
            grows, the decrease in \lya\ damping wing optical depth increases \lya\ transmission.
            At fixed bubble radius, a larger \lya\ velocity offset enhances transmission. 
            Symbols mark the intersection of the \dv\ measurement for each LAE with
            the inferred intergalactic \lya\ transmission. Symbols illustrate three scenarios
            for $T_{igm}$: (1) no ISM attenuation of \lya\ ($\fism = 1$, solid),
            (2) the minimum ISM correction (Col.~9 and 10 of Table~\ref{tab:new_lya}, open), and
            (3) the full ISM correction including both the color excess
            and  velocity offset terms in Eqn.~\ref{eqn:yang2017} (half-filled symbols).
            Modeling interstellar attenuation of \lya\ increases the inferred IGM transmission,
             and we infer larger bubbles.
}
    \label{fig:T_Rion}
     \end{figure}

The observed line profile is the product of the intergalactic attenuation and the emergent 
line profile. Converting that profile from frequency to 
velocity offset, we integrate over the observed line profile and normalize
the result by the emergenet line profile  to obtain the intergalactic transmission
\begin{eqnarray}
T_{igm}(\dv, R_i) \equiv \frac{\int \phi(v) e^{-\tau(v) } dv}{\int \phi(v) dv}.
\end{eqnarray}
Figure~\ref{fig:T_Rion} illustrates how increasing the bubble radius improves the net
\lya\ transmission. Smaller velocity offsets require larger bubbles in order to produce the
observed amount of \lya\ transmission. 

We computed this relation for  each LAE based on the source redshift (Col.~4 of 
Table~\ref{tab:jwst_offset}) and the  \lya\ velocity offset 
{\clm 
(Col.~3 in Table~\ref{tab:new_lya}.
}
The intersection of this curve and the measured \lya\ transmission
defines a bubble radius.

\subsubsection{Measurement of Intergalactic \lya\ Transmission}

We define the IGM transmission by comparing the measured \lya\ luminosity
to the intrinsic \lya\ luminosity of the galaxy,
\begin{eqnarray}
T_{igm} = \frac{L^{obs}(\lya)}{8.7L(\Ha) \fism},
\label{eqn:fism} \end{eqnarray}
where Case~B recombination theory determines the later given a Balmer line luminosity.

\begin{figure}[h]
 \centering
     \includegraphics[scale=0.55,angle=180,trim = 100 0  0 0]{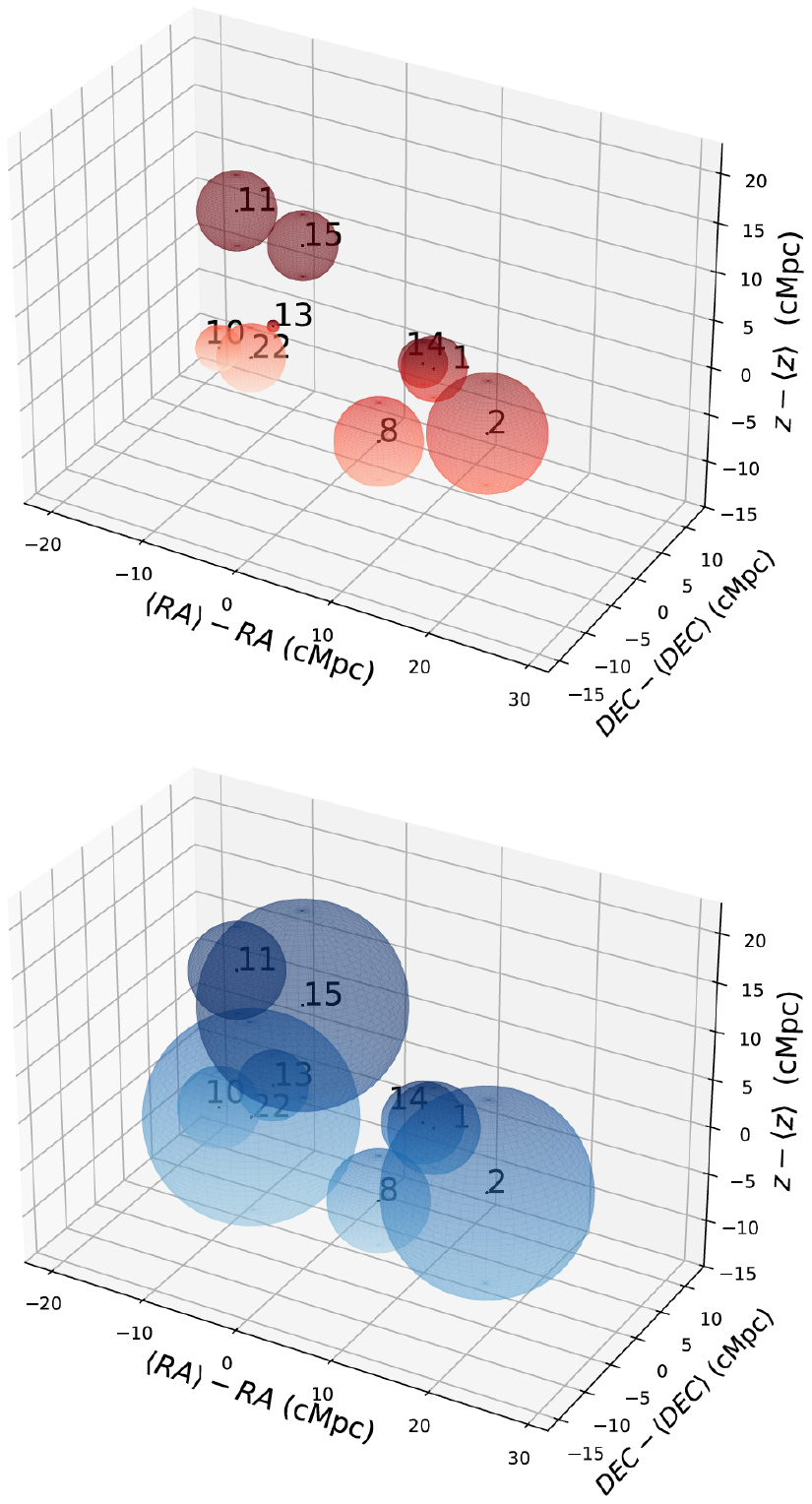}
    \caption{Topology of ionized bubbles in LAGER-z7OD1. Each bubble radius
      represents the distance to the first neutral patch along the line of sight.
      The IGM outside the ionized bubble is assumed to be fully neutral.
      {\it (a)}  Distances based on $T_{igm} = \fesclya$, where attenuation of \lya\
      by the ISM is neglected.  
      {\it (b)} Applying an empirical model for ISM attenuation \citep{Yang2017_esc},
      we obtain higher IGM transmission and infer larger bubbles. A conservative
      estimate of ISM attenuation of \lya\ yields overlapping bubbles. 
}
     \label{fig:myweb_rion}
      \end{figure}

Following previous work on high-redshift LAEs \citep{Mason2020, Witstok2024}, 
we first ignore ISM attenuation. This  is equivalent to taking $\fism \approx 1$
in Equation~\ref{eqn:fesc_lya} and gives $T_{igm} \approx\ \fesclya$. 
The first entry  in Col.~11 of Table~\ref{tab:new_lya} 
lists bubble radii estimated this way. Their sizes range from  $R_{ion} = 70$~kpc 
(LAE-13) to  690 kpc (LAE-2). 
Figure~\ref{fig:myweb_rion} illustrates the ionized volume around each LAE 
based on the simple model described above.  In the limit of $\fism \approx 1$,
the ionized bubbles are just approaching overlap.

Studies of  LAEs in the post-reionization universe offer valuable insight 
into the \lya\ escape fraction from the ISM. The fully ionized IGM transmits 
the \lya\ profile emergent from these lower-redshift galaxies without modifying
its shape. The empirical relation from \citet{Yang2017_esc}  illustrates how
the fraction of \lya\ photons escaping from the ISM  depends on galaxy properties:
\begin{align}
 \log \fism =    -0.437 \times \frac{E(B-V)}{0.1} - 0.483 
  \times \nonumber \\
   \frac{\Delta V(\lya)}{100~{\rm km/s}} + 0.464.
    \label{eqn:yang2017} \end{align}
Since this relation was fitted to Green Pea galaxies, whose spectral properties 
and morphologies are similar to reionization-era LAEs, we expect the dependence on 
\lya\ velocity offset and gas reddening fairly describes the escape of \lya\ photons 
from the ISM of their reionization-era analogs.

{\clm
Eqn.~\ref{eqn:yang2017} is a two-parameter fit because \citet{Yang2017_esc}
did not find a significant correlation between \fesclya\ and velocity offset.
Using the reddening estimates discussed in Sec.~\ref{sec:avstar} and
Sec.~\ref{sec:reddening}, we estimated the instellar escape fraction from
each LAE using Eqn.~\ref{eqn:yang2017}. Notice that while the color excess
term lowers \fism\ in Eqn.~\ref{eqn:yang2017}, the extinction corrected
concomitantly raises \Ha\ luminosity in Eqn.~\ref{eqn:fism}. We emphasize
that their product will always reduce the denominator in Eqn.~\ref{eqn:fism},
thereby raising the inferred IGM transmission, $T_{igm}$. Larger corrections
for dust therefore increase the estimated distances to the first neutral patch.
Col.~10 of Table~\ref{tab:new_lya} lists these larger bubble radii, $R_i^{dust}$.
We emphasize, however, that our current lower limits on reddening and
color excess allow nearly dust-free LAEs.

Inserting our \dv\ measurements into Equation~\ref{eqn:yang2017} with $E(B-V) \approx 0$
defines our {\it dv-bubble} scenario. It illustrates how even the minimum ISM
correction, with no dust, significantly enlarges the inferred ionized volume. 
With  a velocity
offset of $\Delta V \approx 100$ \kms, a dust-poor ISM transmits nearly
all the \lya\ photons. Yet less than half of the \lya\ photons escape from the ISM
when the velocity offset increases 158 \kms. Col.~9 of Table~\ref{tab:new_lya} lists 
these \fism\ estimates. This minimum ISM correction increases the inferred radii of
the bubbles around LAE-13, LAE-22, and LAE-15 by factors of 6,  3, and 3,
respectively, relative to the no-ISM estimate. The second value in Col.~11 of
Table~\ref{tab:new_lya} lists these radii as $R_i^{ism}$. This dust-free ISM
correction increases the bubble radii enough that some individual bubbles overlap,
as illustrated in the bottom panel of Figure~\ref{fig:myweb_rion}. 
}

\subsection{Required Number of Ionizing Photons}  \label{sec:Nph}

In this section, we examine the ionization requirements of
spherical bubbles centered on the individual LAEs.  We initially
assume the bubbles remain in the pre-overlap phase of reionization.
Later, in \S~\ref{sec:overlap}, we discuss how bubble overlap
impacts the ionization budget.

During the pre-overlap phase of reionization, the propagation of the ionization 
front describes the evolution of the bubble radius
\citep[][Equation~3]{Cen2000}:
\begin{eqnarray}
 \frac{dR_i^3}{dt} = 3H(z) R_i^3 + 
  \frac{3 \dot{N}_{ph}}{4 \pi n_H} - 
   C n_H \alpha_B R_i^3,
    \label{eqn:CH2000} \end{eqnarray}
where $C$ is the mean clumping factor of ionized gas within the bubble.
We can simplify this expression because the volume of the ionized bubble is
growing rapidly compared to both the cosmic expansion rate (first term)
and the recombination rate inside the bubble (third term). The terms for
recombination and cosmic expansion only become important for very large bubbles.
For $\dot{N}_{ph}$ in the range $10^{54} -  10^{55}$~s$^{-1}$,
it is a fair approximation when $R_i$ is smaller than 1 to  2.4~Mpc.
In this limit, integration of $\dot{N}_{ph}$ over the starburst duration
yields the required number of ionizing photons to grow an ionized bubble of radius $R_i$:
\begin{eqnarray}
 \log {N}_{ph}  = 69.18 + 3 \log \left [ \frac{R_i}{ 0.500 {\rm ~Mpc}} \right ] + \nonumber \\  
                   3  \log \left [ \frac{1+z}{ 8} \right  ] +
                    \log \barxh \delta,
   \label{eqn:Ndot} \end{eqnarray}
where we have assumed a mean hydrogen density  
$\bar{n}_{H}  = 1.04 \times 10^{-4} {\rm ~cm}^{-3} [(1+z)/8]^3$,
for a local overdensity, $\delta \equiv \rho / \bar{\rho})  $, of one.
This estimate for the required number of ionizing photons is a minimum since
it requires one photon per hydrogen atom.

The ionizing source will clearly need to produce more LyC photons than Eqn.~\ref{eqn:Ndot} requires
because some will be absorbed by interstellar gas. Since the extinction-corrected Balmer line 
luminosities measure the hydrogen recombination rate in the ISM, we can measure the ionizing photon
production rate up to this factor of $(1 - \fesc)^{-1}$. This arguement is independent of whether
stars or AGN produce the LyC photons. Multiplying this rate by the duration of the activity
and \fesc\ then determines the supply of LyC photons. We calculate this quantity here and
examine the implications of setting it equal to required $N_{ph}$ in Eqn.~\ref{eqn:Ndot}.

Let $Q_*(t)$ describe the Lyman continuum
production rate, which may include an AGN contribution in addition to that from massive stars. 
Then  the photoionization rate of the surrounding IGM is
\begin{eqnarray}
 \dot{N}_{ph}(t)  = Q_*(t) \fesc(t),
  \label{eqn:dotNph}  \end{eqnarray}
where \fesc\ describes the Lyman continuum escape fraction from the galaxy.
Case~B recombination within the galaxy produces an \Ha\ luminosity
\begin{eqnarray}
 L(\Ha) =   Q_* (1 - \fesc)  \frac{\alpha_{Ha}^{eff} h \nu_{Ha}}{\alpha_B(H)},
  \label{eqn:ha} \end{eqnarray}
where we adopt the values of the recombination coefficients at temperature 
$T_e = 2 \times 10^4$ ~K and density $n_e = 100$~cm$^{-3}$ \citep{Hummer1995,Osterbrock2006}. 
The measured \Ha\ and \Hb\ line luminosities therefore determine
the recent production rate of ionizing photons in each LAE,
\begin{eqnarray}
 Q_*(H) = 1.07 \times 10^{54} {\rm ~s}^{-1} \frac{L(\Ha)}{10^{42} {\rm ~erg~s}^{-1}}
          \frac{1}{(1 - \fesc)},
  \label{eqn:Q}  \end{eqnarray}
up to a factor $(1 - \fesc)$. We note that
the coefficient increases to $5.15 \times 10^{54} {\rm ~s}^{-1}$ when \Hb\
luminosity needs to be substituted for $L(\Ha)$.

\begin{figure}[h]
 \centering
    \includegraphics[scale=0.55,angle=0,trim = 40 0 0 0]{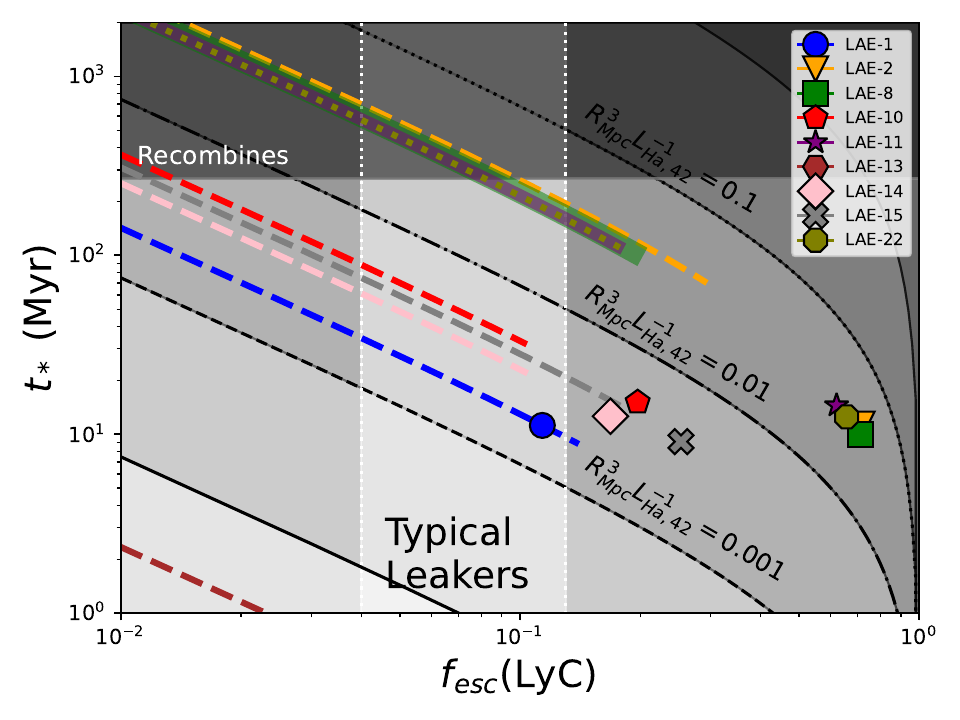}
   \caption{Combinations of starburst age and LyC escape fraction compatible with 
     the {\clm minimum} ionized volume and interstellar \Ha\ luminosity of each LAE. 
     We assume $\barxh \equiv 1$ and $\delta = 1$,  a neutral IGM of average 
     density, and apply Eqn.~\ref{eqn:tstar} at redshift $z = 7$. Only LAE-1
     and LAE-13 (to the lower left) have estimated burst ages (polygons)
     that overlap the track of the same color, which represents the allowed
     range of \fesc. See additional description in text.
            }
       \label{fig:tstar_fesc}       \end{figure}

Inserting Equations~\ref{eqn:Q} and  \ref{eqn:dotNph} into \ref{eqn:Ndot}, we obtain
the timescale required for a LAE to ionize the bubble surrounding it:
\begin{eqnarray}
t_* = 93 {\rm ~Myr} \left [  \frac{L(\Ha)}{10^{42} {\rm ~erg~s}^{-1}} \right ]^{-1}
       \left [ \frac{R_i}{0.493 {\rm ~Mpc}} \right ]^3
        \times \nonumber \\
         \left [ \frac{1 + z }{8} \right ]^{3}  
         \delta \barxh(z) \times
           \frac{1 - \fesc}{\fesc}.
   \label{eqn:tstar} \end{eqnarray}
In words, the \Ha\ luminosity of a specific LAE, along with its estimated bubble 
volume, determine the inverse scaling between the required lifetime of the ionizing
source and the LyC escape fraction. Figure~\ref{fig:tstar_fesc} uses this relation
to illustrate the challenge of ionizing the bubble surrounding each LAE.

The (black) contours in  Figure~\ref{fig:tstar_fesc} represent constant bubble 
volume per unit \Ha\ luminosity.  Colored lines depict the tracks for individual
galaxies. In the limit of no LyC leakage, $\fesc\ \rightarrow 0$, the timescale to 
photoionize a bubble becomes arbitrarily long. In reality, the recombination 
timescale sets the maximum starburst duration of interest, and hence the minimum
\fesc.  Ending each track at $\fesc\ = \fesclya$ defines a minimum source 
lifetime, $\tau_{sB}^{req}$. This upper limit on \fesc\ is admittedly soft
but motivated by conditions in local LyC leakers \citep{Izotov2020a}. Alternatively,
the vertical dotted, white lines select the range of \fesc\ predicted by
radiative transfer simulations \citep{Choustikov2024-lya}.
The required timescales are interesting in part because they differ substantially
among the nine LAEs.

In Figure~\ref{fig:tstar_fesc}, the $R_i^3L_{H\alpha}^{-1}$ tracks for LAE-2, LAE-8, LAE-11,
and LAE-22 all overlap. Terminating each track at $\fesc = \fesclya$, we find minimum
lifetimes from 70 to 113 Myr for their ionizing source. The duration of the most recent
starburst, as estimated from hydrogen Balmer equivalent widths in \S~\ref{sec:sfh}, is
much shorter. A polygon marks the intersection of the extended track with the estimated
duration of a single burst  in Figure~\ref{fig:tstar_fesc}. The polygons for these four galaxies
lie well to the right of their colored lines.

Below them, in contrast, a short burst lasting 9 Myr ionizes the LAE-1 bubble (blue track)
when \fesc\ approaches $\fesclya \approx 0.14$. In the lower left, the LAE-13 bubble is very
easily ionized. The recent activity in the three other LAEs comes close to ionizing their bubbles.
LAE-10, LAE-14, and LAE-15 require only small boosts in LyC leakage. Any combination
of starburst duration and  \fesc\ that increases the supply of LyC photons by a factor
of two relative to the default values of  \fesclya\ and $\tau_{SB}$ works.

In the context of the {\it minimum-bubble} picture then, ionization by the known LAEs alone
could easily be achieved by multiple bursts of star formation.  The duty cycle of bursts 
would need to be about $\approx 33$\% in LAE-8, LAE-11, and LAE-22, i.e. the ratio of 100 Myr 
to the recombination timescale. This frequency of bursts appears to be compatible with 
the \texttt{SPHINX$^{20}$} predictons \citep{Katz2023_sphinx20} and the inferred stochasticity in a
well-studied  $z=7.3$ galaxy \citep{Faisst2024}.

The required number of ionizing photons changes dramatically, however, when we
consider the larger bubbles in the bottom panel of Fig.~\ref{fig:myweb_rion}.
In this {\it dv-bubble} scenario, the ISM provides part of the \lya\ attenuation.
These larger bubbles are not a maximum size. We applied the minimum ISM correction
based on the measured velocity offset and $E_{B-V} \approx 0$. That color excess
is consistent with the lower limits on our measurements. Our best estimates of the
color excess are substantial in several galaxies.  Moderate escape fractions,
$\fesc < 0.25$ for example, are not sufficient for any of the LAEs to ionize
these larger bubbles with a single burst of activity.

If we limit \fesc\ to 10\% \citep{Choustikov2024_escape_physics}, then the question
of whether an LAE can ionize a volume of radius $R_i^{ism}$ depends on its star
formation history. The simple model presented here expresses this requirement as a 
timescale for ionizing photon production (at a constant rate). With no dust then,
but some \lya\ attenuation by the ISM, ionization of the  LAE-1 and LAE-13 bubbles
now require 34 Myr and 43 Myr, respectively. The timescales grow to several hundred
Myr for LAE-8, LAE-10, LAE-11, and LAE-14, and the required duty of bursts is
approaching unity.  It is interesting that LAE-2 and LAE-22 cannot ionize
the larger bubble volume within a recombination timescale. 

Before evaluating the ionization budget further, we examine constraints on the 
star formation histories of each LAE.  We aim to understand whether the recent  burst
or past activity dominates the production of ionizing photons.  For reference,
we quantify the required number of ionizing photons in Col.~11 of Table~\ref{tab:sfh}.
The required duration assumes a constant rate of ionizing photon leakage, 
$\fesc = 0.10$, and no dust. These are conservative, fiducial values. We discuss
the dust correction and its impact further in the next section.

\subsection{Ionizing Sources} \label{sec:sed}

We want to understand whether the LAEs ionize the bubbles that transmit \lya\  emission.
Regions of average IGM density photoionized at $z \approx 9.5$ would not recombine until
redshift $z \sim 7$, so we seek constraints on the ionizing photon production over this
$\approx 270$~Myr period. This measurement is challenging because reionization-era galaxies
are expected to have highly time-variable SFRs \citep{Dekel2023},  and any AGN will likely
have short effective lifetimes \citep{Hopkins2009_BH}. 

In this section, We first use the Balmer equivalent widths to constrain the duration
of the recent starburst.  Then we discuss models with a constant star formation rate
as a proxy for a high duty cycle of bursts. Finally we discuss  non-parametric star formation 
histories fitted to thr spectral energy distributions of the five brightest LAGER-z7OD1 LAEs.
For each of these star formation histories, integration of $Q(\tau)$ over a lookback time of 
300 Myr determines the total number of ionizing photons produced by massive stars.
The results also draw attention to the trade-off between the inferred dust reddening and the 
ionizing photon production efficiency. Finally, we also present new evidence for AGN 
contributing to the ionization budget.

\subsubsection{Starburst Durations}  \label{sec:sfh}

\begin{figure}[h]
 \centering
  \includegraphics[scale=0.5,angle=0,trim = 0 0 0 0]{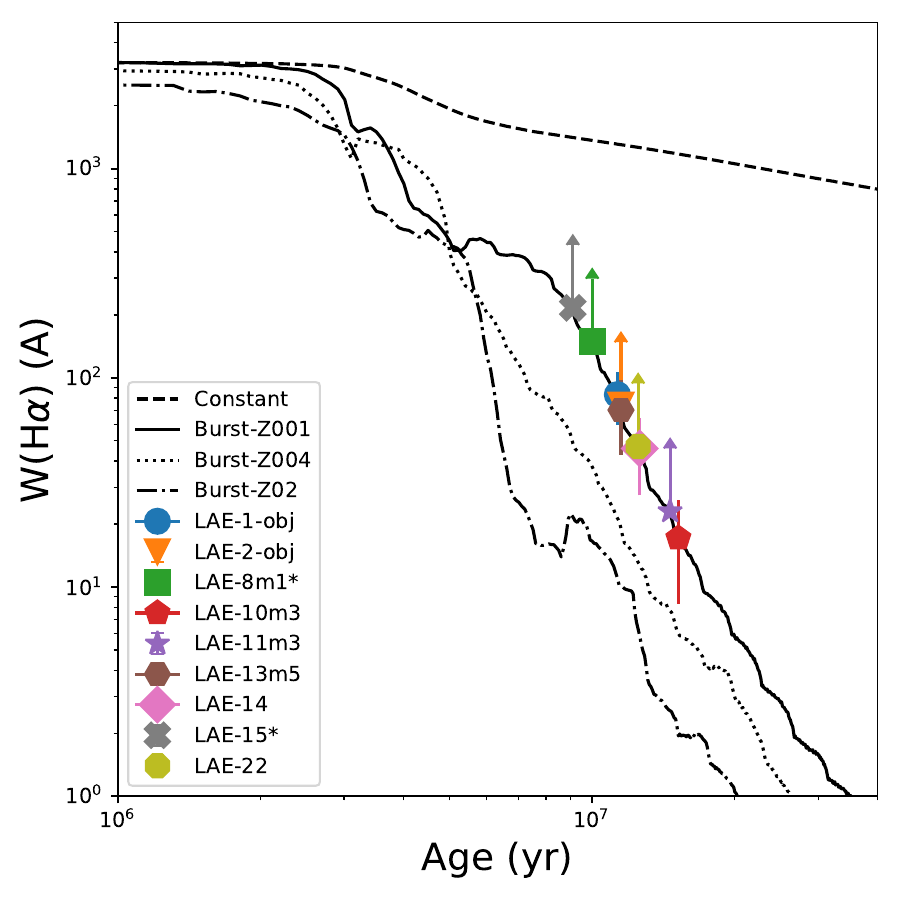}
  \caption{Evolution of \Ha\  equivalent width with age. Population synthesis models 
    illustrate an instantaneous burst (at several metallicities) and a constant SFR 
    \citep[][Starburst~99]{Leitherer1999}.   We plot the rest-frame equivalent width of
    each LAE at the intersection with the Z001 model (5\% solar metallicity).  The
    four measurements with a continuum detection require a recent starburst.
    }
    \label{fig:sb99_ew_age} \end{figure}

Following a short burst of star formation, the strength of hydrogen Balmer lines declines 
relative to the rest-optical continuum as the stellar population ages. Figure~\ref{fig:sb99_ew_age} 
illustrates this evolution using the widely applied Starburst~99  models as a benchmark 
\citep{Leitherer1999}. Comparison to our measurements of W(\Ha) for LAGER-z7OD1 LAEs
suggests burst durations, $\tau_{SB}$,  of roughly   11 to 15~Myr.  The two galaxies 
withouth \Ha\ coverage, LAE-8 and LAE-15, are plotted using  W(\Ha) $ \approx 4.4$ W(\Hb),
which is appropriate for an instantaneous burst model at ages of 6 to 10 Myr. Col.~3 of 
Table~\ref{tab:sfh} lists burst ages for an assumed stellar metallicity of 0.05\zsun.

The low equivalent widths of LAE-1, LAE-13, LAE-14, and LAE-15 relative to the 
continuous star formation model support the expectation of bursty histories.  
For the other five LAEs, however, the absence of a NIRSpec continuum detection 
near the Balmer line yields only a lower limit on W(\Ha). These limits offer
little insight on the star formation history because they are consistent with either
younger starbursts or the continuous star formation rate model.

{\clm
Lowering any track in Fig.~\ref{fig:sb99_ew_age} decreases the estimated starburst duration. 
}
In the other direction, binary evolution
leads to a slower decline in equivalent width with age because mass transfer tends to 
prolong  main sequence lifetimes, especially for massive stars with low metallicity. The 
\texttt{BPASS} v2.2.1 models  \citep[][Binary Population and Spectral Synthesis]{Eldridge2017,Stanway2018}
increase the age at fixed equivalent width by a few Myr relative to the single-star models \citep{Xiao2018}.
The assumed nebular conditions also affect both the line and continuum luminosities. In consideration
of these uncertainties, we scale the total number of hydrogen recombinations in the ISM to 
a fiducial burst duration of 15~Myr (Col.~11 of Table~\ref{tab:sfh}).

\begin{deluxetable*}{lllcllllllll}
\footnotesize
\tablecaption{Star Formation History }
\colnumbers
\tablehead{
\colhead{Galaxy } &
\colhead{W(\Ha)} &
\colhead{$\tau_{SB}$} &
\colhead{SFR$_{H\alpha}$ } &
\colhead{$ {\Muv}$} &
\colhead{$A^*_V$} &
\colhead{$\beta_{UV}$ } &
\colhead{SFR} &
\colhead{$\log(N_{sed})$}   &
\colhead{$R_i^{dust}$}   &
\colhead{$\tau_{SB}^{req}$}   &
\colhead{$\xi_{ion,0}$ }
\\
\colhead{}        &
\colhead{(\AA)}  &
\colhead{(Myr)}  &
\colhead{(\msun/yr)}        &
\colhead{(mag)}   &
\colhead{(mag)} &
\colhead{} &
\colhead{(\msun/yr)}          &
\colhead{(photons)}   &
\colhead{(Mpc)}   &
\colhead{(Myr)}   &
\colhead{(Hz~erg$^{-1}$)}
}
\startdata
LAE-1   &
$83 \pm 24$  &
$11.2^{+0.5  }_{-0.6 }$ &  
$202_{-87}^{+34} $ &
$-21.32 \pm 0.03$     &
$0.74_{-0.11}^{+0.11}$ &
$-1.75_{-0.03}^{+0.02}$ &  
$112_{-3}^{+3}$ &
69.82 &
4.3  &
9 - 34 &
$25.81_{-0.07}^{+0.51}$ 
\\
LAE-2  &
$ > 74$ &
$ < 11.4$ &
$37_{-3}^{+4} $ &
$-21.25 \pm 0.02$  &
$0.12_{-0.02}^{+ 0.01}$ &  
$-2.48_{-0.02}^{+0.02}$ &  
$24.6_{-0.5}^{+0.5} $ &
69.15 &
1.7 &
70 - 1400 &  
$25.73_{-0.04}^{+0.09}$ 
\\
LAE-8   &
$ \ge 149$\tablenotemark{a}  &  
$ \le 10$  &  
$32_{-15}^{+7} $ &
$-21.06 \pm 0.14$  &
$0.62_{-0.10}^{+0.10}$ & 
$-1.9_{-0.1}^{+0.1}$ &  
$63_{-6}^{+7} $ &
69.70 &
1.6 &
100 - 360 &
$25.24_{-0.11}^{+0.35}$ 
\\
LAE-10   &
$17 \pm 9$   &  
$15.1^{+1.5}_{-1.1}$&  
$16_{-2}^{+2} $ &
\nodata & 
\nodata & 
\nodata & 
\nodata &  
\nodata   &
0.48    &
31 - 200 &  
$26.06_{-0.05}^{+0.05}$ 
\\
LAE-11 &
$ > 23$  &  
$ < 14.5$ &
$18_{-5}^{+2} $ &
\nodata & 
\nodata &   
\nodata & 
\nodata & 
\nodata  &
1.2    &
110 - 400 &  
$25.92_{-0.04}^{+0.30}$
\\
LAE-13 &
$70 \pm 28$  &
$11.5^{+1.2  }_{-0.7 }$&  
$77_{-33}^{+9} $ &
\nodata &  
\nodata & 
\nodata & 
\nodata & 
\nodata  &
2.9    &
1 - 43 &  
$25.82_{-0.05}^{+0.51}$
 \\
LAE-14  &
$46 \pm 19 $ &  
$12.6^{+1.1  }_{-1}$&  
$31_{-3}^{+2} $ &
$-21.11 \pm 0.08$  &
$0.13_{-0.07}^{+ 0.10}$ & 
$-2.52_{-0.07}^{+ 0.10}$ &  
$20_{-4}^{+5} $ &
69.35 &
0.58 &
22 - 110 &  
$25.69_{-0.04}^{+0.09}$
\\
LAE-15  &
$ \ge 216 $\tablenotemark{a}  &  
$ \le 9 $ &  
$178_{-115}^{+33} $ &
$-20.93 \pm 0.01$  &
$1.0_{-0.01}^{+0.01}$  &
-1.43 &  
142 &
69.85 &
DNE\tablenotemark{b} &
14 - 780 &  
$25.64_{-0.08}^{+0.57}$
\\
LAE-22  &
$> 47$         &  
$ < 12.5 $ &
$18_{-4}^{+4} $ &
\nodata & 
\nodata & 
\nodata &
\nodata & 
\nodata & 
DNE\tablenotemark{b} &
110 - 6800 &  
$25.73_{-0.12}^{+0.39}$
\\
\enddata
\tablenotetext{a}{Equivalent width of \Ha\ estimated as $W(\Ha) \approx 4.4 W(\Hb)$;
the coefficient represents an instantaneous burst model in Starburst~99  and
is appropriate at ages of 6 to 10 Myr \citep{Leitherer1999}. }
\tablenotetext{b}{Applying the color excess term, $E(B-V) = A_V / 4.05 = 0.25$ \citep{Calzetti2000},
  in Eqn.~\ref{eqn:yang2017}, in addition to the velocity offset term, yields $T_{igm} > 1$.
Specifically, for LAE-15, dividing the extinction corrected \fesclya\ by the very low escape fraction, 
$\fism = 0.031$, yield an estimated IGM transmission $T_{igm} = 2.02$. While for LAE-22, a revised
$\fism 0.09$ translates to $T_{igm} = 1.32$. 
}
\tablecomments{
{\it (Col 2):}    Rest-frame equivalent width.  Lower limits correspond to
                  $3\sigma$ upper limit on the continuum. 
{\it (Col 3):}    Maximum stellar age compatible with W(\Ha) for an instantaneous
                  starburst.
{\it (Col 4):}    Equilibrium star formation rate,
                  $\log SFR(\msunyr) = \log L_{H\alpha}^{corr} ({\rm erg~s}^{-1}) -
                  41.39 - \log(1-\fesc)$, evaluated for $\fesc=0$.  We corrected the
                  \Ha\ luminosity from Table~\ref{tab:jwst_offset} for extinction,
                  where $A(\Ha) = 0.820 A_V^*$ magnitudes  \citep{Calzetti2000}.
                  When no SED is available, we use $A_V = 4.05 E(B-V)^{gas}$.
                  The upper errorbar is the statistical error. The lower errorbar
                  represents the no-dust limit, i.e. the systematic error. 
{\it (Col 5):}    \texttt{BAGPIPES} absolute magnitude at 1500 \AA, uncorrected for
                  reddening, fitted to rest-frame UV and optical photometry; see
                  Sec.~\ref{sec:sed}.
{\it (Col 6):}    Estimated stellar extinction in the visual band.  Values 
                  and statistical errors determined by  \texttt{BAGPIPES} SED fitting, 
                  including the \Hb\ and \Hg\  emission lines. Systematic uncertainties
                  are much larger, as demonstrated by comparison to the color excess
                  measured from the Balmer decrement, which are not correlated with $A_V^*$. 
                  To make this comparison, we used the the \citep{Calzetti2000} reddening
                  curve for galaxies, which suggests the stellar  extinction is
                  $A_V^* \approx 4.05 (\pm 0.80) E(B-V)_*$, and $E(B-V)_* \approx  0.44 E(B-V)_{gas} $
{\it (Col 7):}    UV continuum slope calculated over the wavelength range from 1300 \AA\
                  to 2500 \AA\ using 500 posterior spectra from the \texttt{BAGPIPES} 
                  modeling.  We give the 50th percentile, and the uncertainties denote 
                  the 16th and 84th percentiles.
{\it (Col 8):}    Average star formation rate in past 10 Myr from \texttt{BAGPIPES} SED fit.
{\it (Col 9):}   Number of ionizing photons produced  \texttt{BAGPIPES} non-parametric star 
                  formation history.
{\it (Col 10):}   Bubble radii, as Col.~11 of Table~\ref{tab:new_lya}, but including the color
                  excess term in Eqn.~\ref{eqn:yang2017}. We adopt $E(B-V) =  A_V^* / 4.05$,
                  consistent with a \citet{Calzetti2000} attenuation curve. This
                  correction significantly decreases the estimated interstellar escape fraction  
                  of \lya\ photons, thereby increasing the required transmission (and hence radius)
                  of each ionized bubble. We conservatively assume $A_V^{gas} \approx A_V^*$ for these very
                  young galaxies; taking $E(B-V)^* \approx 0.4 E(B-V)^{gas}$ would further increase
                  the estimated bubble radii.
{\it (Col 11):}   Starburst duration required to ionize a volume defined by the {\it minimum-bubble} 
                  radius (first value) or the {\it dv-bubble} radius (second value). Fiducial
                  values for a  LyC escape fraction of 10\% and no dust. See \S~\ref{sec:Nph}.
{\it (Col 12):}   Ionizing photon production efficiency corrected for dust
                  using $A_V^*$, when SED fit is available, and $E(B-V)^{gas}$ otherwise.
}
\label{tab:sfh}
\end{deluxetable*}

\subsubsection{Constant SFR Approximation}

For a constant SFR scenario, the (extinction corrected) \Ha\ luminosities provide an independent 
measure of the SFR. The number of ionizing photons per unit star formation depends on the initial 
mass function and stellar evolution. The widely used conversion,  $\log SFR(\msunyr) = 
\log L_{H\alpha}({\rm erg~s}^{-1}) - 41.27$  \citep{Murphy2011,Kennicutt2012}, describes a population
of single stars bounded in mass by an upper limit of 100 \msun.  Because binary interactions prolong
massive star lifetimes in a parameter-dependent manner, the timescale to reach an equilbrium 
ionizing-photon luminosity is not as well defined as it is for single-star models. 

We define 
an equilbrium value of the ionizing photon luminosity  as the median $Q(t)$ over a starburst
duration of $\log \tau (Myr) = 8.5$. At a fiducial metallicity of 0.1\zsun, the \texttt{BPASS} v2.2.1 
binary tracks produce $\log Q ({\rm s}^{-1}) = 53.42$ ionizing photons per second per unit SFR 
(in \msunyr). The lower stellar metallicity and the evolution of the binary stars combine to yield 
two times more ionizing photons (per unit SFR) than solar metallicity, single-star models 
\citep{Murphy2011}. Raising the upper mass limit to 300\msun boosts the \texttt{BPASS} 
ionizing photon photon production by a factor of 1.35 relative to this fiducial model.

Combining hydrogen recombination coefficients at $T_e = 2 \times 10^4$~K with the
ionizing photon production rate of the 0.1 \zsun\ stellar population with binaries,
the equilibrium star formation rate becomes
\begin{eqnarray}
\log {\rm SFR}_{H\alpha}(\msunyr) = \log L_{\Ha}({\rm erg~s}^{-1}) - \\ \nonumber
41.39 - \log(1 - \fesc).
\label{eqn:clm_sfr} \end{eqnarray}
Column~4 of Table~\ref{tab:sfh} lists these \Ha\ estimates for the SFR. The correction
for LyC leakage is left as parameter.

\subsubsection{Non-parametric SFHs}  \label{sec:avstar}

As described in Appendix~\ref{sec:bagpipes}, we fit the spectral energy distribution
using the Bayesian Analysis of Galaxies for Physical 
Inference and Parameter EStimation tool, \texttt{BAGPIPES} \citep{Carnall2018}. 
The resulting  non-parametric star formation histories are shown in the right panel of 
Figure~\ref{fig:five_bagpipes}. The starbursts last about 10 Myr,
consistent with the burst duration we estimated from the \Ha\ equivalent width. 
Prior to recent burst, the SFR was at least an order-of-magnitude lower.
However, 10 Myr is less than one-tenth the recombination timescale in the IGM.
Col.~9 of Table~\ref{tab:sfh} lists the total number of ionizing photons produced 
by the \texttt{BAGPIPES} star formation history over the recombination timescale,  
$N_* \equiv \int_0^{300~Myr} Q(\tau) d\tau$, calculated as described in Appendix~\ref{sec:bagpipes}.
The extended star formation history  (300 Myr)  boosts the number of LyC photons by a factor 
of 1.5 for LAE-8. For the other four galaxies,  the most recent starburst is the only 
population with a significant production of ionizing photons. 

When we estimate the SFR from Eqn.~\ref{eqn:clm_sfr}, we obtain a value lower than the true 
SFR if the galaxy leaks LyC photons,  $\fesc > 0$. Assuming that the \texttt{BAGPIPES} model 
finds the true SFR, the difference of these measurements would indicate  the LyC escape fraction, 
\begin{eqnarray}
\fesc \equiv ({\rm SFR}_{SED} - {\rm SFR}_{H\alpha}) / {\rm SFR}_{SED}.
\end{eqnarray}
By this argument, we find  $\fesc \approx 0.65$ for LAE-8.  Comparison of Columns~4 and~8 in 
Table~\ref{tab:sfh} reveals an inconsistency.  The other four LAEs with \texttt{BAGPIPES} fits 
have ${\rm SFR}_{\Ha}$ significantly larger than ${\rm SFR}_{SED}$. Perhaps the \texttt{BAGPIPES}
fitting is simply not constrained well enough by the data, which includes the \Hb, but not the \Ha, 
emission line.  The details of the nebular models used by \texttt{BAGPIPES} were not readily
available; but we note that a low electron temperature, relative to the $T_e = 2 \times 10^4$~K used 
in Equation~\ref{eqn:clm_sfr}, would produce an offset in the direction of the noted
discrepancy.\footnote{  
     Per unit (extinction corrected) $L(\Ha)$,
     a lower electron temperature requires fewer ionizing photons, i.e. a lower SFR. 
     Lowering $T_e$ to $1 \times 10^4$, however, cannot explain the full
     difference between our ${\rm SFR}_{SED}$ and ${\rm SFR}_{\Ha}$ estimates.}
Clearly, the \texttt{BAGPIPES} fitting underestimates the uncertainty on the SFR. Since
the most recent burst appears to dominate the production of ionizing photons, we
adopt a continous star formation history over the starburst duration in \S~\ref{sec:budget}.

\subsubsection{Dust Reddening \& Extinction}  \label{sec:reddening}

Whether at cosmic noon or in the local universe, studies of
LAEs consistently find \fism\ decreases with increasing color
excess E(B-V) \citep{Verhamme2008,Atek2009,Kornei2010,Hayes2011}.
While the Balmer decrements do not require much dust, the measurement uncertainties
{\it allow} significant extinction; see \S~\ref{sec:balmer_ratio}.  
The \texttt{BAGPIPES} fitting confirms nearly dust-free conditions in LAE-2 and LAE-14,
but it favors  extinction corrections which boost the \Ha\ luminosities of LAE-1 and
LAE-15 by nearly a factor of two.

Table~\ref{tab:sfh} lists the fitted visual extinction and the UV spectral slope, $\beta_{UV}$, 
at 1500 \AA. The UV slope increases (flattens) with increasing dust optical depth. The strength 
of the correlation indicates that the UV continuum slope provides the primary constraint on 
the dust optical depth in the \texttt{BAGPIPES} fit. The steep UV spectral slopes and low 
reddening measured for LAE-2 and LAE-14 are consistent with significant LyC leakage
\citep{Chisholm2022}. The \texttt{BAGPIPES} fits LAE-15 and LAE-1 have a less extreme UV 
slope than the other LAEs. This may indicate more dust and higher metallicity, qualitatively 
consistent with their higher stellar masses in Table~\ref{tab:photometry}.

The \texttt{BAGPIPES} fit applies the same dust optical depth to the stars and gas. 
In very young galaxies where the stars are still near their birth clouds,  the distinction 
between stellar and nebular attenuation may be reduced \citep{Reddy2018} relative to the offset 
described by \citet[][]{Calzetti2000}. We noticed
that $A_V^{Bagpipes}$ was significantly larger than $A_V^{gas}$, the visual extinction
measured from the \Ha\ to \Hb\ flux ratio, for LAE-8. Whereas $A_V^{Bagpipes}$ was distinctly
less than $A_V^{gas}$ for LAE-2 and LAE-14. These differences illustrate that the systematic
errors on the extinction exceed the statistical error returned by the \texttt{BAGPIPES} fit.

{\clm
In Sec.~\ref{sec:rion}, we discussed the impact of the corresponding color excess
on the inferred  bubble radius.  Since increasing the dust content reduces ISM
transmission of \lya, a smaller fraction of the observed attenuation is attributed
to the IGM. The bubble radii grow to the $R_i^{dust}$ values listed in Col.~10 of 
Table~\ref{tab:new_lya}. The radii estimated in the limit of a dust-poor ISM, $R_i^{ism}$,
were already large enough to  produce bubble overlap and challenge the ionization budget.
None of the LAEs can individually ionize these larger ($R_i^{dust}$) bubbles.   
In consideration of the significant uncertainties on both $A_V^*$ and $E(B-V)^{gas}$,
our discussion conservatively focuses on the $R_i^{min}$ and $R_i^{ism}$
in \S~\ref{sec:budget}.
}

\subsubsection{Ionizing Photon Production Efficiency} \label{sec:xion}

The UV luminosity provides insight about the amount of star formation on a longer timescale 
than the \Ha\ luminosity does.  At a constant SFR, extending the duration of the starburst 
increases the UV luminosity relative to the ionizing photon luminosity. The latter approaches 
an equilibrium value within 30 Myr, whereas the UV continuum continues to
brighten for 100 Myr. In contrast, following an instantaneous burst of star formation, 
the UV luminosity only brightens for about 3 Myr, by which time the ionizing
photon luminosity is already declining.

\begin{figure}
 \centering
  \includegraphics[scale=0.6,angle=0,trim = 20 0 0 0]{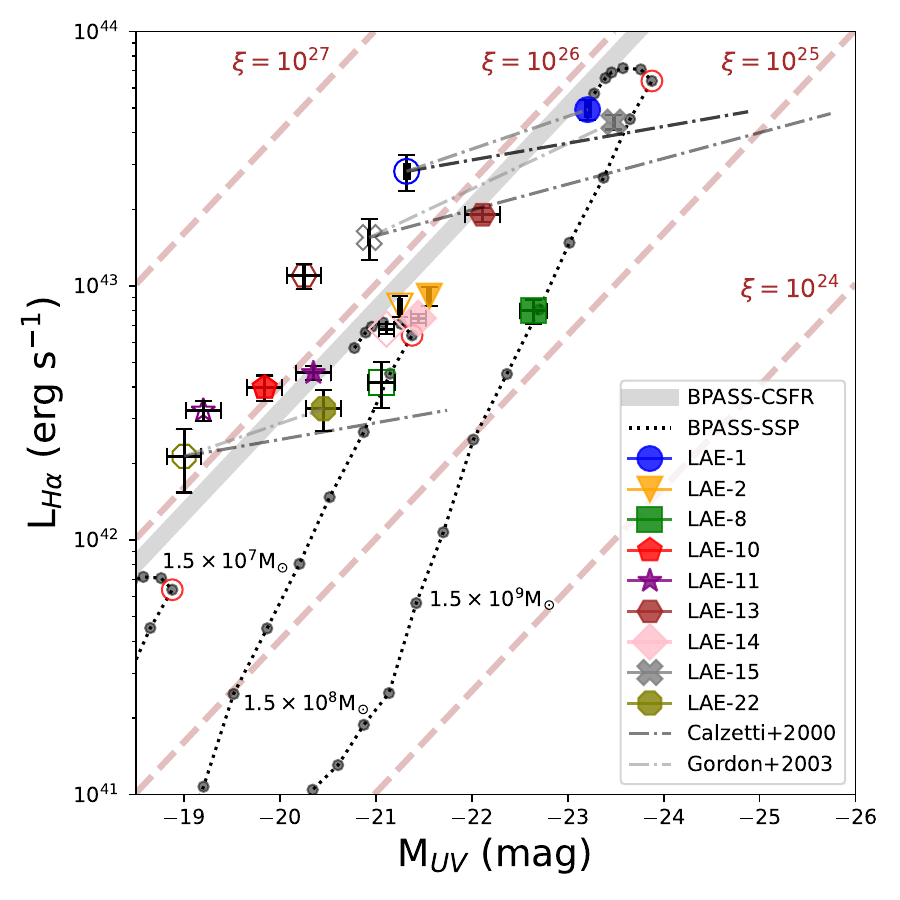}
    \caption{\Ha\ and UV luminosities of z7OD1 LAEs compared to stellar population 
      synthesis models. Small circles at 1, 3, 10, 30, 100, and 300 Myr mark the temporal 
      evolution following an instantaneous burst (black dotted tracks). The red circle selects 
      an age of 3 Myr. Before the reddening correction (filled symbols),  most of the  LAEs lie 
      well to the left of maximum starburst line (thick gray line). Applying the reddening 
      correction (from Col.~6 of Table~\ref{tab:sfh} or Col.~10 of Table~\ref{tab:jwst_offset}) 
      moves the LAEs  (left-filled symbols)  onto the stellar evolutionary tracks. Vectors 
      indicate the sensitivity to the attenuation law. Brown diagonal lines indicate 
      increasing \xion\  from lower right to upper left.  The reddening correction eliminates 
      the need for an extraordinary ionizing photon production  efficiency. 
     }
     \label{fig:sb99_Lha_Muv}
      \end{figure}

Figure~\ref{fig:sb99_Lha_Muv} shows population synthesis models in the $L(\Ha) -M_{UV}$ plane.
The wide, gray diagonal line represents a maximal ionizing photon production efficiency. For
a specified stellar IMF, atmosphere models, and evolutionary tracks determine, an aging stellar 
population evolves away from this limit. For illustration, we adopt  0.1 \zsun\ (z002) metallicity  
tracks with an upper mass limit of 100 \msun, including binary star evolution 
\citep[][BPASS v2.2.1]{Eldridge2017,Stanway2018}.
Evolutionary tracks for stellar populations are shown following an instantaneous burst.

Prior to any reddening correction, only LAE-11 lies within the locus reached
by the evolutionary tracks in Figure~\ref{fig:sb99_Lha_Muv}.  The other LAEs appear 
UV faint for their \Ha\ luminosities. One interpretation of this offset is that
the LAEs are unusually efficient producers of LyC emission. In this case, however,
LAE-1, LAE-13, and LAE-15 require ionizing photon production efficiencies nearly 
an order of magnitude larger than the fiducial BPASS model. Some galaxies do appear 
to produce ionizing photons more efficiently than others \citep{Matthee2023}. 
Previous studies, however, have not found a clear correlation between LyC photon 
production and \lya\ visibility \citep{Tang2023,Saxena2024}. 

Shifting the maximal ionizing photon production efficiency higher would require stellar 
metallicites well below 0.1\zsun, 
our fiducial value. Adopting an upper mass limit of 300\msun, instead of 100\msun, 
shifts the gray line upward by only 0.1~dex. 
Accounting for non-zero values of \fesc\ shifts the models downward in
Figure~\ref{fig:sb99_Lha_Muv};  we ignore this correction here because it
amounts to just 0.022 to 0.10 dex in $\log L(\Ha)$ for \fesc\ $ = 0.05 - 0.20$.

Alternatively, we can interpret the offset in Figure~\ref{fig:sb99_Lha_Muv} as evidence 
of reddening by dust. To illustrate this point, the left-filled symbols plot the LAEs 
after correcting for dust.  The reddening correction brightens the UV luminosity
relative to the ionizing photon luminosity, and all the LAEs shift into the region
reached by stellar evolutionary tracks. For consistency with our \texttt{BAGPIPES} models 
for the five brightest LAEs, we adopt the \citet{Calzetti2000} attenuation law. The
dashed, gray vectors show the reddening correction.

We note, however, the substantial systematic uncertainty in the attenuation law. 
{\clm
At the metallicities of interest, an SMC bar curve \citep[][black, dash-dot]{Gordon2003} 
may be appropriate, and the much higher UV attenuation would lead to significantly lower 
estimates of the ionizing photon efficiency.
}
 On the other hand, if LAEs are too young for AGB stars to have
contributed much dust, the larger grains produced by supernova would flatten
the attenuation law \citep{McKinney2025}, and our dust correction would underestimate
the true ionizing photon production efficiency.

The ionizing photon production efficiency, $\xion \equiv Q / L_{\nu}^{UV}$, 
is related to $\xi_{ion,0}$ by the LyC escape fraction, 
 $\xion = \xi_{ion,0} (1 - \fesc)^{-1}$ by  Eqn.~\ref{eqn:Q}.
We calculated the latter,
\begin{align}
 \xi_{ion,0} = 2.47 \times 10^{25} {\rm ~Hz~erg}^{-1} \frac{L_{H\alpha}}{10^{42} {\rm erg~s}^{-1}}
         \times \nonumber \\
          10^{-0.4 (M_{UV}^{corr} + 20)},
 \label{eqn:xion_lha_muv}  \end{align}
for each LAE. We  corrected  $M_{UV}$ values listed in Table~\ref{tab:jwst_offset} and the Balmer line 
luminosities from Table~\ref{tab:jwst_offset} for dust. We adopted the stellar extinction, $A_V^*$ 
(when a \texttt{BAGPIPES} fit was available), and the color excess ($E_{B-V}^{gas}$ from the Balmer ratio)
otherwise. Table~\ref{tab:sfh} lists the results.

Our estimates of $\xi_{ion,0}$ are consistent with other low-mass, star-forming galaxies,
which exhibit ionizing photon production efficiencies about 0.25 dex
higher than more massive galaxies. The LAEs do not require extraordinary produces of LyC photons.
This is consistent with the \citet{Yung2020c} prediction of very little evolution in \xion\ 
at fixed UV luminosity. However, Figure~\ref{fig:sb99_Lha_Muv} also  emphasizes the importance
of pinning down the reddening correction. The minimum reddening correction ($A_V=0$) would
require high  $\xi_{ion,0}$, as would an attenuation law flatter than \citet{Calzetti2000}.

\subsubsection{Contributions from AGN} \label{sec:agn}

{\clm
Very luminous LAEs are often AGN. \citet{Songaila2018} identified 
$\log L_{Ly\alpha}({\rm erg/s}) \approx 43.50$ as athreshold above which
most LAEs are AGN. The only LAE in LAGER-z7OD1 this bright is LAE-1.
The next most luminous LAEs, LAE-15 and LAE-2, have $\log L_{Ly\alpha}({\rm erg/s}) > 43.30$,
the luminosity above which AGN dominate the \lya\ luminosity function at redshift
$z \approx 2.2$ \citep{Konno2016,Sobral2017}. We argued that LAE-15 may contain an AGN
based on its high O32 ratio (Sec.~\ref{sec:U} and Table~\ref{tab:jwst_offset}).
Based on high \lya\ luminosity alone, our sample contains three AGN candidates. 
}

\begin{figure}
 \centering
  \includegraphics[scale=0.7,angle=0,trim = 10 0 0 0]{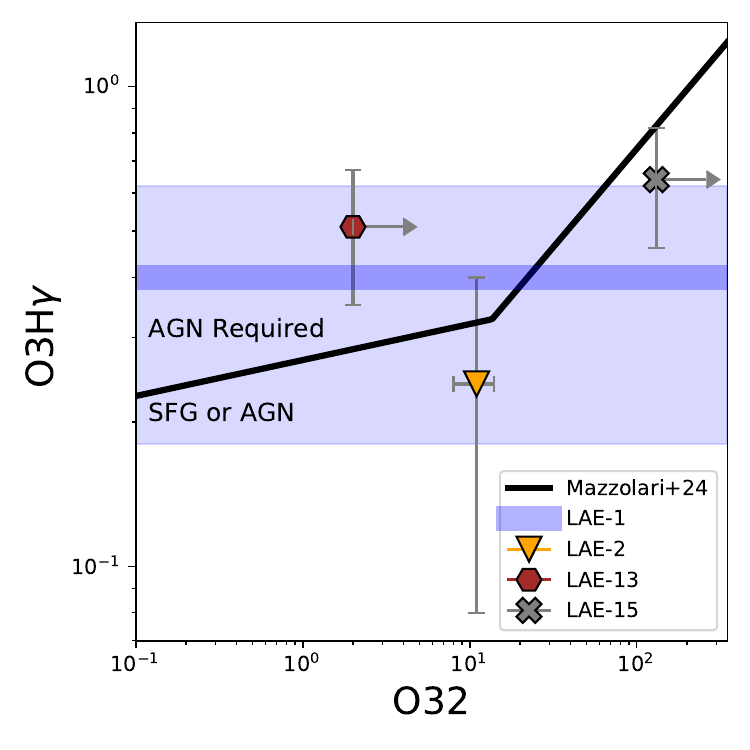}
   \caption{Diagnostic emission-line ratios: $F(\oiii\ \lambda 4364 / F(\Hg)$ vs. 
     $F(\oiii\ \lambda \lambda 4960, 5008 / F(\oii] \lambda \lambda 3727, 30$.
    Above the black line defined by \citet[][Eqn. 3 and 4]{Mazzolari2024},
    an AGN is required to produce the line ratios. Either a star-forming
    galaxy or an AGN can excite nebulae below the black line. The relatively
    strong O3H$\gamma$ ratios of these four galaxies allow an AGN contribution
    to nebular excitation, although none require an AGN due to limited 
    constraints on the O32 ratio.
    }
     \label{fig:new_bpt} \end{figure}

{\clm
The NIRSpec spectra of LAE-1, LAE-2, LAE-13, and LAE-15 detect the temperature-sensitive
\oiii\ $\lambda 4364$ emission line. At low gas-phase metallicity, the standard emission-line ratio 
diagnostics \citep{Baldwin1981,Veilleux1987} fail to distinguish excitation by
AGN and starbursts \citep{Juodzbalis2025}. \citet{Mazzolari2024} argue, however, that strong 
\oiii\ $\lambda 4364$ emission relative to \Hg, hereafter O3H$\gamma$, is sufficient (but not  necessary) to 
identify AGN at low metallicity. Figure~\ref{fig:new_bpt} shows the locations of these four LAEs
in their O3H$\gamma$ vs.~O32 diagnostic diagram.  For LAE-1, we measure O3H$\gamma = 0.40 \pm 0.22$;
but its NIRSpec spectrum does not cover the \oiii\ doublet, so O32 is unconstrained. 
Based on our measured ${\rm O3Hg}$ line ratios 
of  $0.24 \pm 0.16$ and $0.64 \pm 0.18$ for LAE-2 and LAE-15, respectively, both galaxies 
lie on the boundary between the {\it AGN-Required} and the {\it SFR-or-AGN} regions.
While none of the measurements in Fig.~\ref{fig:new_bpt} definitively require an AGN,
this diagram also picks out LAE-13 as an AGN candidate.  The eastern knot of LAE-13
is the strong \oiii\ $\lambda 4364$ emitter. Bright \oiii\ doublet emission contributes to 
that clump's visibly red F150W2-F444W color in Fig.~\ref{fig:nircam}. 
}

{\clm
The spectra in Fig.~\ref{fig:nirspec} clearly show broad wings on the hydrogen Balmer 
lines from four objects - LAE-1, LAE-2, LAE-13, and LAE-15. In LAE-13, only the eastern 
clump with the high O3H$\gamma$ line ratio has broad wings in its spectrum. The spectra of 
LAE-2, LAE-13, and LAE-15, also detect broad wings on the \oiii\ $\lambda \lambda 4960, 5008$ 
lines.   However, the high gas densities in outflows driven by Type~I AGN collisionally 
de-excite the \oiii\ doublet lines. If AGN power these outflows, then they must be Type II AGN.
Recall, however, that the stellar population synthesis models in \S~\ref{sec:sed} 
provided a reasonable fit to the SEDs of the LAEs; no AGN was required.
In a forthcoming paper, we discuss whether a starburst-driven wind is a more likely source
of the broad wings on these emission-line profiles. 
}

{\clm
The identification of AGN candidates does not obviously balance the ionization budget for the bubbles.
We described this challenge in \S~\ref{sec:Nph} using the interstellar recombination rate and 
the LyC escape fraction. This model counts ionizing photons from AGN as well as stars. One difference
would be the lifetime of the ionizing source.
\cite{Hopkins2009_BH} modeled the effective lifetimes of AGN, 
finding $\tau_{eff} \sim 100$ Myr for very active nuclei of moderate mass today. Scaling from the 
Hubble time today to the recombination timescale at $z \approx 7$,  reduces the estimated effective 
AGN lifetime to just $\tau_{eff} \sim 2$ Myr. 
For AGN to solve the ionizing photon deficit in LAE-2, for example, the LyC escape fraction would 
need to exceed $\fesc \approx 0.60$ if the AGN lifetime is shorter than the
estimated duration of the starburst. Including AGN as potential ionizing source therefore adds 
effective AGN lifetimes as well as of star formation histories to the list of possible solutions.
}

\subsection{Balancing the Ionization Budget}  \label{sec:budget}

We used the product of bubble volume and the average density of hydrogen 
(at redshift $z \approx 6.93$) to estimate the ionization requirement. We found
that the recent starbursts in LAE-1 and LAE-13 have produced enough LyC photons
that modest LyC escape fractions enable these galaxies to ionize their bubbles. 
Only a few bursts, comparable in intensity to one observed, would be required 
for LAE-10, LAE-14, and LAE15 to ionized their bubbles. The SED fitting might
not detect a few short bursts spread over several hundred Myr.  In contrast, at
least 10 such bursts would be required to fully ionize the bubbles around LAE-8, LAE-11,
and LAE-22, and the non-parametric star formation histories (\S~\ref{sec:sed}) should
detect these populations. These conclusions reflect the minimum bubble radii, $R_i^{min}$,
derived from an argument which neglects attenuation of \lya\ by the ISM.

Interstellar attenuation of \lya\ increases the inferred IGM transmission, so the required
bubble radii increase to  $R_i^{ism}$.  In \S~\ref{sec:voff}, we estimated the minimum correction
based on  measured \lya\ velocity offset and lower bounds on color excess, which allow
dust-free ISMs. 
{\clm
In this {\it dv-bubble} scenario, the larger $R_i^{ism}$ values shift each LAE to a 
higher $R^3_{Mpc} L_{H\alpha}^{-1}$ track in Fig.~\ref{fig:tstar_fesc}. The intersection
of each track with the duration of that LAE's starburst (from \S~\ref{sec:sfh}) defines a 
required LyC escape fraction, $f_{esc}^{req}(\tau_{SB}) = 0.25 - 0.98$. Even the
median value (0.74) is significantly large than predictions from SPHINX \citep{Choustikov2024_escape_physics}.  
The most recent starburst therefore appears to be insufficient to ionize the volume of each $R_i^{ism}$ bubble.
Multiple starbursts, or additional ionizing sources, are required.
}

Column~10 of Table~\ref{tab:sfh} lists a third, even larger, estimate for the bubble radius,
$R_i^{dust}$. Here we have defined the color excess as $A_V^* / 4.05$ (for the five
LAEs with a \texttt{BAGPIPES} fit); but for LAE-10, LAE-11, LAE-13, and LAE-22,   
we adopt  $E(B-V)^{gas}$ from Col.~10 of Table~\ref{tab:jwst_offset}.  
Applying  Eqn.~\ref{eqn:yang2017}, 
we now estimate interstellar escape fractions of \lya\ $\fism = 0.03 - 0.50$ for 
this {\it dusty-ISM} picture. Dividing \fesclya\ by these low \fism\ values
pushes the required IGM transmission (Eqn.~\ref{eqn:fism}) to unphysically
high values (i.e. $ > 1$) in LAE-15 and LAE-22. 
{\clm
The unphysical values may indicate that the intergalatic hydrogen is entirely ionized along these 
sightlines. Alternatively, it may reflect errors in the ISM correction: 
the large uncertainty on the color-excess correction, and/or the applicability of
a local calibration (Eqn.~\ref{eqn:yang2017}) to the $z \sim 7$ ISM.
}
In the other LAEs, the \fesc\ values
required to ionize the bubbles in a single burst approach unity. If we cap the
escape fraction at \fism, then the required star formation timescales exceed the age
of the universe (for over five of the LAEs).  We therefore explored other ways
to balance the ionization budget.

\subsubsection{Fainter Galaxies}

Does LyC leakage from fainter galaxies  balance the ionizing
photon budget?  In the 1.5 deg$^2$ COSMOS field, for example, \citet{Endsley2021_massive_laes}
identified an overdensity of luminous galaxies at $z = 6.8$ and then detected \lya\ emission from
9 of the 10 galaxies \citep{Endsley2022-bubble}. They argued that these bright galaxies
alone could not ionize the local bubble and suggested that fainter galaxies (down to M$_{UV} = -17$)
provide the additional ionizing photons.  Their solution assumed \fesc\ of  6\% and a
timescale of 200 Myr.

{\clm
The timescales required to ionize a bubble (\S~\ref{sec:Nph}) 
}
typically exceed the duration of the recent starburst. To describe the required boost to the
number of ionizing photons,  we normalize $\tau_{SB}^{req}$ by a burst duration of 15~Myr:
$\mathcal{F}_B  \equiv \tau_{SB}^{req}  / 15 {\rm ~Myr}$. For illustration, we conservatively 
take the product of \xion\ and \fesc\ to be constant and ask  how far down the UV luminosity 
function we need to integrate to increase the UV luminosity density by $\mathcal{F}_B$. We use
the \citet[][]{Bouwens2021} luminosity function at $ z \approx  6.8$ for $\phi(L)$. The integral
of $L \phi(L)$ from some $L_{min}$ up to the UV luminosity each LAE is then normalized by the luminosity
density in bright galaxies, $ \int_{L_{LAE}}^{\infty} L \phi(L) dL$. We convert the $L_{min}$ that
gives a boost of $\mathcal{F}_B$ to an absolute UV magnitude. This limit differs for each bubble
model of course, but the exercise yields new insight.

As an alternative to additional (or longer) starbursts, fainter galaxies
can supply the additional ionizing photons in many of the bubbles. In
LAE-10, LAE-14, and LAE-15, these additional galaxies could be fairly 
bright; integrating down to $M_{UV} = -18.6$, -20.8, and -20.5, respectively,
produces the required $\mathcal{F}_B$. Fainter galaxies would need to
contribute to bubble ionization in LAE-2, LAE-8, LAE-11, and LAE-22.
Galaxies slightly brighter than  $M_{UV} \approx -17$
can provide those ionizations in LAE-2 and LAE-8.  

Interestingly, however, 
faint galaxies cannot meet ionizing photon requirements of the LAE-11 and LAE-22
bubbles. The faint galaxy solution becomes harder  when the $M_{UV}$ of
the LAE is already much fainter than the knee in the luminosity function
\citep[][$M_{UV}^{*} = -21.15$] {Bouwens2021}.  
Unless the assumptions about \xion\ and \fesc\ are changed to
strongly favor leakage from low luminosity galaxies, the required $L_{min}$ values
are unrealistically low. We return to LAE-11 in \S~\ref{sec:overlap} and
argue that the bubble volume has been over estimated.  Further discussion of
LAE-22 requires higher S/N ratio \lya\ spectroscopy;  the
systematic uncertainties are too large at this time.

JWST can detect the fainter galaxies which may ionize the bubbles that 
make the LAEs visible. In the {\it minimum-bubble} scenario, fainter 
cluster members could be detected within most of the bubbles. 
In the {\it  dv-bubble} picture, in contrast, the fainter galaxies solution
could  work in LAE-1, LAE-13, and LAE-14, where surveys reaching
$M_{UV} \le -18.5$ would detect the ionizing sources.
In the other six bubbles,
however, the faint $L_{min}$ values would be difficult to reach with JWST.

\subsubsection{Bubble Overlap in a Protocluster} \label{sec:overlap}

Based on the locations and sizes of ionized bubbles in Fig.~\ref{fig:myweb_rion}, 
bubble overlap seems likely.  The merger of bubbles is predicted to 
accelerate bubble growth \citep{Gnedin2000}. Since we have measured the ionized volume,
we can think of bubble overlap as reducing the required number of ionizing photons. 
The volume common to two bubbles in Fig.~\ref{fig:myweb_rion} need only be
ionized once over the recombination timescale. Denoting the volume common to bubbles 
$A$ and $B$ as $V_{AB}$, the required number of ionizing photons is reduced by a
factor $f_R \approx 1 - V_{AB} / (V_A + V_B)$.  

Fig.~\ref{fig:myweb_rion} suggests possible bubble overlap in
two distinct regions. The western sub-cluster contains LAE-1,
LAE-2, LAE-14, and LAE-8 -- from largest to smallest \Ha\ luminosity.
The eastern sub-cluster includes LAE-13, LAE-10, LAE-11, and LAE-22.
Each of those regions contains one LAE with $\log L(\Ha) ({\rm ~erg/s}) > 43.0$.
The third galaxy this bright, LAE-15, is the farthest away and lies between
the eastern and western sub-clusters on the sky.

In the minimum-bubble picture, the LAE-1 bubble overlaps the LAE-14 bubble. 
The galaxies could already ionize their bubbles with modest LyC escape fractions, 
just 0.07 and 0.17 respectively,  and bubble overlap further reduces the required
number of ionizing photons (by $f_R = 0.77$). In redshift space, LAE-2 is closer to
us than the LAE-1/LAE-14 bubble, and LAE-8 has a lower redshift than LAE-2.  The LAE-2
and LAE-8 bubbles do not overlap other bubbles. Therefore bubble overlap does not explain
their substantial ionizing photon deficits.

In the dv-bubble scenario, the individual bubbles have larger radii. 
The increased overlap between the LAE-1 and LAE-14 bubbles ($f_R = 0.66$) does 
not make up for the factor of 3.0 increase in their combined bubble volumes. 
LAE-1 might still ionize its bubble; it requires a LyC escape fraction
$\fesc \ge\ 0.25$ over 11 Myr.  However, LAE-14 now requires either a high escape fraction
($\fesc \ge\ 0.50$) over 13 Myr;  or, alternatively, a burst duty cycle of 41\% with
a fiducial escape fraction of $\fesc = 0.10$.  
The larger bubbles in the dv-bubble scenario do reduce the ionization deficit elsewhere.
The LAE-2 bubble barely intersects the LAE-1 ($f_R = 0.96$), LAE-14 ($f_R = 0.99$), 
and LAE-8 ($f_R = 0.97$) bubbles.

In the eastern sub-cluster, only the small bubble around the luminous galaxy LAE-13
could be ionized by a single burst in the {\it minimum-bubble} picture.  The LAE-22
bubble barely intersects the LAE-10 bubble ($f_R = 0.99$), so bubble overlap does
explain their ionizing photon deficits. Although the story for LAE-22 is similar to
that of the other LAEs, the Keck/LRIS \lya\ spectrum of LAE-22 has low S/N ratio compared
to the other targets, potentially producing significant systematic errors in the bubble
measurement.

In the dv-bubble scenario, LAE-15 ionizes a large bubble, but it barely intersects the LAE-11 
bubble ($f_R = 0.99$) and does not reach the foregound sub-structure containing LAE-10, LAE-13, 
and LAE-22.  The LAE-13 and LAE-10 bubbles barely touch ($f_R = 0.995$).
The LAE-22 bubble still just barely intersects the LAE-10 bubble ($f_R = 0.97$) but
now also slightly overlap with the LAE-13 bubble ($f_R = 0.97$).  
Bubble overlap does not come close to offsetting the larger bubble volumes,
so the ionizing photon deficits grow.

\subsubsection{Line-of-Sight Fluctuations in Ambient Neutral Fraction} \label{sec:low_neutral}

Our understanding of the reionization history is based in part on the decline in the strength of 
\lya\ emission with increasing redshift.

\begin{figure}[h]
         \includegraphics[scale=0.52,angle=-90,trim = 0 20 0 0]{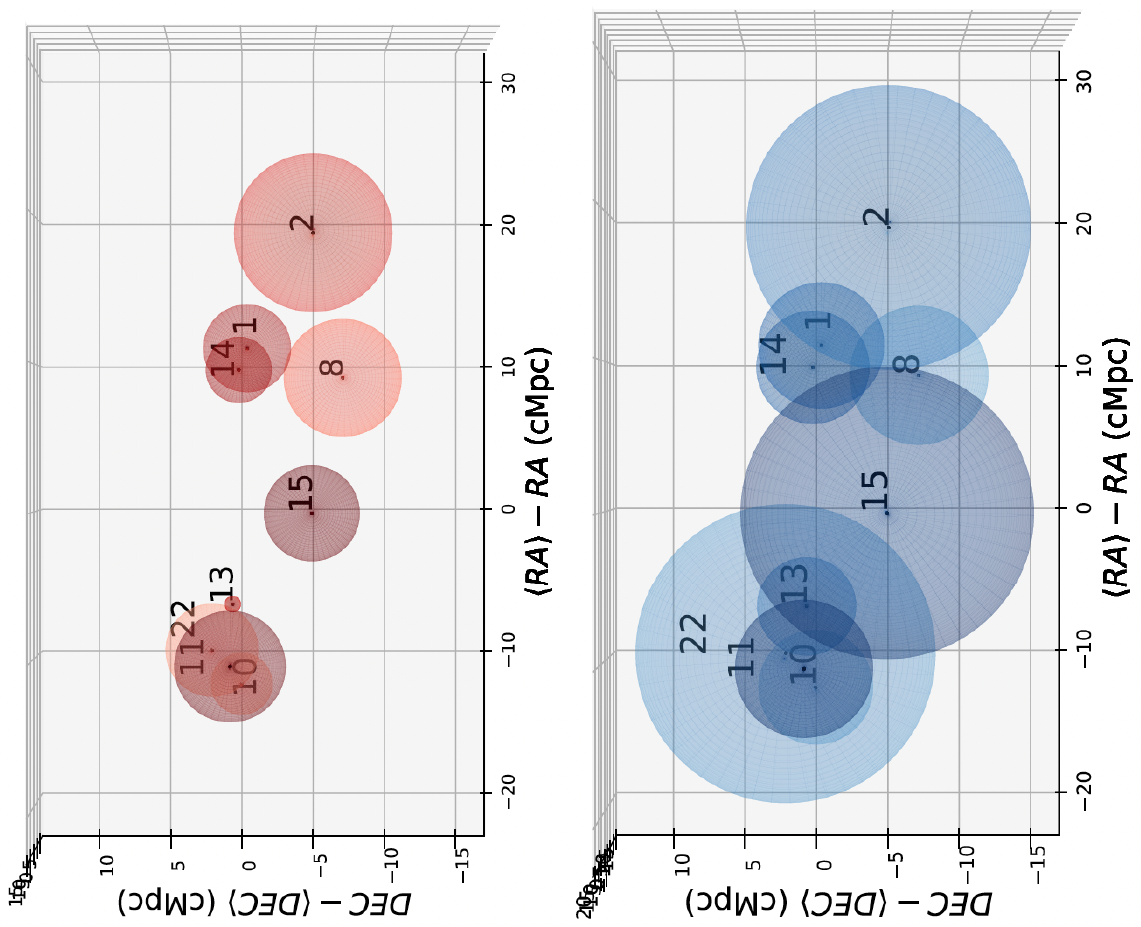}
    \caption{Projection of ionized bubbles on the sky. 
             Emission from LAE-1 passes through the LAE-14 bubble,
             and LAE-10 is seen through the LAE-22 bubble.
      Whereas the volumes ionized by LAE-1 and LAE-14 have a large intersection,
      the LAE-10 and LAE-22 bubbles barely overlap in 3D space
      (in the  minimum bubble illustration). Symbols as in Fig.~\ref{fig:myweb_rion}.
     }
     \label{fig:vmax_fwhm_Rion}
     \end{figure}

Even when bubbles do not intersect in 3D-space, the {\it superposition of those bubbles} on the sky 
can impact \lya\ transmission.  The damping wing optical depth is sensitive to fluctuations in ionization
fraction over Mpc scales \citep{Mesinger2008a}.
A key question is whether the topology of the ionized bubbles within the protocluster
produces significant sightline-to-sightline variation in \lya\ transmission. To explore
this idea further, Figure~\ref{fig:vmax_fwhm_Rion} projects the ionized bubbles onto the sky.

In the western sub-cluster, the bubble from LAE-14 lies in front of LAE-1.
The \lya\ emission from LAE-1 might also pass through the LAE-2 bubble, which
just covers LAE-1 in the {\it dv-bubble} picture. These superpositions do not
balance the ionization budget. The LAEs with the largest deficit, LAE-8 and LAE-2,
do not lie behind any known ionized bubbles.

In the eastern sub-cluster, \lya\ emission from LAE-11 passes through the LAE-10 bubble.
It may then be transmitted through the LAE-13 bubble, which is large enough to cover
LAE-11 on the sky in {\it dv-bubble} picture.  Bubble superposition
therefore reduces the neutral hydrogen fraction along the sightline to LAE-11. As
shown above, accounting for $\bar{X}_{HI,los} < 1$ outside the LAE-11 bubble reduces
the inferred bubble radius and ionization requirement.
As discussed previously, the radius of the LAE-22 bubble carries large uncertainty,
but the LAE-11 emission also passes through our estimated projection for the LAE-22 bubble. 
In the {\it dv-bubble} picture, the LAE-13 \lya\ emission would also pass through the
foreground LAE-22 bubble. We conclude that bubble superposition will most strongly affect \lya\ 
transmission from LAE-11. This result is interesting becuase we found it impossible to
balance the ionization budgets of LAE-11 with faint galaxies or a high burst duty cycle.

In \S~\ref{sec:rion}, we assumed a fully neutral IGM around each ionized bubble.  
A \lya\ photon emitted by a galaxy will propagate through regions 
alternating regions of largely neutral and highly ionized gas. Modeling these variations is beyond
the scope of this paper. To qualitatively demonstrate the impact of foreground  ionized bubbles, however,
we reapplied our simple model using a non-unity neutral fraction (beyond each bubble). This neutral
fraction represents an average neutral fraction over scales of tens of Mpc.

\begin{figure}[h]
 \centering
\includegraphics[scale=0.55,angle=0,trim = 0 0 0 0]{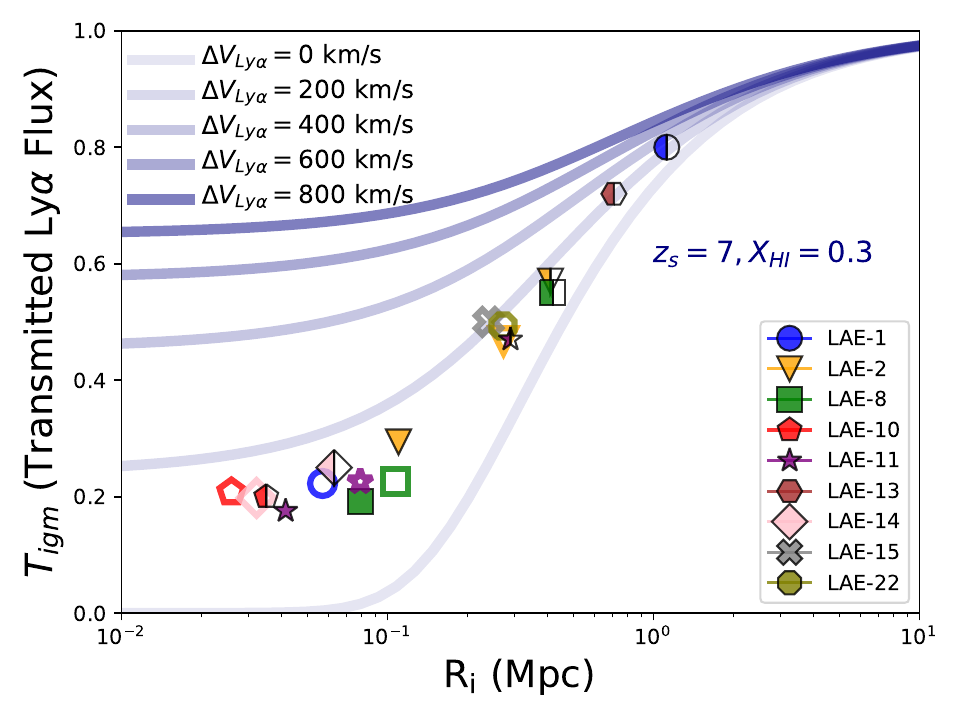}
\caption{Transmission of redshift $z \approx  7$ \lya\ emission by an ionized bubble. Here
         the ambient IGM outside each ionized bubble has a neutral fraction of 0.3. Enhanced transmission
         outside the bubble reduces the required size of the ionized bubble, as can be seen by a 
         comparison to Fig.~\ref{fig:T_Rion} where the neutral fraction is unity outside the bubble.
         Symbols as in Fig.~\ref{fig:T_Rion}. Note the extension of the x-axis to smaller radii.
}
    \label{fig:T_Rion_X03}
     \end{figure}

To illustrate this effect quantitatively, we recomputed the curves shown in 
Fig.~\ref{fig:T_Rion} using a reduced neutral hydrogen fractions outside each bubble.  
Fig.~\ref{fig:T_Rion_X03} shows the result with $\xh = \bar{X}_{HI,los} = 0.3$ instead of unity.
For an emergent \lya\ line with a velocity offset between 200 \kms\ and  400 \kms, our 
inferred bubble radii are reduced by factors between 2 and 3.
Some bubble radii would be reduced to galactic scales ($\le\ 10$ kpc), i.e. no bubble is 
required; their points lie off the left hand edge of Fig.~\ref{fig:T_Rion_X03}. 

This experiment demonstrates that lowering \barxh\ decreases the inferred bubble radii. 
{\clm
The value of the damping wing optical depth will change with variations in neutral fraction along the sightline,
but  our result carries over to more realistic structures  qualitatively. 
}
A reduction in optical
depth (relative to the fully neutral sightline) reduces the inferred bubble radii. 
The 3D topology of ionized pocekts LAGER-z7OD1 explains why faint galaxies and high duty cycles failed
to balance the ionizing photon budget of LAE-11.  The bubble radius is over-estimated because of the 
foreground ionized bubbles within the protocluster.   

{\clm
The fraction of redshift $z \sim 7$  galaxies with detectable \lya\ emission appears to be high in the EGS
field \citep{Chen2025-Stark}
 but quite  low in the UDS \citep{Napolitano2024,Napolitano2025}. 
Toward the  COSMOS field, \citet{Wold2022} suggested reionization is nearly complete by  $z \sim 7$. 
Interpreting their (1-sigma) upper limit, $\barxh = 0.3$, as the average neutral fraction {\it outside}
each bubble would reduce the inferred radii of the ionized bubbles. For illustration, 
the left-filled symbols in Fig.~\ref{fig:T_Rion_X03}  shows the bubble models with the full ISM 
correction including  the color excess term.  While the inferred radii were extremely large 
when $\barxh = 1.0$, the $R_i^{dust}$ radii are 1 Mpc or less when $\barxh = 0.3$.
In the limit of highly ionized sightlines toward all the LAGER-z7OD1 LAEs, bubble radii estimated using 
the full ISM attenuation  (\ref{eqn:yang2017}) yield a topology for the ionized bubbles similar to Fig.~\ref{fig:myweb_rion}.
}

\subsubsection{Departures from Case~B Recombination}

We adopted Case~B recombination values when we estimated \fesclya\ because they
are valid for Lyman continuum leakers at low redshift \citep{Flury2024}. Whether
the picket fence geometry which is used to explain LyC leakage from an ionization
bounded galaxy also decribes LAEs in the reionization era is not yet clear. At
very high redshift,  steep UV spectral slopes identify a population of density-bounded 
galaxies \citep{Topping2024_beta}.  Our  SED fitting results for the  UV-brightest LAEs 
find $\beta > -2.8$, which does not allow such blue UV slopes.  If 
the LAGER-z7OD LAEs are density-bounded galaxies, however, then their intrinsic \lya\ / \Ha\ 
flux ratios will be larger than the Case~B value we adopted \citep{Osterbrock2006}, and 
our estimates of \fesclya\ would decrease.  Under the principle that $\fesc \le\ \fesclya$ 
generally, bubble ionization by the indivdiual LAEs would become harder.

\subsection{Bubble Size and Environment} \label{sec:implications}

In most models of cosmic reionization, ionized bubbles grow from overdense regions 
outwards \citep{Qin2022,Witten2024-NatAs,Lu2024,Neyer2024}. In the early stages, before
individual bubbles begin to overlap other ionized regions, bubble size is predicted to be
strongly connected to environment  \citep{Neyer2024, Lu2024}. Since large ionized bubbles
increase \lya\ transmission, we expect LAEs to select overdense regions at these early times.

Analysis of JWST observations, however, has not revealed a one-to-one correspondence 
between reionization-era LAEs and protoclusters. For example, the redshift $6.6$ LAE COLA1
likely lies in a very unusual environment because the IGM transmits the blueshifted \lya\
peak. Yet a recent NIRCam grism survey identified only a modest overdensity of star-forming
galaxies around COLA1, refuting the hypothesis of a large overdensity and suggesting an
unusual source instead \citep{Torralba-Torregrosa2024}. The Cosmic Evolution Early Release
Science survey identified three $z \approx 7.47-7.75$ LAEs which are unlikely to be the
sole ionizing agents of the bubbles that transmit their \lya\ emission based on indirect
indicators of their LyC escape fractions  \citep{Jung2024_CEERS_laes}. The LAE with the
highest-ionization condition would require a LyC escape fraction $\fesc > 0.5$, and the
other two require  $\fesc \sim 0.7$. In addition, although JWST observations have confirmed a
galaxy overdensity at $z=7.88$  large enough to become a $10^{15}$\msun\ cluster by the
present day, spectroscopy did not detect \lya\ emission from any of the galaxies
\citep[][A2744+z7p9OD]{Morishita2023}.

LAGER-z7OD is one of the largest galaxy overdensities identified so far
in the reionization era. \cite{Hu2021} estimate this protocluster will 
collapse into a cluster with virial mass, $M_h \approx 10^{15}$\msun\
based on the number density of LAEs, which is $ 5.11^{+2.06}_{-1.0}$ 
times higher than the average density. Semi-numerical, cosmological  models
provide further insight about this rare structure because they probe very large
volumes. Such models map galaxy properties onto dark matter halos computed using
merger trees based on the $\Lambda CDM$ cosmological model.

We examined the most massive structures found at $z\sim7$ in one realization 
of the 2-deg$^2$ simulated lightcone presented in \citet{Yung2023}. To identify
structures similar to LAGER-z7OD, we identified massive structures by repeatedly
sampling the lightcone with sub-volumes of 10 by 20 arcmin$^2$ with $\delta z = 0.08$.
Then the subvolumes that contained at least 20 member halos and had a total
mass $M_{h} >  10^{11} $\msun\ were flagged, and large clusters that span over
multiple sub-volumes were linked together as one large cluster. We confirmed
that all these structures become $M_h \approx 10^{15}$\msun\ clusters. 
The seven most-massive protoclusters have just passed turnaround at redshift $z \sim 7$,  
Their 3D structure consists of several subclusters, similar to what we find in LAGER-z7OD1. 
Interestingly, the redshift depth of these protoclusers ranges from $\Delta z = 0.13$ to 0.17,
which  is about twice the line-of-sight depth probed by the LAGER narrowband filter
($\Delta z \approx 0.076$). We conclude that the LAGER-z7OD structure likely spans a
redshift range somewhat larger than that probed by narrowband selection.

\begin{figure}[h]
 \centering
    \includegraphics[scale=0.6,angle=0,trim = 20 0 0 0]{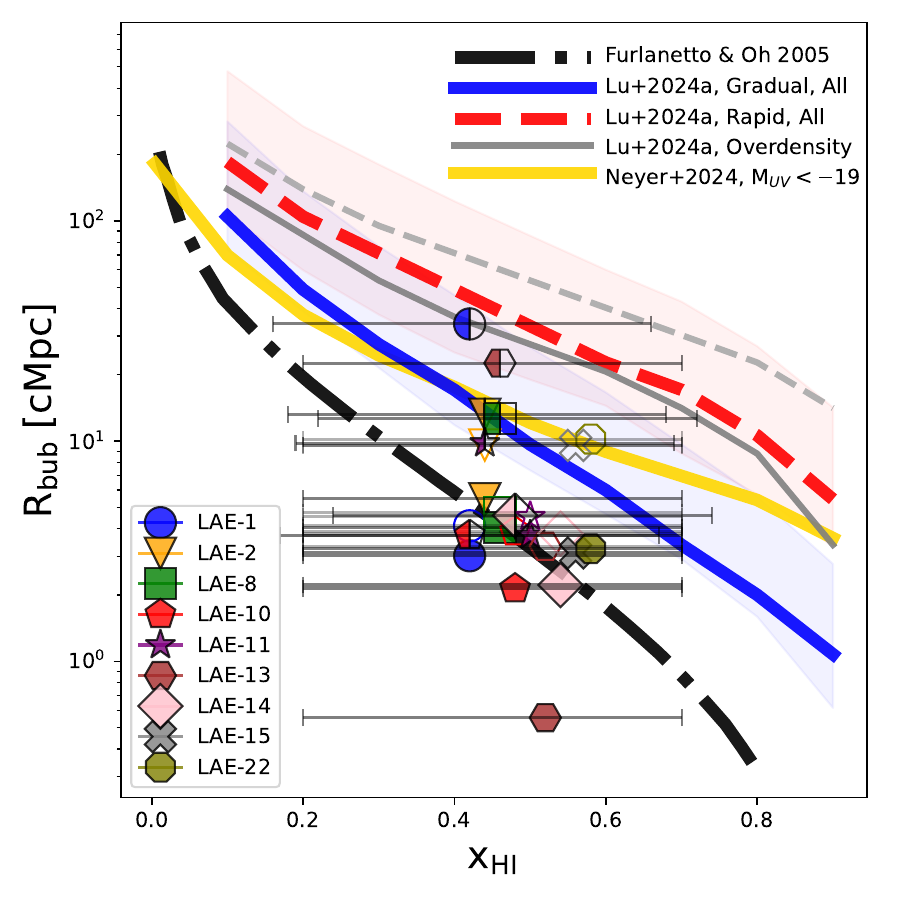}
    \caption{Estimated bubble radii (in comoving units) compared to predictions for
      bubble growth. The temporal evolution in the models is traced by the global neutral 
      fraction.  We plot the LAGER-z7OD1 LAEs at  $ \barxh \sim 0.5$ (with random offsets
      for clarity) and draw wide horizontal error bars to emphasize the uncertainty
      The {\it minimum-bubble} radii (solid symbols) overlap the average radii in the 
      \citet{Furlanetto2005} model at neutral fractions $\barxh\ \approx\  0.5 \pm 0.1$.  
      Our {\it dv-bubble} (open symbols) and $R_i^{dust}$ (left-filled symbols) radii
      shift onto that curve at  $\barxh \approx 0.3$ or   $\barxh < 0.2$, respectively.
      Alternatively, accounting for bubble overlap (blue curves) reproduces these larger
      bubbles at the earlier time when $\barxh \approx 0.5$. Bold lines show median radii; 
      shading indicates the standard deviation about the mean values. For both source models,
      a gray lines denotes the average bubble radius in regions with overdensities of at 
      least 10.
     }
    \label{fig:Rion_Growth}
     \end{figure}

The clustering of discrete ionizing sources, combined with the clumpiness of the IGM,
is what makes reionization inhomogeneous \citep{Furlanetto2005}.  \citet{Furlanetto2005}
provide an analytic model for bubble growth which takes both of these factors into account.
Fig.~\ref{fig:Rion_Growth} compares this evolution to our estimated sizes for {\it individual}
bubbles. Our minimum bubble sizes, i.e. with no ISM attenuation, are consistent with their average
sizes at a neutral fraction of $\barxh \approx 0.5$. The larger radii obtained after
modeling the ISM attenuation of \lya\ shift the bubbles onto the \citet{Furlanetto2005} curve at 
a lower neutral fraction.  This model, however, does not account for bubble overlap.

Evidence for bubble overlap is a a key result of our analysis of LAGER-z7OD1.  Although the
amount of overlap is sensitive to how we model the ISM attenuation of \lya, we consistently
find three distinct pockets of ionized gas in the protocluster.  Adopting the {\it dv-bubble}
scenario for illustration, their characteristic sizes are $14 \pm 0.7$~cMpc in the western sub-cluster,
$11 \pm 3$~cMpc in the eastern subcluster, and $12 \pm 2$~cMpc in the LAE-15 region. These length
scales are quite similar to the predictions of \citet{Lu2024} in the post-overlap phase. They
created large volume (1.6 cGpc)$^3$ simulations using {\it 21 cmfast} \citep{Mesinger2011} and
explored the impact of different reionization histories on the sizes of ionized bubbles. Figure~3
of \citet[][THESAN]{Neyer2024} shows the bubble evolution produced by more accurate radiation transport,
albeit in a much smaller volume simulation. The median bubble size in the THESAN simulations
is indistinguishable from the \citet{Furlanetto2005} curve at $\barxh < 0.6$.  The differences among 
these predictions are comparble in size to the systematic uncertainties in our measurements.

{\clm
The models still provide importanant insight about environmental bias.  Galaxy overdensities
likely ionize the first large bubbles \citep{Furlanetto2004, Lu2024}. Since these cosmic
H~II regions enhance \lya\ transmission \citep{Haiman2002}, reionization-era LAEs may select
overdensities.
}
When restricted to galaxies as bright as LAGER-z7OD1 LAEs, the sizes of THESAN bubbles 
\citep[][Fig.~3 yellow curve for $M_{UV} < -19$]{Neyer2024} grows, shifting upward 
and overlapping the median bubble size in the \citet{Lu2024} gradual reionization model.
That model (blue curve in Fig.~\ref{fig:Rion_Growth}) shifts upward in Fig.~\ref{fig:Rion_Growth}
to the solid, gray line when restricted to overdensities. While these models predict a similar 
environmental shift in the median bubble size, the very wide range in bubble radii at a fixed 
overdensity (or luminosity) makes it difficult to measure by the time $\barxh$ drops to 0.5.
\citet[][Figures~5 and 6]{Neyer2024} predict a much stronger  environmental distinction at 
earlier times when  $\barxh = 0.9$. 

From the model comparison, we conclude that reionization has entered the percolation phase in
LAGER-z7OD1. The slow, early growth of the bubbles has reached bubble overlap, and we expect
dramatic growth in bubble  size in a short period.  In the limit of the full correction
for interstellar attenuation, our large $R_i^{dust}$ values could be interpreted as evidence
for some fully ionized sightlines. We argued that our measurements of \lya\ escape fraction and 
equivalent width favor some IGM attenuation, but the supression relative to local samples is only
a 1 to 2 standard deviation result.  Unless it has already occurred, the ensuing flash 
ionization is about to sweep through the entire protocluster.  
The inside-out growth of the first ionized pockets is also expected 
to quench star formation in low mass halos \citep{Gnedin2000,Zier2025} and contribute to the observed  
dependence of galaxy properties on environment  \citep{Dressler1980, Kodama2001}.

\section{Conclusions} \label{sec:summary}

We presented new JWST/NIRSpec spectroscopy, KECK/LRIS spectroscopy, and NIRCam 
imaging of LAGER-z7OD1, an overdensity of LAEs at redshift $z \approx 7$. 
Using the emission-line spectra of nine LAEs, we mapped the 3D topology of 
ionized gas in the protocluster. We then investigated whether the LAEs
themselves were the primary ionizing agents. We draw the following conclusions.

\begin{itemize}

\item
The LAEs have clumpy UV/optical morphologies. Spectra of individual components 
need to be corrected for aperture losses, and then combined, to generate integrated 
spectra. Since the centroids of the \lya\ nebulae do not match the position of any
individual clump,  accurate measurements of \lya\ emission require wide apertures.

\item
We identified three large pockets of ionized gas, wherein individual bubbles are 
either approaching overlap or have already begun to percolate. These regions
coincide with sub-clusters of LAEs within the overdensity. 
{\clm
This result suggests that other regions of the overdensity will soon be ionized from the outside inwards, 
differentiating the evolution of the protocluster galaxies from those in other environments.
}

\item 
Bubble overlap has generated pockets with radii of at least
of 10 to 14 (comoving) Mpc. This scale follows from a conservative correction 
for interstellar attenuation of \lya. Using our dust measurements and applying
an empirical ISM correction, we demonstrate that the IGM could be entirely ionized
along several sightlines.

\item
{\clm
The star formation histories of the LAEs favor multiple, short bursts
($ \le 15$ Myr) and rule out continuous star formation over the 270~Myr
recombination timescale.  The most luminous LAEs are likely LyC leakers based
on their physical properties. Clumpy morphologies \citep{Martin2024,Witten2024-NatAs},
moderately high O32 ratios \citep{Choustikov2024_escape_physics}, and low \lya\
velocity offsets ($\dv\ = 100 - 250$ \kms) all facilitate Lyman continuum leakage. 
We conclude that five of the nine LAEs are plausibly the primary ionizing agent of
the $R_i^{min}$ bubbles. Whereas none of the LAEs can individually ionize their
$R_i^{dust}$ bubbles.  These results predict the presence of additional ionizing agents,
such as large numbers of fainter galaxies, or unusually efficient Lyman continuum
production in LAGER-z7OD1.
}

\item
{\clm
If the LAEs were dust free, we showed that their \Ha\ luminosities would 
exceed expectations based on the ionizing photon luminosities of starburst
models scaled to their measured  rest-UV luminosities.  This result could
reflect an exceptionally high production efficiency of ionizing photons in
protocluster LAEs, but we emphasize that high efficiences are not required.
Our estimated reddening corrections shift the LAEs into the region of 
the $L(\Ha) - M_{UV}$ plane populated by starburst models, 
thereby eliminating the need for unusual sources of ionizing photons.
}

\item
We identify a candidate AGN within each of the three sub-clusters.
Their O3H$\gamma$ line ratios meet or exceed the pure-AGN threshold recently proposed 
by \citet{Mazzolari2024}. Their \lya\ luminosities are near the threshold above which 
AGN are common \citep[][for redshift $4 < z < 6.6$ LAEs]{Songaila2018}.
In the NIRSpec spectra, however,  broad wings on forbidden lines, as well as the
Balmer lines, indicate outlows driven by the starburst rather than a Type I AGN. 
The extraordinary O32 ratio of LAE-15, along with its possibly time variable \lya\ 
flux, provide the most compelling evidence for active black hole growth.

\item
{\clm
The overlapping bubbles in the $R_i^{dust}$ scenario suggest the protocluster
has completely ionized its environment. In the more conservative {\it dv-bubble}
picture, the individual $R_i^{ism}$ bubbles overlap within each of the three
sub-clusters.  Bubble overlap  invalidates the simple geometry adopted to
estimate ionized volumes from line-of-sight distances, possibly leading
to a significant overestimate of the number of ionizing photons required
by the nine sightlines.  Using the conservative estimatess for bubble radii, 
$R_i^{min}$, we still identify one sightline where bubble superposition 
balances the ionizing photon budget. We conclude that the galaxy overdensity 
plays a very prominent role in \lya\ transmission.
}

\end{itemize}

Insight into the topology of cosmic reionization during the pre-overlap phase will 
likely rely on \lya-emitting galaxies (LAEs) for sometime since the angular scales
of individual
bubbles remain a major challenge for 21-cm intensity mapping \citep{Koopmans2015}. 
Future work should strive to improve upon the approximations made in this work.
These include:

\begin{itemize}

\item
Combining \Ha\ and \lya\ morphologies along with their line profiles 
can improve our understanding of \fesc\ 
\citep{Choustikov2024-lya,Choustikov2024_escape_physics}.  

\item
The red wings detected in the LRIS \lya\ spectra are broader than 
the velocity dispersion among clumps and therefore indicate
multiple scatterings. A better understanding of the physical
origin of the broad wings will make it possible to adopt
more accurate models of the \lya\ line profile emerging from
galaxies. It will be interesting to see whether radiative transfer
which fit the broad, asymmetric  \lya\ wings using outflowing
gas \citep{Gronke_Dijkstra2016,Garel2024} simultaneously reproduce
the broad, but more symmetric, wings on the  optical emission-line profiles.
In cosmological radiation transfer simulations of reionization-era galaxies,
the most pronounced broad wings appear in dust-poor galaxies
with cosmic ray driven outflows \citep[][\texttt{AZAHAR}]{Yuan2025}.

\item
{\clm
The velocity offsets and line widths of the \lya\ components are roughly consistent
with scaling relations in galaxies \citep{Verhamme2018}. We found plausible
signs of intergalactic attenuation: (1)  the offset of the primary \lya\ 
component to the
left of this $\dv - $ FWHM relation, and (2) an offset relative to Green Pea
galaxies in the $\dv - \fesclya$ plane.  We concluded that while {\sc HI} in
the IGM attenuates \lya\ emission, the velocity offset the observed line profile may
not differ much from that of the redshifted profile emergent from the galaxy.
Models of emergent \lya\ line profiles based on galaxy properties may
provide further insight \citep{Hayes2023}.
}

\item
For the five LAEs with well constrained UV spectral slopes, we adopted the 
$A_V$ measurements from SED modeling. Otherwise, we determined the color excess 
from the Balmer decrement (with a large error bar). Our measurement uncertainties
on the Balmer decrement, however, do not currently exclude dust-free galaxies.
{\clm
We anticipate revising these Balmer decrement measurements in the future when NIRSpec
calibration issues are better understood. 
}

\end{itemize}

{\clm
Leveraging information from neighboring sightlines can provide a better 3D 
representation of the topology of the ionized regions in the future.  Identifying
sightlines which do not transmit \lya, for example, can detect bubble boundaries.
Robust techniques are being developed for analyzing large numbers of 
sightlines \citep{Lu2025,Nikolic2025}, and the rising era of all-sky 
surveys will identify large overdensities into the reionization era.
The unequaled  sensitivity of JWST/NIRSpec spectroscopy, however,  will 
continue to be required to resolve the \lya\ and Balmer lines from large 
numbers of objects in those galaxy concentrations.
}

\begin{acknowledgements}

Support for program JWST-GO-01635 was provided by NASA through a grant from the Space Telescope Science Institute,
which is operated by the Association of Universities for Research in Astronomy, Inc., under NASA contract 
NAS 5-03127. LFB acknowledges the support of ANID BASAL project FB210003 and FONDECYT project 1230231.
JRW. acknowledges that support for this work was provided by The Brinson Foundation through a Brinson Prize Fellowship grant.
Some of the data presented herein were obtained at Keck Observatory, which is a private 501(c)3 non-profit organization operated as a scientific partnership among the California Institute of Technology, the University of California, and the National Aeronautics and Space Administration. The Observatory was made possible by the generous financial support of the W. M. Keck Foundation.
The authors wish to recognize and acknowledge the very significant cultural role and reverence that the summit of Maunakea has always had within the Native Hawaiian community. We are most fortunate to have the opportunity to conduct observations from this mountain.
This research was supported in part by NSF PHY-2309135 to Kavli Institute for Theoretical Physics (KITP).
This work made use of v2.2.2 of the Binary Population and Spectral Synthesis (BPASS)
models as last described in \citet{Stanway2018,Eldridge2017}.
\end{acknowledgements}

\begin{center}
AUTHOR CONTRIBUTIONS
\end{center}
WH and IGBW reduced the NIRCam pre-imaging and measured photometry. 
CLM, IGBW, AF, and AK compiled target lists, and CLM designed the MSA configurations.
WH modified the NIRSpec pipeline and reduced those data.   
CLM obtained the 2022 Keck observations, reduced those data, and extracted \lya\ spectra.
CLM and WH made the 2024 Keck observations; WH reduced the data.
IGBW and WH provided \lya\ positions and luminosities from the narrowband images.
CLM measured emission lines, computed aperture corrections, and derived physical properties.
JX fit the morphological structure.  JY measured the {UltraVISTA} photometry and performed \texttt{BEAGLE}
SED fits. WH fit the SEDs and spectra with \texttt{BAGPIPES}.
LYAY contributed the semi-analytic models of protoclusters.
All co-authors discussed the results. 
CLM wrote the manuscript, and all co-authors contributed comments which improved it.

\vspace{5mm}
\noindent
\facilities{JWST(NIRCAM,NIRSPEC), KECKI(LRIS)}

\noindent
\software{
{\sc Astropy} \citep{astropy:2013, astropy:2018, astropy:2022}
{\sc PypeIt}  \citep{pypeit2005,pypeit2020}
}


\noindent
{\it Data Availability:}
The JWST data presented and analyzed in this article can be obtained from the 
Mikulski Archive for Space Telescopes (MAST) at the Space Telescope Science Institute. 
The specific observations can be accessed via \dataset[DOI:10.17909/nz3n-4b83]{}
with Dataset Title {\sc JWSTPID1635}.

\clearpage

\appendix

\restartappendixnumbering

\section{Data Reduction} \label{sec:weida}

\subsection{NIRCam Data Reduction}

We use the STScI JWST Calibration Pipeline\footnote{\url{https://github.com/spacetelescope/jwst}} \citep{Bushouse2022} to reduce the raw NIRCam images.
Our reduction follows the standard JWST pipeline routines, which are divided into three steps: stage 1 detector-level correction (\texttt{calwebb\_detector1}), stage 2 image calibration (\texttt{calwebb\_image2}), and stage 3 mosaic construction (\texttt{calwebb\_image3}). 
These steps remove the instrumental signals (dark current, bias, linearity), determine the average count rate, calibrate astrometry and photometry, and mosaic the individual frames.

We also modify the \texttt{TweakReg} routine in stage 3 reduction to improve the astrometry alignment.
In this section, we describe our custom routines in detail.
The JWST pipeline version 1.10.1 was used with the Calibration References Data System (CRDS) context file of jwst\_1077.

\subsubsection{Snowball Removal}

The snowballs are produced by large cosmic-ray events and can affect hundreds of pixels.
It has the characteristic of a cosmic-ray core with a round bright halo.
During the bad-pixel flagging routine in the JWST pipeline, the cosmic-ray core is flagged as saturated pixels (DQ=2) and the round bright halo is flagged as jump pixels (DQ=4).
Thus, we identify the snowballs by evaluating the shape and area of jump pixels near every saturated pixel.

As suggested in \citet{Bagley2023}, we divide the snowballs into large and small tiers. 
To identify the small snowballs in the short-wavelength channel (long-wavelength channel), we require over 80 (50) pixels to be identified as jump pixels within a box of $15 \times15$ pixels.
To identify the large snowballs, we require over 200 (100) pixels to be identified as jump pixels within a box of $25\times25$.
Since the jump images cannot identify the first exposure in the group, we visually exam those exposures to mask the snowballs.
Further, to avoid the misidentification of "drifting cosmic rays", we require the ellipticity of the snowball to be $<0.5$.
The snowball masks are then grown by a tophat kernel to mask the extended halos.
The kernel sizes are 10 and 20 for the small and large snowballs, respectively.

\subsubsection{Wisp and 1/f Noise Subtraction}
We subtract the wisp structures and 1/f noise following the methods presented in \citet{Bagley2023}.
We adopt the flattened wisp templates (released on 2022 August 26) provided in the JWST User Documentation\footnote{\url{https://jwst-docs.stsci.edu/jwst-near-infrared-camera/nircam-instrument-features-and-caveats/nircam-claws-and-wisps}}.
Since our NIRCam observations are divided into six sets of pointings, we adopt the average scaling factor of each set of pointings to subtract the wisps.
Furthermore, we perform these subtractions during the stage 2 reduction based on the flattened images.
This is because although the wisps and 1/f noise are additive effects, they are best measured on flattened images to mitigate the uncertainty introduced by the spatial variation of the flat field.
In addition, since the wisps can expand to very large scale and may elevate the 1/f noise measurements,  we subtract the 1/f noise after the wisp subtraction.

\subsubsection{Background Subtraction}
Before the stage 3 reduction, we subtract a background for each exposure using \texttt{photutils} \citep{larry_bradley_2023_7946442}.
This step is necessary because the \texttt{SkyMatch} routine in the JWST pipeline cannot successfully match the background of each exposure. 
Since the background of JWST images is very flat, we adopt a constant background for each exposure. 
We use a 3$\sigma$-clipping method to estimate the background while the bright objects are masked.

\subsubsection{Astrometric Alignment and Mosaic}

The astrometry calibration is performed using a modified version of the JWST \texttt{TweakReg} routine.
The astrometry of NIRCam F150W2 and F444W images is calibrated using the same procedure.
We first calibrate the NIRCam F150W2 images using a reference catalog and 
then align the F444W images using the stellar catalog extracted from the F150W2 images.
Instead of the default GAIA reference catalog provided in the JWST pipeline, 
we adopt our custom reference catalog extracted from the HST ACS F814W 
images in the COSMOS field \citep{Koekemoer2007}, which have been registered to GAIA DR2.
Since the GAIA stars are saturated in our NIRCam images,   
we select the isolated, relatively 
compact ($3.5<$ FWHM $<10$ pix), and approximately round galaxies (ellipticity 
$< 0.5$) from ACS images as the reference stars.

Since the FoV of NIRCam is relatively small, only a few reference stars 
can be matched in each NIRCam exposure. Additionally, the small overlaps 
between our six sets of pointings do not contain enough stars to perform relative astrometry.
We therefore run stage 3 reduction twice to perform the relative astrometry 
and absolute astrometry for each set of pointings individually. 
First, we run stage 3 reduction to mosaic the exposures of each set of 
pointings based on the relative positions estimated from the ditherings 
and align the mosaics to the reference catalog. 
We notice that the \texttt{TweakReg} routine calculates the geometry transformation 
based on all the matched objects using the \texttt{Tweakwcs} package. 
However, since our reference catalog has a relatively small size, the geometry 
transformation might be biased by some outliers. 
We set a large ``tweakreg.minobj'' 
value to prevent the relative astrometry in this step.
Then we modify the \texttt{Tweakwcs} package to use the median of matched 
objects to determine the geometry transformation.
We then run stage 3 reduction again to combine the mosaics of each set 
of pointings with the WCS information derived in the first run.
We set the pixel scale of the output mosaic to be $0\farcs03$ per pixel with a drizzle parameter of pixfrac $=1$.

The JWST pipeline version 1.11.4 was used with the Calibration References 
Data System (CRDS) context file of jwst\_1140 to reduce the NIRSpec 
G395H/F290LP data.   We followed the standard procedures used for the 
NIRCam images to remove snowballs and vertical 1/f noise. We do not 
remove the horizontal 1/f noise because it overlaps with the traces of the objects. 
We modify the JWST pipeline to improve the background subtraction, mask
bad pixels, and extract the entire wavelength coverage of the
G395H/F290LP configuration. We describe these modifications to the pipleline
here.

\subsubsection{Spectral Coverage of \Ha}

2D spectrum cutout is extracted during the stage 2 reduction.
The H$\alpha$ emission lines of our targets fall at the wavelength 
of $\sim 52100$ \AA, just beyond the wavelength cutoff (52000 \AA) of the 
JWST pipeline extraction. We therefore modified the reference files to extend the wavelength extraction.
The wavelength cutoff of NIRSpec is determined by the spectrograph flat 
(sflat), fore optics flat (fflat), aperture correction (apcor), photometric 
calibration (photom), and wavelength range (wavelengthrange) reference files. 
We modify the wavelength range of NIRSpec G395H/F290LP combination in the 
wavelengthrange reference file to be 2.87 -- 5.3 \micron\ and extrapolate 
the instrument responses in the sflat, fflat, apcor, and photom reference files to 5.3 \micron.
5.3 \micron\ is chosen to ensure that [N \textsc{ii}] $\lambda$6583 is also extracted.

\subsubsection{Master Background Construction}  \label{sec:master_background}

The background subtraction is performed during the stage 2 reduction.
In the JWST pipeline, the master background is constructed by a weighted 
sum of the 1D surface brightness spectra extracted from the empty slits.
The data quality mask is utilized to reject the bad pixels, cosmic rays, etc.
However, since the data quality mask sometimes does not flag all the bad 
pixels, those bad pixels are also included in the master background and 
result in bright or dark strips in the background-subtracted 2D spectra.
To effectively remove those bad pixels, we modify the \texttt{combine1d} 
routine to use a 2$\sigma$-clipped median to determine the master background for each exposure.
However, since there are only $\sim10$ background slits in each mask, the 
sigma-clipping method cannot clean all the bad pixels, particularly at 
the wavelengths covered by only a few slits. Thus, we adopt a 2$\sigma$-clipped 
median to combine the master backgrounds from 20 exposures. The pipeline fit this spectrum
with a low-order function function of wavelength to generate a master background spectrum.

For compact targets, we compared spectra extracted from two reductions:
(1) master background subtraction, and (2) pixel-to-pixel (nodded) background subtraction.
The extracted spectra agree in the bandpass between \Ha\ and \oiii, but show
some significant differences between observed wavelengths of 3.4 and 3.9 micron.  
We traced this problem to the master background produced by the NIRSpec pipeline. 
{\clm The master background is systematically brighter than the nodded background
  in several discrete bandpasses, each 0.1 to 0.2 micron wide.  Since our spectra
  typically detect only emission lines, and not continuum, the extracted spectra have a
  negative median value (instead of zero) in these regions.
The origin of this error is not yet fully understand. We note that
at longer wavelengths where zodiacal light dominates the background,
the two backgrounds agree; the differences are seen  at shorter wavelengths
where scattered light and stray light dominate the background \citep[][]{Rigby2023}. 
NIRSpec observations, however,  are detector noise limited, and the apparent
difference in backgrounds may actually reflect correlated detector noise.
The IRS$^2$ reference pixel sampling and the NSClean algorithm remove most of
the correlated noise in NIRSpec images \citep{Rauscher2024}. Following these
steps, some of our rate files still show {\it tree ring} artifacts, similar to
the concentric arcs visible in Figure 1 of \citet[][]{Rauscher2013}. The next
phase of this work will determine whether this correlated noise is the source
of the apparent background discrepancy.
}

\subsubsection{Bad pixel masking}

In stage 3 reduction, the 2D spectrum cutouts of each target are rectified 
and combined into the final 2D spectrum. 
We found that a significant fraction of bad pixels are not correctly masked 
in previous steps and thus produce pairs of bad pixels (due to the 2-point 
dithering) in our combined spectrum. Upon visual examination of those bad pixels, 
we noticed that they have a characteristic 
profile of a bright pixel surrounded by a dark (negative) ring.
They also present the same image position in all the 2D spectrum cutouts of each object.
Therefore, we identify the bad pixels by convolving the 2D spectrum cutouts with a kernel:
\begin{equation}
\begin{pmatrix}
-1/8 & -1/8 & -1/8 \\
-1/8 & 1 & -1/8 \\
-1/8 & -1/8 & -1/8
\end{pmatrix}.
\end{equation}
As most of our spectra are dominated by the background which varies smoothly 
across the spatial and wavelength directions and the sum of the kernel 
is 0, the convolution will result in very small values for good pixels.
However, this kernel is very sensitive to bad pixels and will lead to relatively large values.
We adopt a threshold of $10^{-11}$ MJy pix$^{-1}$ to identify bad pixels.
To avoid the emission lines being selected by this method, we request the 
bad pixels be identified in all four exposures (ABBA) of each mask.
Finally, we grow the bad pixel mask by one pixel.

\section{Aperture Corrections} \label{sec:appendix}

\begin{figure}[h]
 \centering
   \includegraphics[scale=0.65,angle=-90,trim = 0 0 220 0]{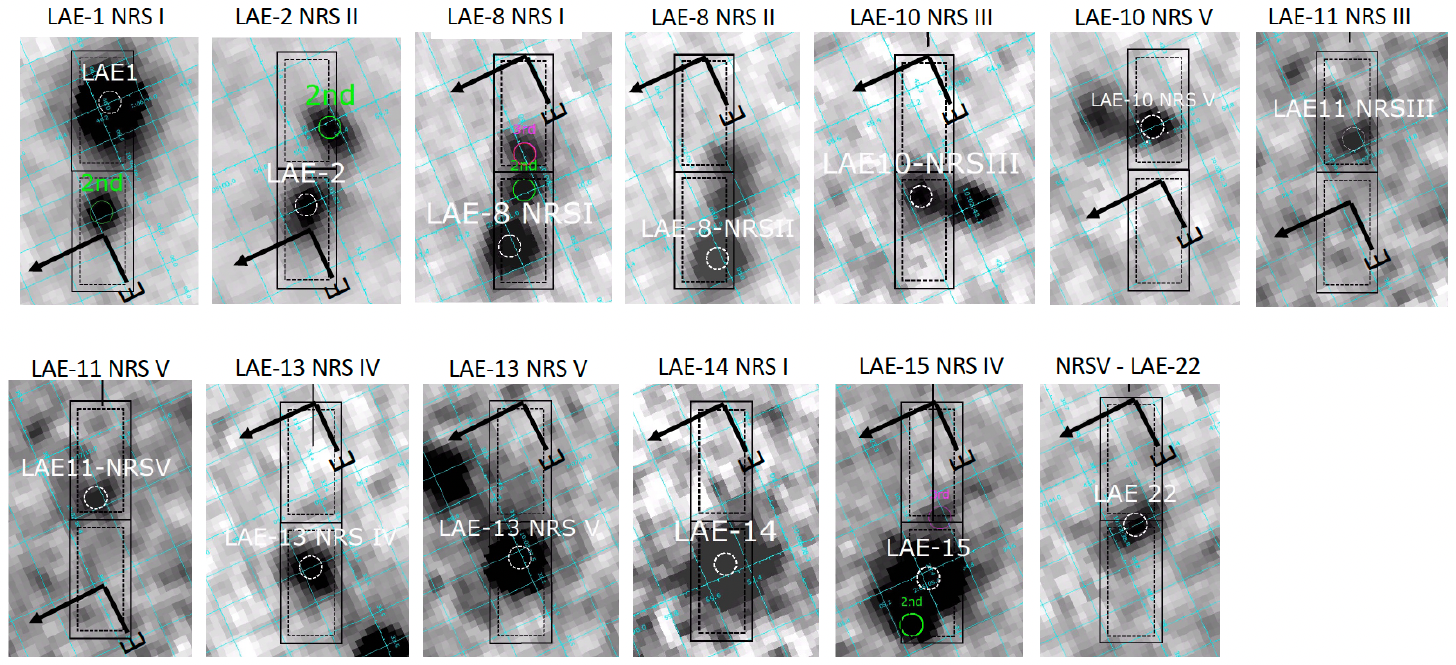}
   \caption{Identification of targets and secondary clumps which were detected spectroscopically.
           Two nearest shutters shown on F444W image. At the nominal platescale of the telescope and 
           NIRSpec foreoptics, the mean angular extent of the shutter open area on the sky is
           0\farcs199 $\times$  0\farcs461 as shown by the dashed, black lines. Black arrow points north.
}
    \label{fig:clumps}
     \end{figure}

Aperture corrections strongly affect our measurement of the total \Ha\ luminosity, and hence
the \lya\ escape fraction, so we outline the steps performed to accurately model them here.
Each slitlet producing a 2D spectrum is comprised of several MSA shutters, and the 
individual clumps land in one or two contiguous shutters. We matched the distinct 
spectra identified along each slitlet to the locations of clumps in the F444W image.
Figure~\ref{fig:clumps} identifies these clumps on two-shutter overlays. We performed 
boxcar extractions of the sub-apertures defined by the individual clumps and applied
aperture corrections based on the pitch of each clump within its shutter. Table~\ref{tab:clumps} 
lists the pitch and world coordinates of each clump that was spectroscopically detected with 
NIRSpec.

First, we defined the target coordinates for spectroscopic follow up based on the locations of clumps 
in the F444W images. Although the spatial resolution is amost a factor of three better in the F150W2
images, our SED modeling suggests that the F444W band contains high equivalent width \oiii\ and \Hb\ emission. 
We noticed that the centroids and sizes of indivdiual clumps sometimes differed from their counterparts in 
the rest-frame near-ultraviolet image, so coordinates needed to be measured near the wavelengths that would be
spectroscopically observed. The nebular structure in \oiii\ and \Hb\  is our best indication 
of the \Ha\ morphology, so we adopted the centroids and position angles of the clumps in the F444W band to
model aperture losses. Larger nebulae, as seen around LAE-1 and LAE-15, will will be strongly attenuated even 
when the galaxy is perfectly centered in a shutter. Individual clumps, moreover, are rarely perfectly centered 
in a shutter. Computing distinct aperture corrections for each clump proves to be
critical since the source of each sub-spectrum has a different shutter pitch.

Next we used the forward-modelling software \texttt{msafit} to compute aperture losses. \texttt{msafit} 
account for the complex geometry, point spread function, and pixellation of the NIRSpec instrument  
\citep{deGraaff2024}.  The software also accounts for bar shadows, the obscuration caused by the
area between the solid and dashed black lines in Figure~\ref{fig:clumps}. We configured with
a 3-shutter slitlet, and used it to convolve the wavelength dependent PSF with a separate S\'{e}sic model for
each clump associated with an LAE.  We fit each clump with a S\'{e}rsic profile using GalFit 
\citep{Peng2002_galfit, Peng2010_galfit}.  Table~\ref{tab:clumps} lists the resulting position angles
and effective radii. Allowing the index $n$ to vary did not affect the aperture corrections
significantly, and we list the results  obtained with $n \equiv\ 1$.  

For individual clumps, the median aperture correction is 2.4 with a range from 2.0 to 4.1.  
Even though many of the individual clumps are spatially extended, the S\'{e}rsic models
increased the median aperture corrections by only a factor of 1.2 relative to a point source model.  
Columns 9 and 10 in Table~\ref{tab:clumps} illustrate this comparison on a clump by clump basis.
Accurate pitches for individual clumps therefore leave little uncertainty about aperture correction.

The pitch describes the fractional displacment of a target within the full shutter. The header
keywords \texttt{(msa\_x, msa\_y)}  in the JWST NIRSpec datamodel give the pitch of our target.  We
assigned this pitch to the world coordinates of our target clump during mask design. To make
the aperture corrections, we compute the pitch of any additional clumps, in the same
shutter or a neighboring shutter, based on the aperture position angle and the offset of that
clump from the target clump in right ascension and declination.  Optical distortions and small
metrology differences produce variations in the open shutter area of up 6.8\%, and \texttt{msafit}
accounts for these shutter-to-shutter variations.  Since the angular sizes of individual shutters
are not publically available, however, we must assume the median angular size of a full MSA shutter,
0\farcs27 $\times$ 0\farcs53, when calculating secondary clump pitches. Our inferred pitches of
the secondary clumps have maximum errors up of 3.9\% and 2.9\% in shutter height and width,
respectively. These errors will not have a significant impact on the aperture corrections because
the corresponding angular shifts of 0\farcs021  and 0\farcs0078 are considerably smaller than
the F444W PSF.

The throughput values in the last column of Table~\ref{tab:clumps} indicate a median aperture
correction of 2.4 at \Ha.  To understand the systematic errors, we compared observations of the
same galaxies made through different masks. The brightest clump in LAE-8, for example, had a
a different shutter pitch on NRS~I and NRS~II, but the aperture corrected fluxes are consistent.
We adopt the NRS~I observation because this spectrum detects Balmer emission from clump 3 which
is largely outside the NRS~II slitlet. The luminosity difference of 0.17~dex is comparable to
the statistical error on the individual measurements. The centering of LAE-11 is reasonable in
the NRS~III shutter and quite poor in NRS~V. We use the NRS~III observation exclusively to derive
galaxy properties but compare their spectral line measurements here in order to illustrate
worst case aperture corrections. We find the \Hb\ flux in the extracted NRS~III spectrum is 1.64
times higher than the flux in the NRS~V spectrum due to the unfavorable pitch of the latter.
The  ratio of throughputs predicted by \texttt{msafit}  is only 1.13, however, so an
aperture-loss-corrected flux for a clump near a shutter boundary might only recover 70\% of
the true flux.  Inspection of Table~\ref{tab:clumps} shows that our primary clumps have much
more favorable shutter pitches than LAE-11 on NRS~V.  The noteworthy exception is LAE-22;
its \Ha\ luminosity may be underestimated by a factor of $\approx 1.4$. The off-center
clumps in LAE-2, LAE-8, and LAE-15 make small contributions to the total Balmer-line
luminosities.

\begin{deluxetable}{lllllllllll}
\tablecaption{Slitlet Throughput for \Ha\ Line Emission\tablenotemark{a} }
\colnumbers
\tablehead{
\colhead{Galaxy \&  Mask} &
\colhead{Clump} &
\colhead{RA} &
\colhead{DEC} &
\colhead{${\rm r_e}$} &
\colhead{q} &
\colhead{PA} &
\colhead{${\rm PA_{xpos}}$ } &
\colhead{(x0, y0) }   &
\colhead{$\epsilon$ (Pt Src) } &
\colhead{$\epsilon$ (S\'{e}rsic) }
\\
\colhead{}        &
\colhead{}        &
\colhead{(J2000)}  &
\colhead{(J2000)}  &
\colhead{(\arcsec)} &
\colhead{} &
\colhead{($^{\circ}$)}   &
\colhead{($^{\circ}$)} &
\colhead{}   &
\colhead{}          &
\colhead{}
}
\startdata
LAE-1 NRS I  &
targ  (1-2)&  
10:02:05.960 &
+02:06:46.14  &  
0.061 &
0.69   &
55.0     &
78.2  & 
(-0.116, -0.070)    &
0.563        &
0.479        
\\
''   &
2nd (1-1)  &  
10:02:05.989 &
+02:06:46.36  &  
0.17 &
0.38   &
40.0   &
73.1       & 
(0.013, -0.156)  &
0.586        &
0.370        
\\
LAE-2 NRS II  &
targ (2-1)  &  
10:01:53.460 &
+02:04:59.68  &
0.11         &
0.59   &
58.3         &
82.7        & 
(0.016, -0.201  &
0.577        &
0.422
\\
''   &
2nd (2-2, 2-3)  &  
10:01:53.442 &
+02:04:59.44  &  
0.058     &
0.60   &
-59.5      &
-35.1      & 
(-0.380, 0.140)        &
0.316        &
0.299
\\
LAE-8  NRS I &
targ (8-1)&  
10:02:09.013 &
+02:04:11.02  &  
0.063     &
0.84   &
61.6        &
84.9        & 
(0.235, 0.135)        &
0.456\tablenotemark{a}        &
0.434\tablenotemark{a}
\\
''  &
3rd (8-3) &  
10:02:08.989 &
+02:04:10.79 &
0.087 &
0.30   &
-55.7  &
-32.5        & 
(-0.017, 0.347)        &
0.478\tablenotemark{a} &
0.392\tablenotemark{a}
\\
LAE-8  NRS II &
targ (8-1)&  
10:02:09.013 &
+02:04:11.02  &  
0.063      &
0.84   &
61.6        &
86.1        & 
(0.016, -0.201)        &
0.689\tablenotemark{a}        &
0.526\tablenotemark{a}
\\
LAE-10 NRS III  &
targ (10-2)    &  
10:02:42.311  &  
+02:06:55.34   &
0.095        &
0.53   &
61.8       &
-12.7        & 
(0.049, -0.287)        &
0.488        &
0.413
\\
LAE-10 NRS V  &
targ (10-1)  &  
10:02:42.320 &  
+02:06:55.14  &
0.056 &
0.55   &
-37.0   &
86.1        & 
(0.105,0.139)        &
0.652\tablenotemark{a}        &
0.524\tablenotemark{a}
\\
LAE-11 NRS III &
targ (11-1)   &  
10:02:39.437 &  
+02:07:12.03  &
0.057       &
0.58   &
-2.1        &
22.1        & 
(-0.116,0.231)        &
0.530        &
0.435
\\
LAE-11 NRS V &
targ (11-1) &  
10:02:39.437 &  
+02:07:12.03  &
0.057       &
0.58   &
-2.1        &
22.1        &  
(0.077, 0.313)        &
0.428        &
0.384
\\
LAE-13 NRS IV   &
targ (13-2)       &  
10:02:33.460        &  
+02:07:09.56     &
0.12        &
0.66   &
-62.2 &
-37.8        & 
(0.011,-0.131)        &
0.585        &
0.372
\\
LAE-13 NRS V &
targ (13-1)  &  
10:02:33.492 &
+02:07:09.41 &  
0.049    &
0.78   &
64.5        &
88.8        & 
(0.020,-0.206)        &
0.577        &
0.495
 \\
LAE-14 NRS I  &
targ (14-1)   &  
10:02:08.257        &  
+02:06:59.49        &
0.079         &
0.77   &
16.7        &
49.0        &  
(0.090,-0.110)        &
0.561        &
0.433
\\
LAE-15 NRS IV  &
targ (15-2)   &  
10:02:23.382 &  
+02:05:04.98  &
0.030       &
0.82   &
-26.2        &
-1.9        & 
(0.208,-0.005)        &
0.576\tablenotemark{a}        &
0.522\tablenotemark{a}
\\
''  &
2nd (15-1)   &  
10:02:23.352 &
+02:05:04.76        &  
0.081         &
0.10   &
39.7        &
64.1        & 
(0.328,0.346)        &
0.294\tablenotemark{a}        &
0.324\tablenotemark{a}
\\
''  &
3rd (15-3)   &  
10:02:23.364 &  
+02:05:04.72  &
0.078      &
0.50   &
24.0       &
48.4        & 
(-0.131,0.457)        &
0.265\tablenotemark{a}        &
0.314\tablenotemark{a}
\\
LAE-22 NRS V &
targ (22-1)        &  
10:02:38.750         &  
+02:07:43.69       &
0.098         &
0.57   &
17.8        &
42.1        &  
(-0.064, 0.466)        &
0.275        &
0.266
\\
\enddata
\tablenotetext{a}{The spectral coverage for LAE-8 (both masks), LAE10 NRS V,  and LAE-15
does not include \Ha.  
We scaled their \Hb\ fluxes by 2.78 to estimate their \Ha\ fluxes. The aperture correction 
listed is computed at the observed wavelength of \Hb\ for these spectra.}
\tablecomments{
{\it (Col 1):}    Name of LAGER LAE and NIRSpec mask number. \\
{\it (Col 2):}    Identification of our spectroscopic target and secondary clumps that fall within
                  a three-shutter slitlet.  GalFit sometimes found additional clumps outside the slitlet;
                  we identify the GalFit clump associated with each spectroscopic target in parentheses.\\
{\it (Col 3):}    Right Ascension. \\
{\it (Col 4):}    Declination. \\
{\it (Col 5):}    The half-light radius of the S\'{e}rsic profile before convolution with the PSF. \\
{\it (Col 6):}    Fitted ratio of minor axis to major axis. \\
{\it (Col 7):}    The position angle of the S\'{e}rsic profile. \\
{\it (Col 8):}    The position angle in \texttt{msafit} is defined relative to the positive
                  x-axis, i.e. in the dispersion direction toward shorter wavelengths. We 
                  calculated it from the sky PA of the S\'{e}rsic model and the slit PA. \\
{\it (Col 9):}    The pitch used in \texttt{msafit}, where the origin is at the center
                  of each shutter. We note that these values are -0.5 less than the keywords
                  \texttt{(msa\_x, msa\_y)} because the latter defines the origin at the 
                  corner of the full shutter.\\
{\it (Col 10):}    Three-shutter throughput for a point-source morphology calculated 
                  from \texttt{msafit}.\\
{\it (Col 11):}   Three-shutter throughput for a S\'{e}rsic profile with index $n \equiv 1$ 
                  calculated from \texttt{msafit}.\\
}
\label{tab:clumps}
\end{deluxetable}


\section{SED Fitting:  Non-parametric Star Formation Histories \& Reddening}  \label{sec:bagpipes}

Table~\ref{tab:photometry} summarizes the broadband photometry.  The NIRCam pre-imaging detects
the nine LAEs in F444W and the very broad F150W2 band. After defining each spectral energy 
distribution (SED) shape based on spectral properties, we estimated the stellar mass in Col.~8 
by fitting the photometry with \texttt{BEAGLE}  \citep{Chevallard2016}, effectively normalizing
the SED. Five of these galaxies are  detected in at least three of the four  UltraVISTA DR6 
bands \citep{McCracken2015}.  Their  UV spectral slope $\beta$, where $F_{\lambda} \propto \lambda^{\beta}$, is well constrained, and we performed SED  fitting using \texttt{BAGPIPES} \citep{Carnall2018}. 

\begin{deluxetable}{lllllllll}
\tablecaption{UltraVISTA and NIRCam Photometry}
\colnumbers
\tablehead{
\colhead{Galaxy } &
\colhead{Y} &
\colhead{J} &
\colhead{H} &
\colhead{Ks} &
\colhead{F150W2} &
\colhead{F444W} &
\colhead{$\log (M_*)$\tablenotemark{b}}    &
\colhead{$\log (M_*)$\tablenotemark{c}}   
\\
\colhead{}     &
\colhead{(nJy)} &
\colhead{(nJy)} &
\colhead{(nJy)} &
\colhead{(nJy)} &
\colhead{(nJy)} &
\colhead{(nJy)} &
\colhead{(\msun)} &  
\colhead{(\msun)}   
}
\startdata
LAE-1  &
$ 176.0 \pm 26.0 $ &  
$ 203.5 \pm 27.9 $ &  
$ 185.0 \pm 21.5 $ &  
$ 195.4 \pm 69.5 $ &  
$ 251.4 \pm 31.4 $ &  
$ 578.1 \pm 39.9 $ &
$9.14_{-0.03}^{+0.06}$ &  
$9.11_{-0.03}^{0.05}$ 
\\
LAE-2\tablenotemark{a}  &
$ 197.3 \pm 14.5 $ &  
$ 170.9 \pm 81.9 $ &  
$ 204.6 \pm 36.9 $ &  
$ 186.4 \pm 41.4 $ &  
$ 182.2 \pm 26.2 $ &  
$ 161.5 \pm 26.1$  &
$8.93_{-0.02}^{+0.03}$ &
$8.43_{-0.01}^{+0.01} $ 
\\
LAE-8  &
$ 140.1 \pm 27.9 $ &  
$ 353.3 \pm  97.3 $ &  
$ 250.3 \pm 37.1$ &  
$ 155.3 \pm 46.4 $ &  
$ 132.0 \pm 28.7 $ &  
$ 235.0 \pm 27.0 $ &
$ 9.44_{-0.05}^{+0.09}$ &  
$9.1_{-0.2}^{+0.2} $ 
\\
LAE-10  &
$ 151.7 \pm 21.9 $ &  
$ < 52.1 $ &  
$ < 87.2$ &  
$ < 227.3 $ &  
$ 59.4 \pm 16.7$ &  
$ 62.08 \pm 22.8 $   &
$ 9.42_{-0.03}^{+0.03} $ &
\nodata 
\\
LAE-11  &
$ 176.0 \pm 44.9 $ &  
$ < 69.7 $ &  
$ < 161.3 $ &  
$ 91.4 \pm 51.1 $ &  
$ 35.5 \pm 29.8 $ &  
$ 72.3 \pm 24.8 $ &
$8.5_{-0.2}^{+0.3}$ &
\nodata 
\\
LAE-13  &
$ 185.1 \pm 158.5 $ &  
$ 153.3 \pm 197.8 $ &  
$ < 264.7 $ &  
$ < 236.2 $ &  
$ 74.3  \pm 53.5 $ &  
$ 163.8 \pm 39.7 $ &
$8.7_{-0.5}^{+0.5}$ &  
\nodata 
\\
LAE-14  &
$ 174.6 \pm 16.8 $ &  
$ 128.7 \pm 106.0 $ &  
$ 186.2 \pm 32.8 $ &  
$ < 92.1 $ &  
$ 82.5 \pm 31.6 $ &  
$ 164.2 \pm 51.4 $ &
$8.84_{-0.12}^{+0.12}$ &
$8.8_{-0.2}^{+0.2} $ 
\\
LAE-15  &
$ 123.7 \pm 19.5 $ &  
$ 144.1 \pm 40.4 $ &  
$ 128.7 \pm 65.9 $ &  
$ < 109.0 $ &  
$ 135.4 \pm 30.3 $ &  
$ 416.7 \pm 42.9$  &
$9.61_{-0.07}^{+0.10}$ &
9.2 
\\
LAE-22  &
$ < 31.4 $ &  
$ < 214.1 $ &  
$ < 48.3 $ &  
$ < 83.3 $ &  
$ 53.6 \pm 29.7 $ &  
$ 128.6 \pm 40.2 $  &
$9.72_{-0.06}^{+0.12}$ &
\nodata 
\\
\enddata
%
\tablenotetext{a}{We note the presence of a small object to the
northwest of LAE-2 which is not resolved by the UltraVISTA imaging. 
This object has a blue F150W2-F444W color, is likely a foreground galaxy,
and may artifically steepens the fitted UV slope.}
\tablenotetext{b}{Stellar mass fitted to photometry using \texttt{BEAGLE} \citep{Chevallard2016}.
We fixed the ionization parameter, dust optical depth, metallicity, and age at
the values estimated from each spectrum. We then fit the photometry. 
}
\tablenotetext{c}{Stellar mass fitted using \texttt{BAGPIPES}  as described in the text.
We give the 50th percentile, and the uncertainties denote the 16th and 84th percentiles.
}

\label{tab:photometry} 
\end{deluxetable}

\begin{figure*}[h]
 \centering
   \includegraphics[scale=0.8,angle=-90,trim = 0 0 50 0]{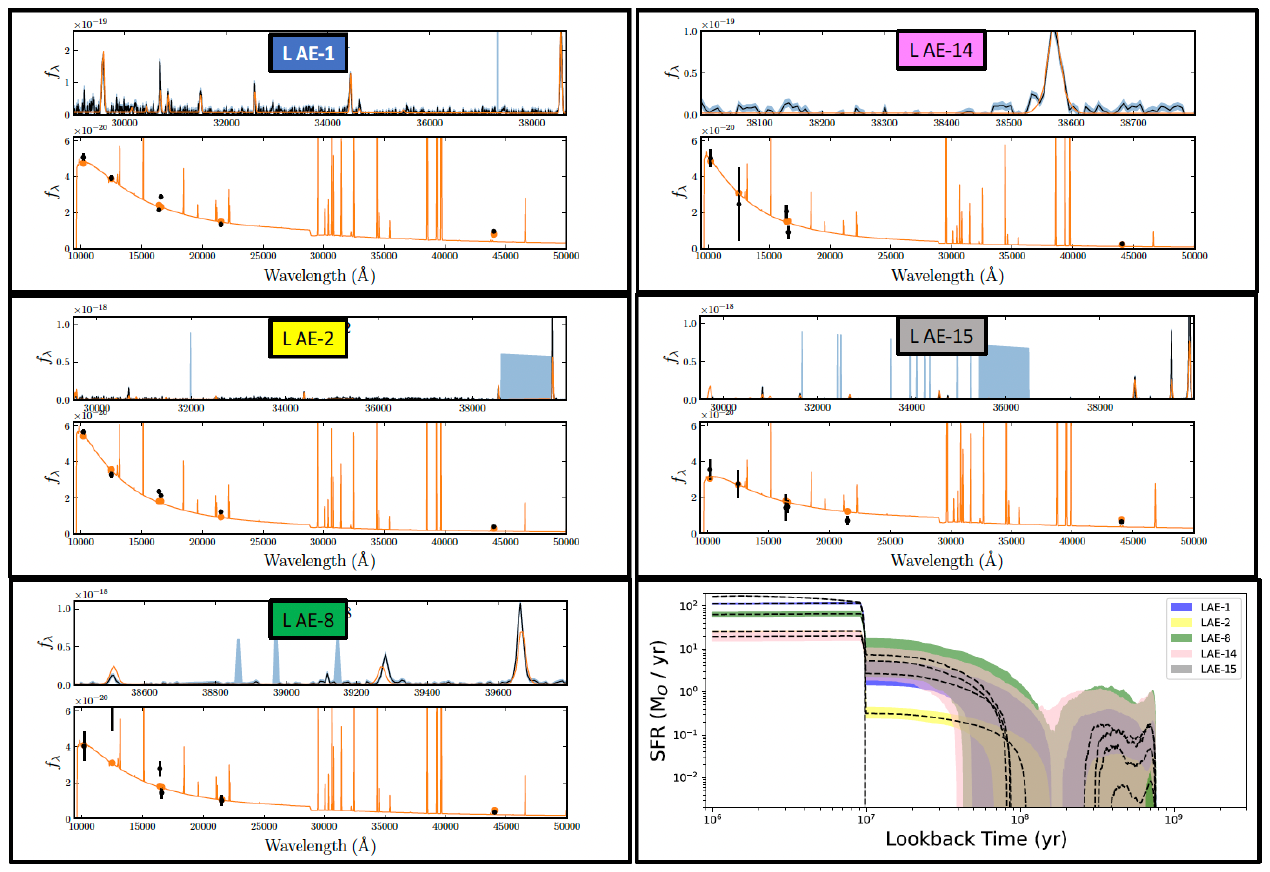}
  \caption{ Fitted \texttt{BAGPIPES} spectral energy distributions. 
      Nonparametric star formation histories are compared in the final panel.
     }
     \label{fig:five_bagpipes}
      \end{figure*}

Nebular \oiii\ emission makes a significant contribution to the F444W photometry, 
so we fit the region of the spectrum covering the \oiii\ doublet, \Hb, and \Hg\
simultaneously with the photometry. These lines lie on the blue side of the chip gap.
The fit did not include \Ha\ because it lands redward of the gap, and \texttt{BAGPIPES} 
requires a continuous spectrum. We fixed the redshifts using our measurements from 
these G395H spectra. We employed the Binary Population and Spectral Synthesis 
v2.2.1 \citep{Eldridge2017,Stanway2018} (BPASS) with a broken power-law initial 
mass function with slopes of $\alpha_1 = -1.3$ for stars with 0.1 -- 0.5 $M_\odot$ 
and $\alpha_2 = -2.35$ for 0.5 -- 100 $M_\odot$ (model `135\_100').
For the star formation history parameterization, we used the Gaussian Process 
model from \texttt{DENSEBASIS} \citep{Iyer2019}, where the star formation 
history is split into three  dynamically adjusted time bins with ages of 10, 40, and 100 Myr.
The stellar metallicity was allowed to span 0.001 -- 1~$Z_\odot$ and tied
to the nebular metallicity. The ionization parameter of the nebular emission
was allowed to vary in the range $\log U = $ $-4$ to $-1$. 
We adopted the \cite{Calzetti2000} dust attenuation law with $A_V$ 
ranging from 0.0 to 4.0~mag.  
We used the uniform priors for the \texttt{BAGPIPES} parameters across the allowed range.
Figure~\ref{fig:five_bagpipes} illustrates the SED fits for five LAEs. 

In this paper, the ionization budget provides the primary motivation for 
deriving non-parametric star formation histories. The final panel of 
Fig.~\ref{fig:five_bagpipes} illustrates the non-parametric star formation histories. 
The non-parametric SED fitting lengthens the star formation history relative to
our fiducial starburst duration of 15~Myr.   
We calculated the number of ionizing photons produced by each star formation history. 
BPASS defines the temporal evolution of the ionizing luminosity following an
instantaneous burst of star formation, relative to mass of $10^6 \msun$ at formation.
We scale this model to the stellar mass produced in each time step of the
\texttt{BAGPIPES} star formation history. As a function of the lookback time, $\tau$, 
we add up the ionizing photon luminosity contributed by 
all previously formed populations,  
\begin{eqnarray}
Q_{SFH}(\tau) = \int_{\tau}^{300~Myr} {\rm SFR}(\tau_i) Q_{SSP}(\tau_i - \tau) d \tau_i,
 \label{eqn:q_integral} \end{eqnarray}
where the argument of  $Q_{SSP}(t)$ is the population age.
We compare the total production
of ionizing photons in these models to the most recent burst in \S~\ref{sec:sed}.

\clearpage

\bibliography{main.bib}

\begin{thebibliography}{}
\expandafter\ifx\csname natexlab\endcsname\relax\def\natexlab#1{#1}\fi
\providecommand{\url}[1]{\href{#1}{#1}}
\providecommand{\dodoi}[1]{doi:~\href{http://doi.org/#1}{\nolinkurl{#1}}}
\providecommand{\doeprint}[1]{\href{http://ascl.net/#1}{\nolinkurl{http://ascl.net/#1}}}
\providecommand{\doarXiv}[1]{\href{https://arxiv.org/abs/#1}{\nolinkurl{https://arxiv.org/abs/#1}}}

\bibitem[{{Adamo} {et~al.}(2024){Adamo}, {Bradley}, {Vanzella}, {Claeyssens},
  {Welch}, {Diego}, {Mahler}, {Oguri}, {Sharon}, {Abdurro'uf}, {Hsiao}, {Xu},
  {Messa}, {Lassen}, {Zackrisson}, {Brammer}, {Coe}, {Kokorev}, {Ricotti},
  {Zitrin}, {Fujimoto}, {Inoue}, {Resseguier}, {Rigby}, {Jim{\'e}nez-Teja},
  {Windhorst}, {Hashimoto}, \& {Tamura}}]{Adamo2024-Nature}
{Adamo}, A., {Bradley}, L.~D., {Vanzella}, E., {et~al.} 2024, \nat, 632, 513,
  \dodoi{10.1038/s41586-024-07703-7}

\bibitem[{{Astropy Collaboration} {et~al.}(2013){Astropy Collaboration},
  {Robitaille}, {Tollerud}, {Greenfield}, {Droettboom}, {Bray}, {Aldcroft},
  {Davis}, {Ginsburg}, {Price-Whelan}, {Kerzendorf}, {Conley}, {Crighton},
  {Barbary}, {Muna}, {Ferguson}, {Grollier}, {Parikh}, {Nair}, {Unther},
  {Deil}, {Woillez}, {Conseil}, {Kramer}, {Turner}, {Singer}, {Fox}, {Weaver},
  {Zabalza}, {Edwards}, {Azalee Bostroem}, {Burke}, {Casey}, {Crawford},
  {Dencheva}, {Ely}, {Jenness}, {Labrie}, {Lim}, {Pierfederici}, {Pontzen},
  {Ptak}, {Refsdal}, {Servillat}, \& {Streicher}}]{astropy:2013}
{Astropy Collaboration}, {Robitaille}, T.~P., {Tollerud}, E.~J., {et~al.} 2013,
  \aap, 558, A33, \dodoi{10.1051/0004-6361/201322068}

\bibitem[{{Astropy Collaboration} {et~al.}(2018){Astropy Collaboration},
  {Price-Whelan}, {Sip{\H{o}}cz}, {G{\"u}nther}, {Lim}, {Crawford}, {Conseil},
  {Shupe}, {Craig}, {Dencheva}, {Ginsburg}, {Vand erPlas}, {Bradley},
  {P{\'e}rez-Su{\'a}rez}, {de Val-Borro}, {Aldcroft}, {Cruz}, {Robitaille},
  {Tollerud}, {Ardelean}, {Babej}, {Bach}, {Bachetti}, {Bakanov}, {Bamford},
  {Barentsen}, {Barmby}, {Baumbach}, {Berry}, {Biscani}, {Boquien}, {Bostroem},
  {Bouma}, {Brammer}, {Bray}, {Breytenbach}, {Buddelmeijer}, {Burke},
  {Calderone}, {Cano Rodr{\'\i}guez}, {Cara}, {Cardoso}, {Cheedella}, {Copin},
  {Corrales}, {Crichton}, {D'Avella}, {Deil}, {Depagne}, {Dietrich}, {Donath},
  {Droettboom}, {Earl}, {Erben}, {Fabbro}, {Ferreira}, {Finethy}, {Fox},
  {Garrison}, {Gibbons}, {Goldstein}, {Gommers}, {Greco}, {Greenfield},
  {Groener}, {Grollier}, {Hagen}, {Hirst}, {Homeier}, {Horton}, {Hosseinzadeh},
  {Hu}, {Hunkeler}, {Ivezi{\'c}}, {Jain}, {Jenness}, {Kanarek}, {Kendrew},
  {Kern}, {Kerzendorf}, {Khvalko}, {King}, {Kirkby}, {Kulkarni}, {Kumar},
  {Lee}, {Lenz}, {Littlefair}, {Ma}, {Macleod}, {Mastropietro}, {McCully},
  {Montagnac}, {Morris}, {Mueller}, {Mumford}, {Muna}, {Murphy}, {Nelson},
  {Nguyen}, {Ninan}, {N{\"o}the}, {Ogaz}, {Oh}, {Parejko}, {Parley}, {Pascual},
  {Patil}, {Patil}, {Plunkett}, {Prochaska}, {Rastogi}, {Reddy Janga},
  {Sabater}, {Sakurikar}, {Seifert}, {Sherbert}, {Sherwood-Taylor}, {Shih},
  {Sick}, {Silbiger}, {Singanamalla}, {Singer}, {Sladen}, {Sooley},
  {Sornarajah}, {Streicher}, {Teuben}, {Thomas}, {Tremblay}, {Turner},
  {Terr{\'o}n}, {van Kerkwijk}, {de la Vega}, {Watkins}, {Weaver}, {Whitmore},
  {Woillez}, {Zabalza}, \& {Astropy Contributors}}]{astropy:2018}
{Astropy Collaboration}, {Price-Whelan}, A.~M., {Sip{\H{o}}cz}, B.~M., {et~al.}
  2018, \aj, 156, 123, \dodoi{10.3847/1538-3881/aabc4f}

\bibitem[{{Astropy Collaboration} {et~al.}(2022){Astropy Collaboration},
  {Price-Whelan}, {Lim}, {Earl}, {Starkman}, {Bradley}, {Shupe}, {Patil},
  {Corrales}, {Brasseur}, {N{"o}the}, {Donath}, {Tollerud}, {Morris},
  {Ginsburg}, {Vaher}, {Weaver}, {Tocknell}, {Jamieson}, {van Kerkwijk},
  {Robitaille}, {Merry}, {Bachetti}, {G{"u}nther}, {Aldcroft},
  {Alvarado-Montes}, {Archibald}, {B{'o}di}, {Bapat}, {Barentsen}, {Baz{'a}n},
  {Biswas}, {Boquien}, {Burke}, {Cara}, {Cara}, {Conroy}, {Conseil}, {Craig},
  {Cross}, {Cruz}, {D'Eugenio}, {Dencheva}, {Devillepoix}, {Dietrich},
  {Eigenbrot}, {Erben}, {Ferreira}, {Foreman-Mackey}, {Fox}, {Freij}, {Garg},
  {Geda}, {Glattly}, {Gondhalekar}, {Gordon}, {Grant}, {Greenfield}, {Groener},
  {Guest}, {Gurovich}, {Handberg}, {Hart}, {Hatfield-Dodds}, {Homeier},
  {Hosseinzadeh}, {Jenness}, {Jones}, {Joseph}, {Kalmbach}, {Karamehmetoglu},
  {Ka{l}uszy{'n}ski}, {Kelley}, {Kern}, {Kerzendorf}, {Koch}, {Kulumani},
  {Lee}, {Ly}, {Ma}, {MacBride}, {Maljaars}, {Muna}, {Murphy}, {Norman},
  {O'Steen}, {Oman}, {Pacifici}, {Pascual}, {Pascual-Granado}, {Patil},
  {Perren}, {Pickering}, {Rastogi}, {Roulston}, {Ryan}, {Rykoff}, {Sabater},
  {Sakurikar}, {Salgado}, {Sanghi}, {Saunders}, {Savchenko}, {Schwardt},
  {Seifert-Eckert}, {Shih}, {Jain}, {Shukla}, {Sick}, {Simpson},
  {Singanamalla}, {Singer}, {Singhal}, {Sinha}, {Sip{H{o}}cz}, {Spitler},
  {Stansby}, {Streicher}, {{{S}}umak}, {Swinbank}, {Taranu}, {Tewary},
  {Tremblay}, {Val-Borro}, {Van Kooten}, {Vasovi{'c}}, {Verma}, {de Miranda
  Cardoso}, {Williams}, {Wilson}, {Winkel}, {Wood-Vasey}, {Xue}, {Yoachim},
  {Zhang}, {Zonca}, \& {Astropy Project Contributors}}]{astropy:2022}
{Astropy Collaboration}, {Price-Whelan}, A.~M., {Lim}, P.~L., {et~al.} 2022,
  \apj, 935, 167, \dodoi{10.3847/1538-4357/ac7c74}

\bibitem[{{Atek} {et~al.}(2009){Atek}, {Kunth}, {Schaerer}, {Hayes},
  {Deharveng}, {{\"O}stlin}, \& {Mas-Hesse}}]{Atek2009}
{Atek}, H., {Kunth}, D., {Schaerer}, D., {et~al.} 2009, \aap, 506, L1,
  \dodoi{10.1051/0004-6361/200912787}

\bibitem[{{Aung} {et~al.}(2024){Aung}, {Mandelker}, {Dekel}, {Nagai},
  {Semenov}, \& {van den Bosch}}]{Aung2024}
{Aung}, H., {Mandelker}, N., {Dekel}, A., {et~al.} 2024, \mnras, 532, 2965,
  \dodoi{10.1093/mnras/stae1673}

\bibitem[{{Aver} {et~al.}(2015){Aver}, {Olive}, \& {Skillman}}]{Aver2015}
{Aver}, E., {Olive}, K.~A., \& {Skillman}, E.~D. 2015, \jcap, 2015, 011,
  \dodoi{10.1088/1475-7516/2015/07/011}

\bibitem[{{Bagley} {et~al.}(2023){Bagley}, {Finkelstein}, {Koekemoer},
  {Ferguson}, {Arrabal Haro}, {Dickinson}, {Kartaltepe}, {Papovich},
  {P{\'e}rez-Gonz{\'a}lez}, {Pirzkal}, {Somerville}, {Willmer}, {Yang}, {Yung},
  {Fontana}, {Grazian}, {Grogin}, {Hirschmann}, {Kewley}, {Kirkpatrick},
  {Kocevski}, {Lotz}, {Medrano}, {Morales}, {Pentericci}, {Ravindranath},
  {Trump}, {Wilkins}, {Calabr{\`o}}, {Cooper}, {Costantin}, {de la Vega},
  {Hilbert}, {Hutchison}, {Larson}, {Lucas}, {McGrath}, {Ryan}, {Wang}, \&
  {Wuyts}}]{Bagley2023}
{Bagley}, M.~B., {Finkelstein}, S.~L., {Koekemoer}, A.~M., {et~al.} 2023,
  \apjl, 946, L12, \dodoi{10.3847/2041-8213/acbb08}

\bibitem[{{Baldwin} {et~al.}(1981){Baldwin}, {Phillips}, \&
  {Terlevich}}]{Baldwin1981}
{Baldwin}, J.~A., {Phillips}, M.~M., \& {Terlevich}, R. 1981, \pasp, 93, 5,
  \dodoi{10.1086/130766}

\bibitem[{{Becker} {et~al.}(2015){Becker}, {Bolton}, {Madau}, {Pettini},
  {Ryan-Weber}, \& {Venemans}}]{Becker2015}
{Becker}, G.~D., {Bolton}, J.~S., {Madau}, P., {et~al.} 2015, \mnras, 447,
  3402, \dodoi{10.1093/mnras/stu2646}

\bibitem[{{B{\"o}ker} {et~al.}(2023){B{\"o}ker}, {Beck}, {Birkmann},
  {Giardino}, {Keyes}, {Kumari}, {Muzerolle}, {Rawle}, {Zeidler}, {Abul-Huda},
  {Alves de Oliveira}, {Arribas}, {Bechtold}, {Bhatawdekar}, {Bonaventura},
  {Bunker}, {Cameron}, {Carniani}, {Charlot}, {Curti}, {Espinoza}, {Ferruit},
  {Franx}, {Jakobsen}, {Karakla}, {L{\'o}pez-Caniego}, {L{\"u}tzgendorf},
  {Maiolino}, {Manjavacas}, {Marston}, {Moseley}, {Ogle}, {Perna},
  {Pe{\~n}a-Guerrero}, {Pirzkal}, {Plesha}, {Proffitt}, {Rauscher}, {Rix},
  {Rodr{\'\i}guez del Pino}, {Rustamkulov}, {Sabbi}, {Sing}, {Sirianni}, {te
  Plate}, {{\'U}beda}, {Wahlgren}, {Wislowski}, {Wu}, \& {Willott}}]{Boker2023}
{B{\"o}ker}, T., {Beck}, T.~L., {Birkmann}, S.~M., {et~al.} 2023, \pasp, 135,
  038001, \dodoi{10.1088/1538-3873/acb846}

\bibitem[{{Bolan} {et~al.}(2022){Bolan}, {Lemaux}, {Mason}, {Brada{\v{c}}},
  {Treu}, {Strait}, {Pelliccia}, {Pentericci}, \& {Malkan}}]{Bolan2022}
{Bolan}, P., {Lemaux}, B.~C., {Mason}, C., {et~al.} 2022, \mnras, 517, 3263,
  \dodoi{10.1093/mnras/stac1963}

\bibitem[{{Bouwens} {et~al.}(2014){Bouwens}, {Illingworth}, {Oesch},
  {Labb{\'e}}, {van Dokkum}, {Trenti}, {Franx}, {Smit}, {Gonzalez}, \&
  {Magee}}]{Bouwens2014}
{Bouwens}, R.~J., {Illingworth}, G.~D., {Oesch}, P.~A., {et~al.} 2014, \apj,
  793, 115, \dodoi{10.1088/0004-637X/793/2/115}

\bibitem[{{Bouwens} {et~al.}(2021){Bouwens}, {Oesch}, {Stefanon},
  {Illingworth}, {Labb{\'e}}, {Reddy}, {Atek}, {Montes}, {Naidu},
  {Nanayakkara}, {Nelson}, \& {Wilkins}}]{Bouwens2021}
{Bouwens}, R.~J., {Oesch}, P.~A., {Stefanon}, M., {et~al.} 2021, \aj, 162, 47,
  \dodoi{10.3847/1538-3881/abf83e}

\bibitem[{Bradley {et~al.}(2023)Bradley, Sip{\H o}cz, Robitaille, Tollerud,
  Vin{\'{\i}}cius, Deil, Barbary, Wilson, Busko, Donath, G{\"u}nther, Cara,
  Lim, Me{\ss}linger, Conseil, Bostroem, Droettboom, Bray, Bratholm, Barentsen,
  Craig, Rathi, Pascual, Perren, Georgiev, de~Val-Borro, Kerzendorf, Bach,
  Quint, \& Souchereau}]{larry_bradley_2023_7946442}
Bradley, L., Sip{\H o}cz, B., Robitaille, T., {et~al.} 2023, astropy/photutils:
  1.8.0, 1.8.0,  Zenodo, \dodoi{10.5281/zenodo.7946442}

\bibitem[{{Bunker} {et~al.}(2023){Bunker}, {Saxena}, {Cameron}, {Willott},
  {Curtis-Lake}, {Jakobsen}, {Carniani}, {Smit}, {Maiolino}, {Witstok},
  {Curti}, {D'Eugenio}, {Jones}, {Ferruit}, {Arribas}, {Charlot}, {Chevallard},
  {Giardino}, {de Graaff}, {Looser}, {L{\"u}tzgendorf}, {Maseda}, {Rawle},
  {Rix}, {Del Pino}, {Alberts}, {Egami}, {Eisenstein}, {Endsley}, {Hainline},
  {Hausen}, {Johnson}, {Rieke}, {Rieke}, {Robertson}, {Shivaei}, {Stark},
  {Sun}, {Tacchella}, {Tang}, {Williams}, {Willmer}, {Baker}, {Baum},
  {Bhatawdekar}, {Bowler}, {Boyett}, {Chen}, {Circosta}, {Helton}, {Ji},
  {Kumari}, {Lyu}, {Nelson}, {Parlanti}, {Perna}, {Sandles}, {Scholtz},
  {Suess}, {Topping}, {{\"U}bler}, {Wallace}, \& {Whitler}}]{Bunker2023}
{Bunker}, A.~J., {Saxena}, A., {Cameron}, A.~J., {et~al.} 2023, \aap, 677, A88,
  \dodoi{10.1051/0004-6361/202346159}

\bibitem[{{Bushouse} {et~al.}(2022){Bushouse}, {Eisenhamer}, {Dencheva},
  {Davies}, {Greenfield}, {Morrison}, {Hodge}, {Simon}, {Grumm}, {Droettboom},
  {Slavich}, {Sosey}, {Pauly}, {Miller}, {Jedrzejewski}, {Hack}, {Davis},
  {Crawford}, {Law}, {Gordon}, {Regan}, {Cara}, {MacDonald}, {Bradley},
  {Shanahan}, \& {Jamieson}}]{Bushouse2022}
{Bushouse}, H., {Eisenhamer}, J., {Dencheva}, N., {et~al.} 2022,
  {spacetelescope/jwst: JWST 1.6.2}, 1.6.2,  Zenodo,
  \dodoi{10.5281/zenodo.6984366}

\bibitem[{{Calzetti} {et~al.}(2000){Calzetti}, {Armus}, {Bohlin}, {Kinney},
  {Koornneef}, \& {Storchi-Bergmann}}]{Calzetti2000}
{Calzetti}, D., {Armus}, L., {Bohlin}, R.~C., {et~al.} 2000, \apj, 533, 682,
  \dodoi{10.1086/308692}

\bibitem[{{Carnall} {et~al.}(2018){Carnall}, {McLure}, {Dunlop}, \&
  {Dav{\'e}}}]{Carnall2018}
{Carnall}, A.~C., {McLure}, R.~J., {Dunlop}, J.~S., \& {Dav{\'e}}, R. 2018,
  \mnras, 480, 4379, \dodoi{10.1093/mnras/sty2169}

\bibitem[{{Casey} {et~al.}(2023){Casey}, {Kartaltepe}, {Drakos}, {Franco},
  {Harish}, {Paquereau}, {Ilbert}, {Rose}, {Cox}, {Nightingale}, {Robertson},
  {Silverman}, {Koekemoer}, {Massey}, {McCracken}, {Rhodes}, {Akins}, {Allen},
  {Amvrosiadis}, {Arango-Toro}, {Bagley}, {Bongiorno}, {Capak}, {Champagne},
  {Chartab}, {Ch{\'a}vez Ortiz}, {Chworowsky}, {Cooke}, {Cooper}, {Darvish},
  {Ding}, {Faisst}, {Finkelstein}, {Fujimoto}, {Gentile}, {Gillman}, {Gould},
  {Gozaliasl}, {Hayward}, {He}, {Hemmati}, {Hirschmann}, {Jahnke}, {Jin},
  {Khostovan}, {Kokorev}, {Lambrides}, {Laigle}, {Larson}, {Leung}, {Liu},
  {Liaudat}, {Long}, {Magdis}, {Mahler}, {Mainieri}, {Manning}, {Maraston},
  {Martin}, {McCleary}, {McKinney}, {McPartland}, {Mobasher}, {Pattnaik},
  {Renzini}, {Rich}, {Sanders}, {Sattari}, {Scognamiglio}, {Scoville}, {Sheth},
  {Shuntov}, {Sparre}, {Suzuki}, {Talia}, {Toft}, {Trakhtenbrot}, {Urry},
  {Valentino}, {Vanderhoof}, {Vardoulaki}, {Weaver}, {Whitaker}, {Wilkins},
  {Yang}, \& {Zavala}}]{Casey2023_web}
{Casey}, C.~M., {Kartaltepe}, J.~S., {Drakos}, N.~E., {et~al.} 2023, \apj, 954,
  31, \dodoi{10.3847/1538-4357/acc2bc}

\bibitem[{{Cen} \& {Haiman}(2000)}]{Cen2000}
{Cen}, R., \& {Haiman}, Z. 2000, \apjl, 542, L75, \dodoi{10.1086/312937}

\bibitem[{{Chen} {et~al.}(2025){Chen}, {Stark}, {Mason}, {Tang}, {Whitler},
  {Lu}, \& {Topping}}]{Chen2025-Stark}
{Chen}, Z., {Stark}, D.~P., {Mason}, C.~A., {et~al.} 2025, arXiv e-prints,
  arXiv:2505.24080, \dodoi{10.48550/arXiv.2505.24080}

\bibitem[{{Chevallard} \& {Charlot}(2016)}]{Chevallard2016}
{Chevallard}, J., \& {Charlot}, S. 2016, \mnras, 462, 1415,
  \dodoi{10.1093/mnras/stw1756}

\bibitem[{{Chiang} {et~al.}(2017){Chiang}, {Overzier}, {Gebhardt}, \&
  {Henriques}}]{Chiang2017}
{Chiang}, Y.-K., {Overzier}, R.~A., {Gebhardt}, K., \& {Henriques}, B. 2017,
  \apjl, 844, L23, \dodoi{10.3847/2041-8213/aa7e7b}

\bibitem[{{Chisholm} {et~al.}(2022){Chisholm}, {Saldana-Lopez}, {Flury},
  {Schaerer}, {Jaskot}, {Amor{\'\i}n}, {Atek}, {Finkelstein}, {Fleming},
  {Ferguson}, {Fern{\'a}ndez}, {Giavalisco}, {Hayes}, {Heckman}, {Henry}, {Ji},
  {Marques-Chaves}, {Mauerhofer}, {McCandliss}, {Oey}, {{\"O}stlin},
  {Rutkowski}, {Scarlata}, {Thuan}, {Trebitsch}, {Wang}, {Worseck}, \&
  {Xu}}]{Chisholm2022}
{Chisholm}, J., {Saldana-Lopez}, A., {Flury}, S., {et~al.} 2022, \mnras, 517,
  5104, \dodoi{10.1093/mnras/stac2874}

\bibitem[{{Choustikov} {et~al.}(2024{\natexlab{a}}){Choustikov}, {Katz},
  {Saxena}, {Cameron}, {Devriendt}, {Slyz}, {Rosdahl}, {Blaizot}, \&
  {Michel-Dansac}}]{Choustikov2024_escape_physics}
{Choustikov}, N., {Katz}, H., {Saxena}, A., {et~al.} 2024{\natexlab{a}},
  \mnras, 529, 3751, \dodoi{10.1093/mnras/stae776}

\bibitem[{{Choustikov} {et~al.}(2024{\natexlab{b}}){Choustikov}, {Katz},
  {Saxena}, {Garel}, {Devriendt}, {Slyz}, {Kimm}, {Blaizot}, \&
  {Rosdahl}}]{Choustikov2024-lya}
---. 2024{\natexlab{b}}, \mnras, 532, 2463, \dodoi{10.1093/mnras/stae1586}

\bibitem[{{de Graaff} {et~al.}(2024){de Graaff}, {Rix}, {Carniani}, {Suess},
  {Charlot}, {Curtis-Lake}, {Arribas}, {Baker}, {Boyett}, {Bunker}, {Cameron},
  {Chevallard}, {Curti}, {Eisenstein}, {Franx}, {Hainline}, {Hausen}, {Ji},
  {Johnson}, {Jones}, {Maiolino}, {Maseda}, {Nelson}, {Parlanti}, {Rawle},
  {Robertson}, {Tacchella}, {{\"U}bler}, {Williams}, {Willmer}, \&
  {Willott}}]{deGraaff2024}
{de Graaff}, A., {Rix}, H.-W., {Carniani}, S., {et~al.} 2024, \aap, 684, A87,
  \dodoi{10.1051/0004-6361/202347755}

\bibitem[{{Dekel} {et~al.}(2023){Dekel}, {Sarkar}, {Birnboim}, {Mandelker}, \&
  {Li}}]{Dekel2023}
{Dekel}, A., {Sarkar}, K.~C., {Birnboim}, Y., {Mandelker}, N., \& {Li}, Z.
  2023, \mnras, 523, 3201, \dodoi{10.1093/mnras/stad1557}

\bibitem[{{Dijkstra}(2014)}]{Dijkstra2014}
{Dijkstra}, M. 2014, \pasa, 31, e040, \dodoi{10.1017/pasa.2014.33}

\bibitem[{{Dijkstra} {et~al.}(2011){Dijkstra}, {Mesinger}, \&
  {Wyithe}}]{Dijkstra2011}
{Dijkstra}, M., {Mesinger}, A., \& {Wyithe}, J. S.~B. 2011, \mnras, 414, 2139,
  \dodoi{10.1111/j.1365-2966.2011.18530.x}

\bibitem[{{Dressler}(1980)}]{Dressler1980}
{Dressler}, A. 1980, \apj, 236, 351, \dodoi{10.1086/157753}

\bibitem[{{Eldridge} {et~al.}(2017){Eldridge}, {Stanway}, {Xiao}, {McClelland
  }, {Taylor}, {Ng}, {Greis}, \& {Bray}}]{Eldridge2017}
{Eldridge}, J.~J., {Stanway}, E.~R., {Xiao}, L., {et~al.} 2017, \pasa, 34,
  e058, \dodoi{10.1017/pasa.2017.51}

\bibitem[{{Endsley} \& {Stark}(2022)}]{Endsley2022-bubble}
{Endsley}, R., \& {Stark}, D.~P. 2022, \mnras, 511, 6042,
  \dodoi{10.1093/mnras/stac524}

\bibitem[{{Endsley} {et~al.}(2021){Endsley}, {Stark}, {Charlot}, {Chevallard},
  {Robertson}, {Bouwens}, \& {Stefanon}}]{Endsley2021_massive_laes}
{Endsley}, R., {Stark}, D.~P., {Charlot}, S., {et~al.} 2021, \mnras, 502, 6044,
  \dodoi{10.1093/mnras/stab432}

\bibitem[{{Faisst} \& {Morishita}(2024)}]{Faisst2024}
{Faisst}, A.~L., \& {Morishita}, T. 2024, \apj, 971, 47,
  \dodoi{10.3847/1538-4357/ad58e2}

\bibitem[{{Finkelstein} {et~al.}(2019){Finkelstein}, {D'Aloisio},
  {Paardekooper}, {Ryan}, {Behroozi}, {Finlator}, {Livermore}, {Upton
  Sanderbeck}, {Dalla Vecchia}, \& {Khochfar}}]{Finkelstein2019}
{Finkelstein}, S.~L., {D'Aloisio}, A., {Paardekooper}, J.-P., {et~al.} 2019,
  \apj, 879, 36, \dodoi{10.3847/1538-4357/ab1ea8}

\bibitem[{{Flury} {et~al.}(2022){Flury}, {Jaskot}, {Ferguson}, {Worseck},
  {Makan}, {Chisholm}, {Saldana-Lopez}, {Schaerer}, {McCandliss}, {Xu}, {Wang},
  {Oey}, {Ford}, {Heckman}, {Ji}, {Giavalisco}, {Amor{\'\i}n}, {Atek},
  {Blaizot}, {Borthakur}, {Carr}, {Castellano}, {Barros}, {Dickinson},
  {Finkelstein}, {Fleming}, {Fontanot}, {Garel}, {Grazian}, {Hayes}, {Henry},
  {Mauerhofer}, {Micheva}, {Ostlin}, {Papovich}, {Pentericci}, {Ravindranath},
  {Rosdahl}, {Rutkowski}, {Santini}, {Scarlata}, {Teplitz}, {Thuan},
  {Trebitsch}, {Vanzella}, \& {Verhamme}}]{Flury2022}
{Flury}, S.~R., {Jaskot}, A.~E., {Ferguson}, H.~C., {et~al.} 2022, \apj, 930,
  126, \dodoi{10.3847/1538-4357/ac61e4}

\bibitem[{{Flury} {et~al.}(2024){Flury}, {Jaskot}, {Saldana-Lopez}, {Oey},
  {Chisholm}, {Amor{\'\i}n}, {Bait}, {Borthakur}, {Carr}, {Ferguson},
  {Giavalisco}, {Hayes}, {Heckman}, {Henry}, {Ji}, {Komarova}, {Leclercq}, {Le
  Reste}, {McCandliss}, {Marques-Chaves}, {{\"O}stlin}, {Pentericci},
  {Ravindranath}, {Rutkowski}, {Scarlata}, {Schaerer}, {Thuan}, {Trebitsch},
  {Vanzella}, {Verhamme}, {Wang}, {Worseck}, \& {Xu}}]{Flury2024}
{Flury}, S.~R., {Jaskot}, A.~E., {Saldana-Lopez}, A., {et~al.} 2024, arXiv
  e-prints, arXiv:2409.12118, \dodoi{10.48550/arXiv.2409.12118}

\bibitem[{{Furlanetto} \& {Oh}(2005)}]{Furlanetto2005}
{Furlanetto}, S.~R., \& {Oh}, S.~P. 2005, \mnras, 363, 1031,
  \dodoi{10.1111/j.1365-2966.2005.09505.x}

\bibitem[{{Furlanetto} {et~al.}(2004){Furlanetto}, {Zaldarriaga}, \&
  {Hernquist}}]{Furlanetto2004}
{Furlanetto}, S.~R., {Zaldarriaga}, M., \& {Hernquist}, L. 2004, \apj, 613, 1,
  \dodoi{10.1086/423025}

\bibitem[{{Garel} {et~al.}(2024){Garel}, {Michel-Dansac}, {Verhamme},
  {Mauerhofer}, {Katz}, {Blaizot}, {Leclercq}, \& {Salvignol}}]{Garel2024}
{Garel}, T., {Michel-Dansac}, L., {Verhamme}, A., {et~al.} 2024, \aap, 691,
  A213, \dodoi{10.1051/0004-6361/202450654}

\bibitem[{{Gnedin}(2000)}]{Gnedin2000}
{Gnedin}, N.~Y. 2000, \apj, 535, 530, \dodoi{10.1086/308876}

\bibitem[{{Gnedin} \& {Madau}(2022)}]{Gnedin_Madau2022}
{Gnedin}, N.~Y., \& {Madau}, P. 2022, Living Reviews in Computational
  Astrophysics, 8, 3, \dodoi{10.1007/s41115-022-00015-5}

\bibitem[{{Gnedin} \& {Prada}(2004)}]{Gnedin2004}
{Gnedin}, N.~Y., \& {Prada}, F. 2004, \apjl, 608, L77, \dodoi{10.1086/422390}

\bibitem[{{Gordon} {et~al.}(2003){Gordon}, {Clayton}, {Misselt}, {Land olt}, \&
  {Wolff}}]{Gordon2003}
{Gordon}, K.~D., {Clayton}, G.~C., {Misselt}, K.~A., {Land olt}, A.~U., \&
  {Wolff}, M.~J. 2003, \apj, 594, 279, \dodoi{10.1086/376774}

\bibitem[{{Greene} {et~al.}(2024){Greene}, {Labbe}, {Goulding}, {Furtak},
  {Chemerynska}, {Kokorev}, {Dayal}, {Volonteri}, {Williams}, {Wang}, {Setton},
  {Burgasser}, {Bezanson}, {Atek}, {Brammer}, {Cutler}, {Feldmann}, {Fujimoto},
  {Glazebrook}, {de Graaff}, {Khullar}, {Leja}, {Marchesini}, {Maseda},
  {Matthee}, {Miller}, {Naidu}, {Nanayakkara}, {Oesch}, {Pan}, {Papovich},
  {Price}, {van Dokkum}, {Weaver}, {Whitaker}, \& {Zitrin}}]{Greene2024}
{Greene}, J.~E., {Labbe}, I., {Goulding}, A.~D., {et~al.} 2024, \apj, 964, 39,
  \dodoi{10.3847/1538-4357/ad1e5f}

\bibitem[{{Greig} {et~al.}(2017){Greig}, {Mesinger}, {Haiman}, \&
  {Simcoe}}]{Greig2017}
{Greig}, B., {Mesinger}, A., {Haiman}, Z., \& {Simcoe}, R.~A. 2017, \mnras,
  466, 4239, \dodoi{10.1093/mnras/stw3351}

\bibitem[{{Gronke} \& {Dijkstra}(2016)}]{Gronke_Dijkstra2016}
{Gronke}, M., \& {Dijkstra}, M. 2016, \apj, 826, 14,
  \dodoi{10.3847/0004-637X/826/1/14}

\bibitem[{{Haiman}(2002)}]{Haiman2002}
{Haiman}, Z. 2002, \apjl, 576, L1, \dodoi{10.1086/343101}

\bibitem[{{Harikane} {et~al.}(2022){Harikane}, {Ono}, {Ouchi}, {Liu},
  {Sawicki}, {Shibuya}, {Behroozi}, {He}, {Shimasaku}, {Arnouts}, {Coupon},
  {Fujimoto}, {Gwyn}, {Huang}, {Inoue}, {Kashikawa}, {Komiyama}, {Matsuoka}, \&
  {Willott}}]{Harikane2022}
{Harikane}, Y., {Ono}, Y., {Ouchi}, M., {et~al.} 2022, \apjs, 259, 20,
  \dodoi{10.3847/1538-4365/ac3dfc}

\bibitem[{{Harikane} {et~al.}(2023){Harikane}, {Zhang}, {Nakajima}, {Ouchi},
  {Isobe}, {Ono}, {Hatano}, {Xu}, \& {Umeda}}]{Harikane2023}
{Harikane}, Y., {Zhang}, Y., {Nakajima}, K., {et~al.} 2023, \apj, 959, 39,
  \dodoi{10.3847/1538-4357/ad029e}

\bibitem[{{Harikane} {et~al.}(2024){Harikane}, {Inoue}, {Ellis}, {Ouchi},
  {Nakazato}, {Yoshida}, {Ono}, {Sun}, {Sato}, {Fujimoto}, {Kashikawa},
  {McLeod}, {Perez-Gonzalez}, {Sawicki}, {Sugahara}, {Xu}, {Yamanaka},
  {Carnall}, {Cullen}, {Dunlop}, {Egami}, {Grogin}, {Isobe}, {Koekemoer},
  {Laporte}, {Lee}, {Magee}, {Matsuo}, {Matsuoka}, {Mawatari}, {Nakajima},
  {Nakane}, {Tamura}, {Umeda}, \& {Yanagisawa}}]{Harikane2024_clumpy_compact}
{Harikane}, Y., {Inoue}, A.~K., {Ellis}, R.~S., {et~al.} 2024, arXiv e-prints,
  arXiv:2406.18352, \dodoi{10.48550/arXiv.2406.18352}

\bibitem[{{Hayes} {et~al.}(2011){Hayes}, {Schaerer}, {{\"O}stlin}, {Mas-Hesse},
  {Atek}, \& {Kunth}}]{Hayes2011}
{Hayes}, M., {Schaerer}, D., {{\"O}stlin}, G., {et~al.} 2011, \apj, 730, 8,
  \dodoi{10.1088/0004-637X/730/1/8}

\bibitem[{{Hayes} \& {Scarlata}(2023)}]{Hayes2023}
{Hayes}, M.~J., \& {Scarlata}, C. 2023, \apjl, 954, L14,
  \dodoi{10.3847/2041-8213/acee6a}

\bibitem[{{Henry} {et~al.}(2015){Henry}, {Scarlata}, {Martin}, \&
  {Erb}}]{Henry2015}
{Henry}, A., {Scarlata}, C., {Martin}, C.~L., \& {Erb}, D. 2015, \apj, 809, 19,
  \dodoi{10.1088/0004-637X/809/1/19}

\bibitem[{{Hopkins} \& {Hernquist}(2009)}]{Hopkins2009_BH}
{Hopkins}, P.~F., \& {Hernquist}, L. 2009, \apj, 698, 1550,
  \dodoi{10.1088/0004-637X/698/2/1550}

\bibitem[{{Hu} {et~al.}(2019){Hu}, {Wang}, {Zheng}, {Malhotra}, {Rhoads},
  {Infante}, {Barrientos}, {Yang}, {Jiang}, {Kang}, {Perez}, {Wold}, {Hibon},
  {Jiang}, {Khostovan}, {Valdes}, {Walker}, {Galaz}, {Coughlin}, {Harish},
  {Kong}, {Pharo}, \& {Zheng}}]{Hu2019}
{Hu}, W., {Wang}, J., {Zheng}, Z.-Y., {et~al.} 2019, \apj, 886, 90,
  \dodoi{10.3847/1538-4357/ab4cf4}

\bibitem[{{Hu} {et~al.}(2021){Hu}, {Wang}, {Infante}, {Rhoads}, {Zheng},
  {Yang}, {Malhotra}, {Barrientos}, {Jiang}, {Gonz{\'a}lez-L{\'o}pez},
  {Prieto}, {Perez}, {Hibon}, {Galaz}, {Coughlin}, {Harish}, {Kong}, {Kang},
  {Khostovan}, {Pharo}, {Valdes}, {Wold}, {Walker}, \& {Zheng}}]{Hu2021}
{Hu}, W., {Wang}, J., {Infante}, L., {et~al.} 2021, Nature Astronomy, 5, 485,
  \dodoi{10.1038/s41550-020-01291-y}

\bibitem[{{Hu} {et~al.}(2023){Hu}, {Martin}, {Gronke}, {Gazagnes}, {Hayes},
  {Chisholm}, {Heckman}, {Mingozzi}, {Roy}, {Senchyna}, {Xu}, {Berg}, {James},
  {Stark}, {Arellano-C{\'o}rdova}, {Henry}, {Jaskot}, {Kumari}, {Parker},
  {Scarlata}, {Wofford}, {Amor{\'\i}n}, {Leonhardes-Barboza}, {Brinchmann},
  {Carr}, \& {Aloisi}}]{Hu2023}
{Hu}, W., {Martin}, C.~L., {Gronke}, M., {et~al.} 2023, \apj, 956, 39,
  \dodoi{10.3847/1538-4357/aceefd}

\bibitem[{{Iliev} {et~al.}(2006){Iliev}, {Mellema}, {Pen}, {Merz}, {Shapiro},
  \& {Alvarez}}]{Iliev2006}
{Iliev}, I.~T., {Mellema}, G., {Pen}, U.~L., {et~al.} 2006, \mnras, 369, 1625,
  \dodoi{10.1111/j.1365-2966.2006.10502.x}

\bibitem[{{Itoh} {et~al.}(2018){Itoh}, {Ouchi}, {Zhang}, {Inoue}, {Mawatari},
  {Shibuya}, {Harikane}, {Ono}, {Kusakabe}, {Shimasaku}, {Fujimoto}, {Iwata},
  {Kajisawa}, {Kashikawa}, {Kawanomoto}, {Komiyama}, {Lee}, {Nagao}, \&
  {Taniguchi}}]{Itoh2018}
{Itoh}, R., {Ouchi}, M., {Zhang}, H., {et~al.} 2018, \apj, 867, 46,
  \dodoi{10.3847/1538-4357/aadfe4}

\bibitem[{{Iyer} {et~al.}(2019){Iyer}, {Gawiser}, {Faber}, {Ferguson},
  {Kartaltepe}, {Koekemoer}, {Pacifici}, \& {Somerville}}]{Iyer2019}
{Iyer}, K.~G., {Gawiser}, E., {Faber}, S.~M., {et~al.} 2019, \apj, 879, 116,
  \dodoi{10.3847/1538-4357/ab2052}

\bibitem[{{Izotov} {et~al.}(2016){Izotov}, {Schaerer}, {Thuan}, {Worseck},
  {Guseva}, {Orlitov{\'a}}, \& {Verhamme}}]{Izotov2016-leakers}
{Izotov}, Y.~I., {Schaerer}, D., {Thuan}, T.~X., {et~al.} 2016, \mnras, 461,
  3683, \dodoi{10.1093/mnras/stw1205}

\bibitem[{{Izotov} {et~al.}(2020){Izotov}, {Schaerer}, {Worseck}, {Verhamme},
  {Guseva}, {Thuan}, {Orlitov{\'a}}, \& {Fricke}}]{Izotov2020a}
{Izotov}, Y.~I., {Schaerer}, D., {Worseck}, G., {et~al.} 2020, \mnras, 491,
  468, \dodoi{10.1093/mnras/stz3041}

\bibitem[{{Izotov} {et~al.}(2024){Izotov}, {Thuan}, {Guseva}, {Schaerer},
  {Worseck}, \& {Verhamme}}]{Izotov2024}
{Izotov}, Y.~I., {Thuan}, T.~X., {Guseva}, N.~G., {et~al.} 2024, \mnras, 527,
  281, \dodoi{10.1093/mnras/stad3151}

\bibitem[{{Izotov} {et~al.}(2018){Izotov}, {Worseck}, {Schaerer}, {Guseva},
  {Thuan}, {Fricke}, \& {Orlitov{\'a}}}]{Izotov2018b}
{Izotov}, Y.~I., {Worseck}, G., {Schaerer}, D., {et~al.} 2018, \mnras, 478,
  4851, \dodoi{10.1093/mnras/sty1378}

\bibitem[{{Jaskot} {et~al.}(2019){Jaskot}, {Dowd}, {Oey}, {Scarlata}, \&
  {McKinney}}]{Jaskot2019}
{Jaskot}, A.~E., {Dowd}, T., {Oey}, M.~S., {Scarlata}, C., \& {McKinney}, J.
  2019, \apj, 885, 96, \dodoi{10.3847/1538-4357/ab3d3b}

\bibitem[{{Jaskot} \& {Oey}(2013)}]{Jaskot2013}
{Jaskot}, A.~E., \& {Oey}, M.~S. 2013, \apj, 766, 91,
  \dodoi{10.1088/0004-637X/766/2/91}

\bibitem[{{Jaskot} {et~al.}(2024){Jaskot}, {Silveyra}, {Plantinga}, {Flury},
  {Hayes}, {Chisholm}, {Heckman}, {Pentericci}, {Schaerer}, {Trebitsch},
  {Verhamme}, {Carr}, {Ferguson}, {Ji}, {Giavalisco}, {Henry},
  {Marques-Chaves}, {{\"O}stlin}, {Saldana-Lopez}, {Scarlata}, {Worseck}, \&
  {Xu}}]{Jaskot2024_LzLCS}
{Jaskot}, A.~E., {Silveyra}, A.~C., {Plantinga}, A., {et~al.} 2024, \apj, 972,
  92, \dodoi{10.3847/1538-4357/ad58b9}

\bibitem[{{Jones} {et~al.}(2024){Jones}, {Bunker}, {Saxena}, {Witstok},
  {Stark}, {Arribas}, {Baker}, {Bhatawdekar}, {Bowler}, {Boyett}, {Cameron},
  {Carniani}, {Charlot}, {Chevallard}, {Curti}, {Curtis-Lake}, {Eisenstein},
  {Hainline}, {Hausen}, {Ji}, {Johnson}, {Kumari}, {Looser}, {Maiolino},
  {Maseda}, {Parlanti}, {Rix}, {Robertson}, {Sandles}, {Scholtz}, {Smit},
  {Tacchella}, {{\"U}bler}, {Williams}, \& {Willott}}]{Jones2024}
{Jones}, G.~C., {Bunker}, A.~J., {Saxena}, A., {et~al.} 2024, \aap, 683, A238,
  \dodoi{10.1051/0004-6361/202347099}

\bibitem[{{Jones} {et~al.}(2025){Jones}, {Bunker}, {Saxena}, {Arribas},
  {Bhatawdekar}, {Boyett}, {Cameron}, {Carniani}, {Charlot}, {Curtis-Lake},
  {Hainline}, {Johnson}, {Kumari}, {Maseda}, {Rix}, {Robertson}, {Tacchella},
  {{\"U}bler}, {Williams}, {Willott}, {Witstok}, \& {Zhu}}]{Jones2025}
---. 2025, \mnras, 536, 2355, \dodoi{10.1093/mnras/stae2670}

\bibitem[{{Jung} {et~al.}(2024){Jung}, {Finkelstein}, {Arrabal Haro},
  {Dickinson}, {Ferguson}, {Hutchison}, {Kartaltepe}, {Larson}, {Simons},
  {Papovich}, {Park}, {Pentericci}, {Trump}, {Amor{\'\i}n}, {Backhaus},
  {Bagley}, {Casey}, {Cheng}, {Cleri}, {Cooper}, {Cooper}, {Gardner},
  {Gawiser}, {Grazian}, {Hathi}, {Hirschmann}, {Koekemoer}, {Lucas},
  {Mobasher}, {Pirzkal}, {Ravindranath}, {Straughn}, {Yung}, \& {de la
  Vega}}]{Jung2024_CEERS_laes}
{Jung}, I., {Finkelstein}, S.~L., {Arrabal Haro}, P., {et~al.} 2024, \apj, 967,
  73, \dodoi{10.3847/1538-4357/ad3913}

\bibitem[{{Juod{\v{z}}balis} {et~al.}(2025){Juod{\v{z}}balis}, {Maiolino},
  {Baker}, {Lake}, {Scholtz}, {D'Eugenio}, {Trefoloni}, {Isobe}, {Tacchella},
  {Bunker}, {Carniani}, {Charlot}, {Jones}, {Parlanti}, {Perna}, {Rinaldi},
  {Robertson}, {{\"U}bler}, {Venturi}, \& {Willott}}]{Juodzbalis2025}
{Juod{\v{z}}balis}, I., {Maiolino}, R., {Baker}, W.~M., {et~al.} 2025, arXiv
  e-prints, arXiv:2504.03551, \dodoi{10.48550/arXiv.2504.03551}

\bibitem[{{Kageura} {et~al.}(2025){Kageura}, {Ouchi}, {Nakane}, {Umeda},
  {Harikane}, {Yoshiura}, {Nakajima}, {Yajima}, \& {Thai}}]{Kageura2025}
{Kageura}, Y., {Ouchi}, M., {Nakane}, M., {et~al.} 2025, arXiv e-prints,
  arXiv:2501.05834, \dodoi{10.48550/arXiv.2501.05834}

\bibitem[{{Kakiichi} \& {Gronke}(2021)}]{Kakiichi2021}
{Kakiichi}, K., \& {Gronke}, M. 2021, \apj, 908, 30,
  \dodoi{10.3847/1538-4357/abc2d9}

\bibitem[{{Katz} {et~al.}(2023){Katz}, {Rosdahl}, {Kimm}, {Blaizot},
  {Choustikov}, {Farcy}, {Garel}, {Haehnelt}, {Michel-Dansac}, \&
  {Ocvirk}}]{Katz2023_sphinx20}
{Katz}, H., {Rosdahl}, J., {Kimm}, T., {et~al.} 2023, The Open Journal of
  Astrophysics, 6, 44, \dodoi{10.21105/astro.2309.03269}

\bibitem[{{Kennicutt} \& {Evans}(2012)}]{Kennicutt2012}
{Kennicutt}, R.~C., \& {Evans}, N.~J. 2012, \araa, 50, 531,
  \dodoi{10.1146/annurev-astro-081811-125610}

\bibitem[{{Kikuta} {et~al.}(2023){Kikuta}, {Ouchi}, {Shibuya}, {Liang},
  {Umeda}, {Matsumoto}, {Shimasaku}, {Harikane}, {Ono}, {Inoue}, {Yamanaka},
  {Kusakabe}, {Momose}, {Kashikawa}, {Matsuda}, \& {Lee}}]{Kikuta2023}
{Kikuta}, S., {Ouchi}, M., {Shibuya}, T., {et~al.} 2023, \apjs, 268, 24,
  \dodoi{10.3847/1538-4365/ace4cb}

\bibitem[{{Kocevski} {et~al.}(2024){Kocevski}, {Finkelstein}, {Barro},
  {Taylor}, {Calabr{\`o}}, {Laloux}, {Buchner}, {Trump}, {Leung}, {Yang},
  {Dickinson}, {P{\'e}rez-Gonz{\'a}lez}, {Pacucci}, {Inayoshi}, {Somerville},
  {McGrath}, {Akins}, {Bagley}, {Bisigello}, {Bowler}, {Carnall}, {Casey},
  {Cheng}, {Cleri}, {Costantin}, {Cullen}, {Davis}, {Donnan}, {Dunlop},
  {Ellis}, {Ferguson}, {Fujimoto}, {Fontana}, {Giavalisco}, {Grazian},
  {Grogin}, {Hathi}, {Hirschmann}, {Huertas-Company}, {Holwerda},
  {Illingworth}, {Juneau}, {Kartaltepe}, {Koekemoer}, {Li}, {Lucas}, {Magee},
  {Mason}, {McLeod}, {McLure}, {Napolitano}, {Papovich}, {Pirzkal},
  {Rodighiero}, {Santini}, {Wilkins}, \& {Yung}}]{Kocevski2024}
{Kocevski}, D.~D., {Finkelstein}, S.~L., {Barro}, G., {et~al.} 2024, arXiv
  e-prints, arXiv:2404.03576, \dodoi{10.48550/arXiv.2404.03576}

\bibitem[{{Kodama} {et~al.}(2001){Kodama}, {Smail}, {Nakata}, {Okamura}, \&
  {Bower}}]{Kodama2001}
{Kodama}, T., {Smail}, I., {Nakata}, F., {Okamura}, S., \& {Bower}, R.~G. 2001,
  \apjl, 562, L9, \dodoi{10.1086/338100}

\bibitem[{{Koekemoer} {et~al.}(2007){Koekemoer}, {Aussel}, {Calzetti}, {Capak},
  {Giavalisco}, {Kneib}, {Leauthaud}, {Le F{\`e}vre}, {McCracken}, {Massey},
  {Mobasher}, {Rhodes}, {Scoville}, \& {Shopbell}}]{Koekemoer2007}
{Koekemoer}, A.~M., {Aussel}, H., {Calzetti}, D., {et~al.} 2007, \apjs, 172,
  196, \dodoi{10.1086/520086}

\bibitem[{{Konno} {et~al.}(2016){Konno}, {Ouchi}, {Nakajima}, {Duval},
  {Kusakabe}, {Ono}, \& {Shimasaku}}]{Konno2016}
{Konno}, A., {Ouchi}, M., {Nakajima}, K., {et~al.} 2016, \apj, 823, 20,
  \dodoi{10.3847/0004-637X/823/1/20}

\bibitem[{{Koopmans} {et~al.}(2015){Koopmans}, {Pritchard}, {Mellema},
  {Aguirre}, {Ahn}, {Barkana}, {van Bemmel}, {Bernardi}, {Bonaldi}, {Briggs},
  {de Bruyn}, {Chang}, {Chapman}, {Chen}, {Ciardi}, {Dayal}, {Ferrara},
  {Fialkov}, {Fiore}, {Ichiki}, {Illiev}, {Inoue}, {Jelic}, {Jones}, {Lazio},
  {Maio}, {Majumdar}, {Mack}, {Mesinger}, {Morales}, {Parsons}, {Pen},
  {Santos}, {Schneider}, {Semelin}, {de Souza}, {Subrahmanyan}, {Takeuchi},
  {Vedantham}, {Wagg}, {Webster}, {Wyithe}, {Datta}, \& {Trott}}]{Koopmans2015}
{Koopmans}, L., {Pritchard}, J., {Mellema}, G., {et~al.} 2015, in Advancing
  Astrophysics with the Square Kilometre Array (AASKA14), 1,
  \dodoi{10.22323/1.215.0001}

\bibitem[{{Kornei} {et~al.}(2010){Kornei}, {Shapley}, {Erb}, {Steidel},
  {Reddy}, {Pettini}, \& {Bogosavljevi{\'c}}}]{Kornei2010}
{Kornei}, K.~A., {Shapley}, A.~E., {Erb}, D.~K., {et~al.} 2010, \apj, 711, 693,
  \dodoi{10.1088/0004-637X/711/2/693}

\bibitem[{{Kostyuk} \& {Ciardi}(2024)}]{Kostyuk2024}
{Kostyuk}, I., \& {Ciardi}, B. 2024, arXiv e-prints, arXiv:2412.04348,
  \dodoi{10.48550/arXiv.2412.04348}

\bibitem[{{Leitherer} {et~al.}(1999){Leitherer}, {Schaerer}, {Goldader},
  {Delgado}, {Robert}, {Kune}, {de Mello}, {Devost}, \&
  {Heckman}}]{Leitherer1999}
{Leitherer}, C., {Schaerer}, D., {Goldader}, J.~D., {et~al.} 1999, \apjs, 123,
  3, \dodoi{10.1086/313233}

\bibitem[{{Loeb} \& {Rybicki}(1999)}]{Loeb1999}
{Loeb}, A., \& {Rybicki}, G.~B. 1999, \apj, 524, 527, \dodoi{10.1086/307844}

\bibitem[{{Lu} {et~al.}(2024){Lu}, {Mason}, {Hutter}, {Mesinger}, {Qin},
  {Stark}, \& {Endsley}}]{Lu2024}
{Lu}, T.-Y., {Mason}, C.~A., {Hutter}, A., {et~al.} 2024, \mnras, 528, 4872,
  \dodoi{10.1093/mnras/stae266}

\bibitem[{{Lu} {et~al.}(2025){Lu}, {Mason}, {Mesinger}, {Prelogovi{\'c}},
  {Nikoli{\'c}}, {Hutter}, {Gagnon-Hartman}, {Tang}, {Qin}, \&
  {Kakiichi}}]{Lu2025}
{Lu}, T.-Y., {Mason}, C.~A., {Mesinger}, A., {et~al.} 2025, \aap, 697, A69,
  \dodoi{10.1051/0004-6361/202452912}

\bibitem[{{Madau} {et~al.}(2024){Madau}, {Giallongo}, {Grazian}, \&
  {Haardt}}]{Madau2024}
{Madau}, P., {Giallongo}, E., {Grazian}, A., \& {Haardt}, F. 2024, \apj, 971,
  75, \dodoi{10.3847/1538-4357/ad5ce8}

\bibitem[{{Malhotra} \& {Rhoads}(2006)}]{Malhotra2006}
{Malhotra}, S., \& {Rhoads}, J.~E. 2006, \apjl, 647, L95,
  \dodoi{10.1086/506983}

\bibitem[{{Martin} {et~al.}(2024){Martin}, {Peng}, \& {Li}}]{Martin2024}
{Martin}, C.~L., {Peng}, Z., \& {Li}, Y. 2024, \apj, 966, 190,
  \dodoi{10.3847/1538-4357/ad34ac}

\bibitem[{{Mascia} {et~al.}(2025){Mascia}, {Pentericci}, {Llerena},
  {Calabr{\`o}}, {Matthee}, {Flury}, {Pacucci}, {Jaskot}, {Amor{\'\i}n},
  {Bhatawdekar}, {Castellano}, {Cleri}, {Costantin}, {Davis}, {Di Cesare},
  {Dickinson}, {Fontana}, {Guo}, {Giavalisco}, {Holwerda}, {Hu},
  {Huertas-Company}, {Jung}, {Kartaltepe}, {Kashino}, {Koekemoer}, {Lucas},
  {Lotz}, {Napolitano}, {Jogee}, \& {Wilkins}}]{Mascia2025}
{Mascia}, S., {Pentericci}, L., {Llerena}, M., {et~al.} 2025, arXiv e-prints,
  arXiv:2501.08268, \dodoi{10.48550/arXiv.2501.08268}

\bibitem[{{Mason} \& {Gronke}(2020)}]{Mason2020}
{Mason}, C.~A., \& {Gronke}, M. 2020, \mnras

\bibitem[{{Mason} {et~al.}(2018){Mason}, {Treu}, {Dijkstra}, {Mesinger},
  {Trenti}, {Pentericci}, {de Barros}, \& {Vanzella}}]{Mason2018a}
{Mason}, C.~A., {Treu}, T., {Dijkstra}, M., {et~al.} 2018, \apj, 856, 2,
  \dodoi{10.3847/1538-4357/aab0a7}

\bibitem[{{Matthee} {et~al.}(2023){Matthee}, {Mackenzie}, {Simcoe}, {Kashino},
  {Lilly}, {Bordoloi}, \& {Eilers}}]{Matthee2023}
{Matthee}, J., {Mackenzie}, R., {Simcoe}, R.~A., {et~al.} 2023, \apj, 950, 67,
  \dodoi{10.3847/1538-4357/acc846}

\bibitem[{{Matthee} {et~al.}(2024){Matthee}, {Naidu}, {Brammer}, {Chisholm},
  {Eilers}, {Goulding}, {Greene}, {Kashino}, {Labbe}, {Lilly}, {Mackenzie},
  {Oesch}, {Weibel}, {Wuyts}, {Xiao}, {Bordoloi}, {Bouwens}, {van Dokkum},
  {Illingworth}, {Kramarenko}, {Maseda}, {Mason}, {Meyer}, {Nelson}, {Reddy},
  {Shivaei}, {Simcoe}, \& {Yue}}]{Matthee2024}
{Matthee}, J., {Naidu}, R.~P., {Brammer}, G., {et~al.} 2024, \apj, 963, 129,
  \dodoi{10.3847/1538-4357/ad2345}

\bibitem[{{Mazzolari} {et~al.}(2024){Mazzolari}, {{\"U}bler}, {Maiolino}, {Ji},
  {Nakajima}, {Feltre}, {Scholtz}, {D'Eugenio}, {Curti}, {Mignoli}, \&
  {Marconi}}]{Mazzolari2024}
{Mazzolari}, G., {{\"U}bler}, H., {Maiolino}, R., {et~al.} 2024, \aap, 691,
  A345, \dodoi{10.1051/0004-6361/202450407}

\bibitem[{{McCracken} {et~al.}(2015){McCracken}, {Wolk}, {Colombi},
  {Kilbinger}, {Ilbert}, {Peirani}, {Coupon}, {Dunlop}, {Milvang-Jensen},
  {Caputi}, {Aussel}, {B{\'e}thermin}, \& {Le F{\`e}vre}}]{McCracken2015}
{McCracken}, H.~J., {Wolk}, M., {Colombi}, S., {et~al.} 2015, \mnras, 449, 901,
  \dodoi{10.1093/mnras/stv305}

\bibitem[{{McKinney} {et~al.}(2025){McKinney}, {Cooper}, {Casey}, {Mu{\~n}oz},
  {Akins}, {Lambrides}, \& {Long}}]{McKinney2025}
{McKinney}, J., {Cooper}, O.~R., {Casey}, C.~M., {et~al.} 2025, \apjl, 985,
  L21, \dodoi{10.3847/2041-8213/add15d}

\bibitem[{{McQuinn} {et~al.}(2007){McQuinn}, {Hernquist}, {Zaldarriaga}, \&
  {Dutta}}]{McQuinn2007}
{McQuinn}, M., {Hernquist}, L., {Zaldarriaga}, M., \& {Dutta}, S. 2007, \mnras,
  381, 75, \dodoi{10.1111/j.1365-2966.2007.12085.x}

\bibitem[{{Mesinger} {et~al.}(2011){Mesinger}, {Furlanetto}, \&
  {Cen}}]{Mesinger2011}
{Mesinger}, A., {Furlanetto}, S., \& {Cen}, R. 2011, \mnras, 411, 955,
  \dodoi{10.1111/j.1365-2966.2010.17731.x}

\bibitem[{{Mesinger} \& {Furlanetto}(2008)}]{Mesinger2008a}
{Mesinger}, A., \& {Furlanetto}, S.~R. 2008, \mnras, 385, 1348,
  \dodoi{10.1111/j.1365-2966.2007.12836.x}

\bibitem[{{Morishita} {et~al.}(2023){Morishita}, {Roberts-Borsani}, {Treu},
  {Brammer}, {Mason}, {Trenti}, {Vulcani}, {Wang}, {Acebron}, {Bah{\'e}},
  {Bergamini}, {Boyett}, {Bradac}, {Calabr{\`o}}, {Castellano}, {Chen}, {De
  Lucia}, {Filippenko}, {Fontana}, {Glazebrook}, {Grillo}, {Henry}, {Jones},
  {Kelly}, {Koekemoer}, {Leethochawalit}, {Lu}, {Marchesini}, {Mascia},
  {Mercurio}, {Merlin}, {Metha}, {Nanayakkara}, {Nonino}, {Paris},
  {Pentericci}, {Rosati}, {Santini}, {Strait}, {Vanzella}, {Windhorst}, \&
  {Xie}}]{Morishita2023}
{Morishita}, T., {Roberts-Borsani}, G., {Treu}, T., {et~al.} 2023, \apjl, 947,
  L24, \dodoi{10.3847/2041-8213/acb99e}

\bibitem[{{Mukherjee} {et~al.}(2024){Mukherjee}, {Zafar}, {Nanayakkara},
  {Gupta}, {Gurung-Lopez}, {Battisti}, {Wisnioski}, {Foster}, {Mendel},
  {Harborne}, {Lagos}, {Kodama}, {Croom}, {Thater}, {Webb}, {Barsanti},
  {Sweet}, {Prathap}, {Valenzuela}, {Mailvaganam}, \& {Carrillo
  Martinez}}]{Mukherjee2024}
{Mukherjee}, T., {Zafar}, T., {Nanayakkara}, T., {et~al.} 2024, arXiv e-prints,
  arXiv:2410.17684, \dodoi{10.48550/arXiv.2410.17684}

\bibitem[{{Murphy} {et~al.}(2011){Murphy}, {Condon}, {Schinnerer}, {Kennicutt},
  {Calzetti}, {Armus}, {Helou}, {Turner}, {Aniano}, {Beir{\~a}o}, {Bolatto},
  {Brandl}, {Croxall}, {Dale}, {Donovan Meyer}, {Draine}, {Engelbracht},
  {Hunt}, {Hao}, {Koda}, {Roussel}, {Skibba}, \& {Smith}}]{Murphy2011}
{Murphy}, E.~J., {Condon}, J.~J., {Schinnerer}, E., {et~al.} 2011, \apj, 737,
  67, \dodoi{10.1088/0004-637X/737/2/67}

\bibitem[{{Naidu} {et~al.}(2020){Naidu}, {Tacchella}, {Mason}, {Bose}, {Oesch},
  \& {Conroy}}]{Naidu2020}
{Naidu}, R.~P., {Tacchella}, S., {Mason}, C.~A., {et~al.} 2020, \apj, 892, 109,
  \dodoi{10.3847/1538-4357/ab7cc9}

\bibitem[{{Nakane} {et~al.}(2024){Nakane}, {Ouchi}, {Nakajima}, {Harikane},
  {Ono}, {Umeda}, {Isobe}, {Zhang}, \& {Xu}}]{Nakane2024}
{Nakane}, M., {Ouchi}, M., {Nakajima}, K., {et~al.} 2024, \apj, 967, 28,
  \dodoi{10.3847/1538-4357/ad38c2}

\bibitem[{{Napolitano} {et~al.}(2024){Napolitano}, {Pentericci}, {Santini},
  {Calabr{\`o}}, {Mascia}, {Llerena}, {Castellano}, {Dickinson}, {Finkelstein},
  {Amor{\'\i}n}, {Arrabal Haro}, {Bagley}, {Bhatawdekar}, {Cleri}, {Davis},
  {Gardner}, {Gawiser}, {Giavalisco}, {Hathi}, {Holwerda}, {Hu}, {Jung},
  {Kartaltepe}, {Koekemoer}, {Larson}, {Merlin}, {Mobasher}, {Papovich},
  {Park}, {Pirzkal}, {Trump}, {Wilkins}, \& {Yung}}]{Napolitano2024}
{Napolitano}, L., {Pentericci}, L., {Santini}, P., {et~al.} 2024, \aap, 688,
  A106, \dodoi{10.1051/0004-6361/202449644}

\bibitem[{{Napolitano} {et~al.}(2025){Napolitano}, {Pentericci}, {Dickinson},
  {Arrabal Haro}, {Taylor}, {Calabr{\`o}}, {Bhagwat}, {Santini},
  {Arevalo-Gonzalez}, {Begley}, {Castellano}, {Ciardi}, {Donnan}, {Dottorini},
  {Dunlop}, {Finkelstein}, {Fontana}, {Giavalisco}, {Hirschmann}, {Jung},
  {Koekemoer}, {Kokorev}, {Llerena}, {Lucas}, {Mascia}, {Merlin},
  {P{\'e}rez-Gonz{\'a}lez}, {Stanton}, {Tripodi}, {Wang}, \&
  {Weiner}}]{Napolitano2025}
{Napolitano}, L., {Pentericci}, L., {Dickinson}, M., {et~al.} 2025, arXiv
  e-prints, arXiv:2508.14171, \dodoi{10.48550/arXiv.2508.14171}

\bibitem[{{Neufeld}(1991)}]{Neufeld1991}
{Neufeld}, D.~A. 1991, \apjl, 370, L85, \dodoi{10.1086/185983}

\bibitem[{{Neyer} {et~al.}(2024){Neyer}, {Smith}, {Kannan}, {Vogelsberger},
  {Garaldi}, {Gal{\'a}rraga-Espinosa}, {Borrow}, {Hernquist}, {Pakmor}, \&
  {Springel}}]{Neyer2024}
{Neyer}, M., {Smith}, A., {Kannan}, R., {et~al.} 2024, \mnras, 531, 2943,
  \dodoi{10.1093/mnras/stae1325}

\bibitem[{{Nikoli{\'c}} {et~al.}(2025){Nikoli{\'c}}, {Mesinger}, {Mason}, {Lu},
  {Tang}, {Prelogovi{\'c}}, {Gagnon-Hartman}, \& {Stark}}]{Nikolic2025}
{Nikoli{\'c}}, I., {Mesinger}, A., {Mason}, C.~A., {et~al.} 2025, arXiv
  e-prints, arXiv:2501.07980, \dodoi{10.48550/arXiv.2501.07980}

\bibitem[{{Osterbrock} \& {Ferland}(2006)}]{Osterbrock2006}
{Osterbrock}, D.~E., \& {Ferland}, G.~J. 2006, {Astrophysics of gaseous nebulae
  and active galactic nuclei}

\bibitem[{{Ouchi} {et~al.}(2010){Ouchi}, {Shimasaku}, {Furusawa}, {Saito},
  {Yoshida}, {Akiyama}, {Ono}, {Yamada}, {Ota}, {Kashikawa}, {Iye}, {Kodama},
  {Okamura}, {Simpson}, \& {Yoshida}}]{Ouchi2010}
{Ouchi}, M., {Shimasaku}, K., {Furusawa}, H., {et~al.} 2010, \apj, 723, 869,
  \dodoi{10.1088/0004-637X/723/1/869}

\bibitem[{{Ouchi} {et~al.}(2018){Ouchi}, {Harikane}, {Shibuya}, {Shimasaku},
  {Taniguchi}, {Konno}, {Kobayashi}, {Kajisawa}, {Nagao}, {Ono}, {Inoue},
  {Umemura}, {Mori}, {Hasegawa}, {Higuchi}, {Komiyama}, {Matsuda}, {Nakajima},
  {Saito}, \& {Wang}}]{Ouchi2018}
{Ouchi}, M., {Harikane}, Y., {Shibuya}, T., {et~al.} 2018, \pasj, 70, S13,
  \dodoi{10.1093/pasj/psx074}

\bibitem[{{Peng} {et~al.}(2002){Peng}, {Ho}, {Impey}, \&
  {Rix}}]{Peng2002_galfit}
{Peng}, C.~Y., {Ho}, L.~C., {Impey}, C.~D., \& {Rix}, H.-W. 2002, \aj, 124,
  266, \dodoi{10.1086/340952}

\bibitem[{{Peng} {et~al.}(2010){Peng}, {Ho}, {Impey}, \&
  {Rix}}]{Peng2010_galfit}
---. 2010, \aj, 139, 2097, \dodoi{10.1088/0004-6256/139/6/2097}

\bibitem[{{Peng} {et~al.}(2025){Peng}, {Martin}, {Chen}, {Fielding}, {Xu},
  {Heckman}, {Ramambason}, {Li}, {Carr}, {Hu}, {Chen}, {Scarlata}, \&
  {Henry}}]{Peng2025}
{Peng}, Z., {Martin}, C.~L., {Chen}, Z., {et~al.} 2025, \apj, 981, 171,
  \dodoi{10.3847/1538-4357/ada606}

\bibitem[{{Pentericci} {et~al.}(2014){Pentericci}, {Vanzella}, {Fontana},
  {Castellano}, {Treu}, {Mesinger}, {Dijkstra}, {Grazian}, {Brada{\v{c}}},
  {Conselice}, {Cristiani}, {Dunlop}, {Galametz}, {Giavalisco}, {Giallongo},
  {Koekemoer}, {McLure}, {Maiolino}, {Paris}, \& {Santini}}]{Pentericci2014}
{Pentericci}, L., {Vanzella}, E., {Fontana}, A., {et~al.} 2014, \apj, 793, 113,
  \dodoi{10.1088/0004-637X/793/2/113}

\bibitem[{{Perrin} {et~al.}(2015){Perrin}, {Long}, {Sivaramakrishnan},
  {Lajoie}, {Elliot}, {Pueyo}, \& {Albert}}]{Perrin2015}
{Perrin}, M.~D., {Long}, J., {Sivaramakrishnan}, A., {et~al.} 2015, {WebbPSF:
  James Webb Space Telescope PSF Simulation Tool}, Astrophysics Source Code
  Library, record ascl:1504.007

\bibitem[{{Planck Collaboration} {et~al.}(2020){Planck Collaboration},
  {Aghanim}, {Akrami}, {Ashdown}, {Aumont}, {Baccigalupi}, {Ballardini},
  {Banday}, {Barreiro}, {Bartolo}, {Basak}, {Battye}, {Benabed}, {Bernard},
  {Bersanelli}, {Bielewicz}, {Bock}, {Bond}, {Borrill}, {Bouchet}, {Boulanger},
  {Bucher}, {Burigana}, {Butler}, {Calabrese}, {Cardoso}, {Carron},
  {Challinor}, {Chiang}, {Chluba}, {Colombo}, {Combet}, {Contreras}, {Crill},
  {Cuttaia}, {de Bernardis}, {de Zotti}, {Delabrouille}, {Delouis}, {Di
  Valentino}, {Diego}, {Dor{\'e}}, {Douspis}, {Ducout}, {Dupac}, {Dusini},
  {Efstathiou}, {Elsner}, {En{\ss}lin}, {Eriksen}, {Fantaye}, {Farhang},
  {Fergusson}, {Fernandez-Cobos}, {Finelli}, {Forastieri}, {Frailis},
  {Fraisse}, {Franceschi}, {Frolov}, {Galeotta}, {Galli}, {Ganga},
  {G{\'e}nova-Santos}, {Gerbino}, {Ghosh}, {Gonz{\'a}lez-Nuevo}, {G{\'o}rski},
  {Gratton}, {Gruppuso}, {Gudmundsson}, {Hamann}, {Handley}, {Hansen},
  {Herranz}, {Hildebrandt}, {Hivon}, {Huang}, {Jaffe}, {Jones}, {Karakci},
  {Keih{\"a}nen}, {Keskitalo}, {Kiiveri}, {Kim}, {Kisner}, {Knox},
  {Krachmalnicoff}, {Kunz}, {Kurki-Suonio}, {Lagache}, {Lamarre}, {Lasenby},
  {Lattanzi}, {Lawrence}, {Le Jeune}, {Lemos}, {Lesgourgues}, {Levrier},
  {Lewis}, {Liguori}, {Lilje}, {Lilley}, {Lindholm}, {L{\'o}pez-Caniego},
  {Lubin}, {Ma}, {Mac{\'\i}as-P{\'e}rez}, {Maggio}, {Maino}, {Mandolesi},
  {Mangilli}, {Marcos-Caballero}, {Maris}, {Martin}, {Martinelli},
  {Mart{\'\i}nez-Gonz{\'a}lez}, {Matarrese}, {Mauri}, {McEwen}, {Meinhold},
  {Melchiorri}, {Mennella}, {Migliaccio}, {Millea}, {Mitra},
  {Miville-Desch{\^e}nes}, {Molinari}, {Montier}, {Morgante}, {Moss}, {Natoli},
  {N{\o}rgaard-Nielsen}, {Pagano}, {Paoletti}, {Partridge}, {Patanchon},
  {Peiris}, {Perrotta}, {Pettorino}, {Piacentini}, {Polastri}, {Polenta},
  {Puget}, {Rachen}, {Reinecke}, {Remazeilles}, {Renzi}, {Rocha}, {Rosset},
  {Roudier}, {Rubi{\~n}o-Mart{\'\i}n}, {Ruiz-Granados}, {Salvati}, {Sandri},
  {Savelainen}, {Scott}, {Shellard}, {Sirignano}, {Sirri}, {Spencer},
  {Sunyaev}, {Suur-Uski}, {Tauber}, {Tavagnacco}, {Tenti}, {Toffolatti},
  {Tomasi}, {Trombetti}, {Valenziano}, {Valiviita}, {Van Tent}, {Vibert},
  {Vielva}, {Villa}, {Vittorio}, {Wandelt}, {Wehus}, {White}, {White},
  {Zacchei}, \& {Zonca}}]{Planck2020}
{Planck Collaboration}, {Aghanim}, N., {Akrami}, Y., {et~al.} 2020, \aap, 641,
  A6, \dodoi{10.1051/0004-6361/201833910}

\bibitem[{{Prochaska} {et~al.}(2020{\natexlab{a}}){Prochaska}, {Hennawi},
  {Westfall}, {Cooke}, {Wang}, {Hsyu}, {Davies}, \& {Farina}}]{pypeit2005}
{Prochaska}, J.~X., {Hennawi}, J.~F., {Westfall}, K.~B., {et~al.}
  2020{\natexlab{a}}, arXiv e-prints, arXiv:2005.06505.
\newblock \doarXiv{2005.06505}

\bibitem[{{Prochaska} {et~al.}(2020{\natexlab{b}}){Prochaska}, {Hennawi},
  {Cooke}, {Westfall}, {Wang}, {EmAstro}, {Tiffanyhsyu}, {Wasserman},
  {Villaume}, {Marijana777}, {Schindler}, {Young}, {Simha}, {Wilde}, {Tejos},
  {Isbell}, {Fl{\"o}rs}, {Sandford}, {Vasovi{\'c}}, {Betts}, \&
  {Holden}}]{pypeit2020}
{Prochaska}, J.~X., {Hennawi}, J., {Cooke}, R., {et~al.} 2020{\natexlab{b}},
  {pypeit/PypeIt: Release 1.0.0}, v1.0.0,  Zenodo,
  \dodoi{10.5281/zenodo.3743493}

\bibitem[{{Qin} {et~al.}(2022){Qin}, {Wyithe}, {Oesch}, {Illingworth},
  {Leonova}, {Mutch}, \& {Naidu}}]{Qin2022}
{Qin}, Y., {Wyithe}, J. S.~B., {Oesch}, P.~A., {et~al.} 2022, \mnras, 510,
  3858, \dodoi{10.1093/mnras/stab3733}

\bibitem[{{Rauscher}(2024)}]{Rauscher2024}
{Rauscher}, B.~J. 2024, \pasp, 136, 015001, \dodoi{10.1088/1538-3873/ad1b36}

\bibitem[{{Rauscher} {et~al.}(2013){Rauscher}, {Arendt}, {Fixsen},
  {Greenhouse}, {Lander}, {Lindler}, {Loose}, {Moseley}, {Mott}, {Wen},
  {Wilson}, \& {Xenophontos}}]{Rauscher2013}
{Rauscher}, B.~J., {Arendt}, R.~G., {Fixsen}, D.~J., {et~al.} 2013, in Society
  of Photo-Optical Instrumentation Engineers (SPIE) Conference Series, Vol.
  8860, UV/Optical/IR Space Telescopes and Instruments: Innovative Technologies
  and Concepts VI, ed. H.~A. {MacEwen} \& J.~B. {Breckinridge}, 886005,
  \dodoi{10.1117/12.2025053}

\bibitem[{{Reddy} {et~al.}(2018){Reddy}, {Shapley}, {Sanders}, {Kriek}, {Coil},
  {Shivaei}, {Freeman}, {Mobasher}, {Siana}, {Azadi}, {Fetherolf}, {Fornasini},
  {Leung}, {Price}, {Zick}, \& {Barro}}]{Reddy2018}
{Reddy}, N.~A., {Shapley}, A.~E., {Sanders}, R.~L., {et~al.} 2018, \apj, 869,
  92, \dodoi{10.3847/1538-4357/aaed1e}

\bibitem[{{Rigby} {et~al.}(2023){Rigby}, {Lightsey}, {Garc{\'\i}a Mar{\'\i}n},
  {Bowers}, {Smith}, {Glasse}, {McElwain}, {Rieke}, {Chary}, {Liu}, {Clampin},
  {Kimble}, {Kinzel}, {Laidler}, {Mehalick}, {Noriega-Crespo}, {Shivaei},
  {Skelton}, {Stark}, {Temim}, {Wei}, \& {Willott}}]{Rigby2023}
{Rigby}, J.~R., {Lightsey}, P.~A., {Garc{\'\i}a Mar{\'\i}n}, M., {et~al.} 2023,
  \pasp, 135, 048002, \dodoi{10.1088/1538-3873/acbcf4}

\bibitem[{{Robertson}(2022)}]{Robertson2022}
{Robertson}, B.~E. 2022, \araa, 60, 121,
  \dodoi{10.1146/annurev-astro-120221-044656}

\bibitem[{{Rockosi} {et~al.}(2010){Rockosi}, {Stover}, {Kibrick}, {Lockwood},
  {Peck}, {Cowley}, {Bolte}, {Adkins}, {Alcott}, {Allen}, {Brown}, {Cabak},
  {Deich}, {Hilyard}, {Kassis}, {Lanclos}, {Lewis}, {Pfister}, {Phillips},
  {Robinson}, {Saylor}, {Thompson}, {Ward}, {Wei}, \& {Wright}}]{Rockosi2010}
{Rockosi}, C., {Stover}, R., {Kibrick}, R., {et~al.} 2010, in Society of
  Photo-Optical Instrumentation Engineers (SPIE) Conference Series, Vol. 7735,
  Ground-based and Airborne Instrumentation for Astronomy III, ed. I.~S.
  {McLean}, S.~K. {Ramsay}, \& H.~{Takami}, 77350R, \dodoi{10.1117/12.856818}

\bibitem[{{Santos}(2004)}]{Santos2004}
{Santos}, M.~R. 2004, \mnras, 349, 1137,
  \dodoi{10.1111/j.1365-2966.2004.07594.x}

\bibitem[{{Saxena} {et~al.}(2023){Saxena}, {Robertson}, {Bunker}, {Endsley},
  {Cameron}, {Charlot}, {Simmonds}, {Tacchella}, {Witstok}, {Willott},
  {Carniani}, {Curtis-Lake}, {Ferruit}, {Jakobsen}, {Arribas}, {Chevallard},
  {Curti}, {D'Eugenio}, {De Graaff}, {Jones}, {Looser}, {Maseda}, {Rawle},
  {Rix}, {Del Pino}, {Smit}, {{\"U}bler}, {Eisenstein}, {Hainline}, {Hausen},
  {Johnson}, {Rieke}, {Williams}, {Willmer}, {Baker}, {Bhatawdekar}, {Bowler},
  {Boyett}, {Chen}, {Egami}, {Ji}, {Kumari}, {Nelson}, {Perna}, {Sandles},
  {Scholtz}, \& {Shivaei}}]{Saxena2023}
{Saxena}, A., {Robertson}, B.~E., {Bunker}, A.~J., {et~al.} 2023, \aap, 678,
  A68, \dodoi{10.1051/0004-6361/202346245}

\bibitem[{{Saxena} {et~al.}(2024){Saxena}, {Bunker}, {Jones}, {Stark},
  {Cameron}, {Witstok}, {Arribas}, {Baker}, {Baum}, {Bhatawdekar}, {Bowler},
  {Boyett}, {Carniani}, {Charlot}, {Chevallard}, {Curti}, {Curtis-Lake},
  {Eisenstein}, {Endsley}, {Hainline}, {Helton}, {Johnson}, {Kumari}, {Looser},
  {Maiolino}, {Rieke}, {Rix}, {Robertson}, {Sandles}, {Simmonds}, {Smit},
  {Tacchella}, {Williams}, {Willmer}, \& {Willott}}]{Saxena2024}
{Saxena}, A., {Bunker}, A.~J., {Jones}, G.~C., {et~al.} 2024, \aap, 684, A84,
  \dodoi{10.1051/0004-6361/202347132}

\bibitem[{{Shapley} {et~al.}(2003){Shapley}, {Steidel}, {Pettini}, \&
  {Adelberger}}]{Shapley2003}
{Shapley}, A.~E., {Steidel}, C.~C., {Pettini}, M., \& {Adelberger}, K.~L. 2003,
  \apj, 588, 65, \dodoi{10.1086/373922}

\bibitem[{{Shen} {et~al.}(2024){Shen}, {Papovich}, {Matharu}, {Pirzkal}, {Hu},
  {Berg}, {Bagley}, {Backhaus}, {Cleri}, {Dickinson}, {Finkelstein}, {Hathi},
  {Huertas-Company}, {Hutchison}, {Giavalisco}, {Grogin}, {Jaskot}, {Jung},
  {Kartaltepe}, {Koekemoer}, {Lotz}, {P{\'e}rez-Gonz{\'a}lez}, {Rothberg},
  {Simons}, {Vanderhoof}, \& {Yung}}]{Shen2024}
{Shen}, L., {Papovich}, C., {Matharu}, J., {et~al.} 2024, arXiv e-prints,
  arXiv:2410.23349, \dodoi{10.48550/arXiv.2410.23349}

\bibitem[{{Sobacchi} \& {Mesinger}(2015)}]{Sobacchi2015}
{Sobacchi}, E., \& {Mesinger}, A. 2015, \mnras, 453, 1843,
  \dodoi{10.1093/mnras/stv1751}

\bibitem[{{Sobral} {et~al.}(2017){Sobral}, {Matthee}, {Best}, {Stroe},
  {R{\"o}ttgering}, {Oteo}, {Smail}, {Morabito}, \&
  {Paulino-Afonso}}]{Sobral2017}
{Sobral}, D., {Matthee}, J., {Best}, P., {et~al.} 2017, \mnras, 466, 1242,
  \dodoi{10.1093/mnras/stw3090}

\bibitem[{{Songaila} {et~al.}(2024){Songaila}, {Cowie}, {Barger}, {Hu}, \&
  {Taylor}}]{Songaila2024}
{Songaila}, A., {Cowie}, L.~L., {Barger}, A.~J., {Hu}, E.~M., \& {Taylor},
  A.~J. 2024, \apj, 971, 136, \dodoi{10.3847/1538-4357/ad5674}

\bibitem[{{Songaila} {et~al.}(2018){Songaila}, {Hu}, {Barger}, {Cowie},
  {Hasinger}, {Rosenwasser}, \& {Waters}}]{Songaila2018}
{Songaila}, A., {Hu}, E.~M., {Barger}, A.~J., {et~al.} 2018, \apj, 859, 91,
  \dodoi{10.3847/1538-4357/aac021}

\bibitem[{{Stanway} \& {Eldridge}(2018)}]{Stanway2018}
{Stanway}, E.~R., \& {Eldridge}, J.~J. 2018, \mnras, 479, 75,
  \dodoi{10.1093/mnras/sty1353}

\bibitem[{{Storey} \& {Hummer}(1995)}]{Hummer1995}
{Storey}, P.~J., \& {Hummer}, D.~G. 1995, \mnras, 272, 41,
  \dodoi{10.1093/mnras/272.1.41}

\bibitem[{{Tacchella} {et~al.}(2023){Tacchella}, {Eisenstein}, {Hainline},
  {Johnson}, {Baker}, {Helton}, {Robertson}, {Suess}, {Chen}, {Nelson},
  {Pusk{\'a}s}, {Sun}, {Alberts}, {Egami}, {Hausen}, {Rieke}, {Rieke},
  {Shivaei}, {Williams}, {Willmer}, {Bunker}, {Cameron}, {Carniani}, {Charlot},
  {Curti}, {Curtis-Lake}, {Looser}, {Maiolino}, {Maseda}, {Rawle}, {Rix},
  {Smit}, {{\"U}bler}, {Willott}, {Witstok}, {Baum}, {Bhatawdekar}, {Boyett},
  {Danhaive}, {de Graaff}, {Endsley}, {Ji}, {Lyu}, {Sandles}, {Saxena},
  {Scholtz}, {Topping}, \& {Whitler}}]{Tacchella2023}
{Tacchella}, S., {Eisenstein}, D.~J., {Hainline}, K., {et~al.} 2023, \apj, 952,
  74, \dodoi{10.3847/1538-4357/acdbc6}

\bibitem[{{Tang} {et~al.}(2025){Tang}, {Stark}, {Mason}, {Gelli}, {Chen}, \&
  {Topping}}]{Tang2025-z13}
{Tang}, M., {Stark}, D.~P., {Mason}, C.~A., {et~al.} 2025, arXiv e-prints,
  arXiv:2507.08245, \dodoi{10.48550/arXiv.2507.08245}

\bibitem[{{Tang} {et~al.}(2024){Tang}, {Stark}, {Topping}, {Mason}, \&
  {Ellis}}]{Tang2024}
{Tang}, M., {Stark}, D.~P., {Topping}, M.~W., {Mason}, C., \& {Ellis}, R.~S.
  2024, \apj, 975, 208, \dodoi{10.3847/1538-4357/ad7eb7}

\bibitem[{{Tang} {et~al.}(2023){Tang}, {Stark}, {Chen}, {Mason}, {Topping},
  {Endsley}, {Senchyna}, {Plat}, {Lu}, {Whitler}, {Robertson}, \&
  {Charlot}}]{Tang2023}
{Tang}, M., {Stark}, D.~P., {Chen}, Z., {et~al.} 2023, \mnras, 526, 1657,
  \dodoi{10.1093/mnras/stad2763}

\bibitem[{{Tasitsiomi}(2006)}]{Tasitsiomi2006}
{Tasitsiomi}, A. 2006, \apj, 645, 792, \dodoi{10.1086/504460}

\bibitem[{{Topping} {et~al.}(2024){Topping}, {Stark}, {Endsley}, {Whitler},
  {Hainline}, {Johnson}, {Robertson}, {Tacchella}, {Chen}, {Alberts}, {Baker},
  {Bunker}, {Carniani}, {Charlot}, {Chevallard}, {Curtis-Lake}, {DeCoursey},
  {Egami}, {Eisenstein}, {Ji}, {Maiolino}, {Williams}, {Willmer}, {Willott}, \&
  {Witstok}}]{Topping2024_beta}
{Topping}, M.~W., {Stark}, D.~P., {Endsley}, R., {et~al.} 2024, \mnras, 529,
  4087, \dodoi{10.1093/mnras/stae800}

\bibitem[{{Torralba-Torregrosa} {et~al.}(2024){Torralba-Torregrosa}, {Matthee},
  {Naidu}, {Mackenzie}, {Pezzulli}, {Hutter}, {Arnalte-Mur},
  {Gurung-L{\'o}pez}, {Tacchella}, {Oesch}, {Kashino}, {Conroy}, \&
  {Sobral}}]{Torralba-Torregrosa2024}
{Torralba-Torregrosa}, A., {Matthee}, J., {Naidu}, R.~P., {et~al.} 2024, \aap,
  689, A44, \dodoi{10.1051/0004-6361/202450318}

\bibitem[{{Umeda} {et~al.}(2024){Umeda}, {Ouchi}, {Nakajima}, {Harikane},
  {Ono}, {Xu}, {Isobe}, \& {Zhang}}]{Umeda2024}
{Umeda}, H., {Ouchi}, M., {Nakajima}, K., {et~al.} 2024, \apj, 971, 124,
  \dodoi{10.3847/1538-4357/ad554e}

\bibitem[{{Veilleux} \& {Osterbrock}(1987)}]{Veilleux1987}
{Veilleux}, S., \& {Osterbrock}, D.~E. 1987, \apjs, 63, 295,
  \dodoi{10.1086/191166}

\bibitem[{{Verhamme} {et~al.}(2017){Verhamme}, {Orlitov{\'a}}, {Schaerer},
  {Izotov}, {Worseck}, {Thuan}, \& {Guseva}}]{Verhamme2017}
{Verhamme}, A., {Orlitov{\'a}}, I., {Schaerer}, D., {et~al.} 2017, \aap, 597,
  A13, \dodoi{10.1051/0004-6361/201629264}

\bibitem[{{Verhamme} {et~al.}(2008){Verhamme}, {Schaerer}, {Atek}, \&
  {Tapken}}]{Verhamme2008}
{Verhamme}, A., {Schaerer}, D., {Atek}, H., \& {Tapken}, C. 2008, \aap, 491,
  89, \dodoi{10.1051/0004-6361:200809648}

\bibitem[{{Verhamme} {et~al.}(2006){Verhamme}, {Schaerer}, \&
  {Maselli}}]{Verhamme2006}
{Verhamme}, A., {Schaerer}, D., \& {Maselli}, A. 2006, \aap, 460, 397,
  \dodoi{10.1051/0004-6361:20065554}

\bibitem[{{Verhamme} {et~al.}(2018){Verhamme}, {Garel}, {Ventou}, {Contini},
  {Bouch{\'e}}, {Herenz}, {Richard}, {Bacon}, {Schmidt}, {Maseda}, {Marino},
  {Brinchmann}, {Cantalupo}, {Caruana}, {Cl{\'e}ment}, {Diener}, {Drake},
  {Hashimoto}, {Inami}, {Kerutt}, {Kollatschny}, {Leclercq}, {Patr{\'\i}cio},
  {Schaye}, {Wisotzki}, \& {Zabl}}]{Verhamme2018}
{Verhamme}, A., {Garel}, T., {Ventou}, E., {et~al.} 2018, \mnras, 478, L60,
  \dodoi{10.1093/mnrasl/sly058}

\bibitem[{{Verner} {et~al.}(1996){Verner}, {Ferland}, {Korista}, \&
  {Yakovlev}}]{Verner1996}
{Verner}, D.~A., {Ferland}, G.~J., {Korista}, K.~T., \& {Yakovlev}, D.~G. 1996,
  \apj, 465, 487, \dodoi{10.1086/177435}

\bibitem[{Virtanen {et~al.}(2020)Virtanen, Gommers, Oliphant, Haberland, Reddy,
  Cournapeau, Burovski, Peterson, Weckesser, Bright, {van der Walt}, Brett,
  Wilson, Millman, Mayorov, Nelson, Jones, Kern, Larson, Carey, Polat, Feng,
  Moore, {VanderPlas}, Laxalde, Perktold, Cimrman, Henriksen, Quintero, Harris,
  Archibald, Ribeiro, Pedregosa, {van Mulbregt}, \& {SciPy 1.0
  Contributors}}]{SciPy-NMeth2020}
Virtanen, P., Gommers, R., Oliphant, T.~E., {et~al.} 2020, Nature Methods, 17,
  261, \dodoi{10.1038/s41592-019-0686-2}

\bibitem[{{Vitte} {et~al.}(2025){Vitte}, {Verhamme}, {Hibon}, {Leclercq},
  {Alcalde Pampliega}, {Kerutt}, {Kusakabe}, {Matthee}, {Guo}, {Bacon},
  {Maseda}, {Richard}, {Pharo}, {Schaye}, {Boogaard}, {Nanayakkara}, \&
  {Contini}}]{Vitte2025}
{Vitte}, E., {Verhamme}, A., {Hibon}, P., {et~al.} 2025, \aap, 694, A100,
  \dodoi{10.1051/0004-6361/202450426}

\bibitem[{{Witstok} {et~al.}(2024{\natexlab{a}}){Witstok}, {Smit}, {Saxena},
  {Jones}, {Helton}, {Sun}, {Maiolino}, {Kumari}, {Stark}, {Bunker}, {Arribas},
  {Baker}, {Bhatawdekar}, {Boyett}, {Cameron}, {Carniani}, {Charlot},
  {Chevallard}, {Curti}, {Curtis-Lake}, {Eisenstein}, {Endsley}, {Hainline},
  {Ji}, {Johnson}, {Looser}, {Nelson}, {Perna}, {Rix}, {Robertson}, {Sandles},
  {Scholtz}, {Simmonds}, {Tacchella}, {{\"U}bler}, {Williams}, {Willmer}, \&
  {Willott}}]{Witstok2024}
{Witstok}, J., {Smit}, R., {Saxena}, A., {et~al.} 2024{\natexlab{a}}, \aap,
  682, A40, \dodoi{10.1051/0004-6361/202347176}

\bibitem[{{Witstok} {et~al.}(2024{\natexlab{b}}){Witstok}, {Maiolino}, {Smit},
  {Jones}, {Bunker}, {Helton}, {Johnson}, {Tacchella}, {Saxena}, {Arribas},
  {Bhatawdekar}, {Boyett}, {Cameron}, {Cargile}, {Carniani}, {Charlot},
  {Chevallard}, {Curti}, {Curtis-Lake}, {D'Eugenio}, {Eisenstein}, {Hainline},
  {Hausen}, {Kumari}, {Laseter}, {Maseda}, {Rieke}, {Robertson}, {Scholtz},
  {Shivaei}, {Williams}, {Willmer}, \& {Willott}}]{Witstok2024-z9}
{Witstok}, J., {Maiolino}, R., {Smit}, R., {et~al.} 2024{\natexlab{b}}, arXiv
  e-prints, arXiv:2404.05724, \dodoi{10.48550/arXiv.2404.05724}

\bibitem[{{Witstok} {et~al.}(2025){Witstok}, {Jakobsen}, {Maiolino}, {Helton},
  {Johnson}, {Robertson}, {Tacchella}, {Cameron}, {Smit}, {Bunker}, {Saxena},
  {Sun}, {Alberts}, {Arribas}, {Baker}, {Bhatawdekar}, {Boyett}, {Cargile},
  {Carniani}, {Charlot}, {Chevallard}, {Curti}, {Curtis-Lake}, {D'Eugenio},
  {Eisenstein}, {Hainline}, {Jones}, {Kumari}, {Maseda},
  {P{\'e}rez-Gonz{\'a}lez}, {Rinaldi}, {Scholtz}, {{\"U}bler}, {Williams},
  {Willmer}, {Willott}, \& {Zhu}}]{Witstok2024-z13}
{Witstok}, J., {Jakobsen}, P., {Maiolino}, R., {et~al.} 2025, \nat, 639, 897,
  \dodoi{10.1038/s41586-025-08779-5}

\bibitem[{{Witten} {et~al.}(2024){Witten}, {Laporte}, {Martin-Alvarez},
  {Sijacki}, {Yuan}, {Haehnelt}, {Baker}, {Dunlop}, {Ellis}, {Grogin},
  {Illingworth}, {Katz}, {Koekemoer}, {Magee}, {Maiolino}, {McClymont},
  {P{\'e}rez-Gonz{\'a}lez}, {Pusk{\'a}s}, {Roberts-Borsani}, {Santini}, \&
  {Simmonds}}]{Witten2024-NatAs}
{Witten}, C., {Laporte}, N., {Martin-Alvarez}, S., {et~al.} 2024, Nature
  Astronomy, 8, 384, \dodoi{10.1038/s41550-023-02179-3}

\bibitem[{{Witten} {et~al.}(2023){Witten}, {Laporte}, \&
  {Katz}}]{Witten2023_low_fesc}
{Witten}, C. E.~C., {Laporte}, N., \& {Katz}, H. 2023, \apj, 944, 61,
  \dodoi{10.3847/1538-4357/acac9d}

\bibitem[{{Wold} {et~al.}(2022){Wold}, {Malhotra}, {Rhoads}, {Wang}, {Hu},
  {Perez}, {Zheng}, {Khostovan}, {Walker}, {Barrientos},
  {Gonz{\'a}lez-L{\'o}pez}, {Harish}, {Infante}, {Jiang}, {Pharo},
  {Moya-Sierralta}, {Bauer}, {Galaz}, {Valdes}, \& {Yang}}]{Wold2022}
{Wold}, I. G.~B., {Malhotra}, S., {Rhoads}, J., {et~al.} 2022, \apj, 927, 36,
  \dodoi{10.3847/1538-4357/ac4997}

\bibitem[{{Wyithe} \& {Loeb}(2007)}]{Wyithe2007}
{Wyithe}, J. S.~B., \& {Loeb}, A. 2007, \mnras, 375, 1034,
  \dodoi{10.1111/j.1365-2966.2007.11366.x}

\bibitem[{{Xiao} {et~al.}(2018){Xiao}, {Stanway}, \& {Eldridge}}]{Xiao2018}
{Xiao}, L., {Stanway}, E.~R., \& {Eldridge}, J.~J. 2018, \mnras, 477, 904,
  \dodoi{10.1093/mnras/sty646}

\bibitem[{{Yajima} {et~al.}(2018){Yajima}, {Sugimura}, \&
  {Hasegawa}}]{Yajima2018}
{Yajima}, H., {Sugimura}, K., \& {Hasegawa}, K. 2018, \mnras, 477, 5406,
  \dodoi{10.1093/mnras/sty997}

\bibitem[{{Yang} {et~al.}(2017){Yang}, {Malhotra}, {Gronke}, {Rhoads},
  {Leitherer}, {Wofford}, {Jiang}, {Dijkstra}, {Tilvi}, \&
  {Wang}}]{Yang2017_esc}
{Yang}, H., {Malhotra}, S., {Gronke}, M., {et~al.} 2017, \apj, 844, 171,
  \dodoi{10.3847/1538-4357/aa7d4d}

\bibitem[{{Yuan} {et~al.}(2025){Yuan}, {Martin-Alvarez}, {Haehnelt}, {Garel},
  {Keating}, {Witstok}, \& {Sijacki}}]{Yuan2025}
{Yuan}, Y., {Martin-Alvarez}, S., {Haehnelt}, M.~G., {et~al.} 2025, \mnras,
  542, 762, \dodoi{10.1093/mnras/staf1252}

\bibitem[{{Yung} {et~al.}(2020){Yung}, {Somerville}, {Popping}, \&
  {Finkelstein}}]{Yung2020c}
{Yung}, L.~Y.~A., {Somerville}, R.~S., {Popping}, G., \& {Finkelstein}, S.~L.
  2020, \mnras, 494, 1002, \dodoi{10.1093/mnras/staa714}

\bibitem[{{Yung} {et~al.}(2023){Yung}, {Somerville}, {Finkelstein}, {Behroozi},
  {Dav{\'e}}, {Ferguson}, {Gardner}, {Popping}, {Malhotra}, {Papovich},
  {Rhoads}, {Bagley}, {Hirschmann}, \& {Koekemoer}}]{Yung2023}
{Yung}, L.~Y.~A., {Somerville}, R.~S., {Finkelstein}, S.~L., {et~al.} 2023,
  \mnras, 519, 1578, \dodoi{10.1093/mnras/stac3595}

\bibitem[{{Zheng} {et~al.}(2017){Zheng}, {Wang}, {Rhoads}, {Infante},
  {Malhotra}, {Hu}, {Walker}, {Jiang}, {Jiang}, {Hibon}, {Gonzalez}, {Kong},
  {Zheng}, {Galaz}, \& {Barrientos}}]{Zheng2017}
{Zheng}, Z.-Y., {Wang}, J., {Rhoads}, J., {et~al.} 2017, \apjl, 842, L22,
  \dodoi{10.3847/2041-8213/aa794f}

\bibitem[{{Zier} {et~al.}(2025){Zier}, {Kannan}, {Smith}, {Puchwein},
  {Vogelsberger}, {Borrow}, {Garaldi}, {Keating}, {McClymont}, {Shen}, \&
  {Hernquist}}]{Zier2025}
{Zier}, O., {Kannan}, R., {Smith}, A., {et~al.} 2025, arXiv e-prints,
  arXiv:2503.02927, \dodoi{10.48550/arXiv.2503.02927}

\end{thebibliography}
\bibliographystyle{aasjournal}

\allauthors

\end{document}